# THZ METAMATERIALS AND PLASMONICS USING 2D MATERIALS & HIGHLY CONDUCTIVE TRANSPARENT OXIDES

by

Sara Arezoomandan

A dissertation submitted to the faculty of
The University of Utah
in partial fulfillment of the requirements for the degree of

Doctor of Philosophy

Department of Electrical Engineering

The University of Utah

November 2017

## The University of Utah Graduate School

# STATEMENT OF DISSERTATION APPROVAL

The dissertation of **Sara Arezoomandan**

has been approved by the following supervisory committee members:

| | | |
|---|---|---|
| **Berardi Sensale Rodriguez** | , Chair | **11/29/2017**<br>Date Approved |
| **Ajay Nahata** | , Member | **12/13/2017**<br>Date Approved |
| **David Schurig** | , Member | **11/29/2017**<br>Date Approved |
| **Steve Blair** | , Member | **11/29/2017**<br>Date Approved |
| **Ashotosh Tiwari** | , Member | **11/29/2017**<br>Date Approved |

and by **Florian Solzbacher** , Chair/Dean of

the Department/College/School of **Electrical and Computer Engineering**

and by David B. Kieda, Dean of The Graduate School.

ABSTRACT

Driven by a myriad of potential applications such as communications, medical imaging, security, spectroscopy, and so on, terahertz (THz) technology has emerged as a rapidly growing technological field during the last three decades. However, since conventional materials typically used in microwave and optical frequencies are lossy or do not effectively respond at these frequencies, it is essential to find or develop novel materials that are suitable for device applications in the THz range. Therefore, there is wide interest in the community in employing novel naturally-occurring materials, such as 2D materials, as well as in designing artificial metamaterial structures for THz applications. Here, we combined both of these approaches so to develop reconfigurable THz devices capable of providing amplitude modulation, phase modulation, and resonance frequency tuning.

First, graphene is employed as the reconfigurable element in metamaterial phase modulators. For this purpose, we propose the use of unit-cells with deep-subwavelength dimensions, which can have multiple advantaged for beam shaping applications. The analyzed metamaterials have one of the smallest unit-cell to wavelength ratio reported or proposed to-date at THz frequencies. By systematic analysis of the geometrical tradeoffs in these devices it is found that there is an optimal unit cell dimension, corresponding roughly to $\sim\lambda/20$, which can deliver the best performance. In addition to this, we explored other applications of graphene in metamaterial devices, including amplitude modulation and resonance-shifting. These studies motivated us to analyze on what is the most suitable role of graphene from a THz device perspective: is graphene a good plasmonic material? Or it is better suited as a reconfigurable material providing tunability to otherwise passive metallic structures? Our studies show that the Drude scattering time in graphene is an important parameter in this regard. In order to attain strong plasmonic resonances graphene samples with $\tau >> 1$ps are required, which is challenging in large area CVD samples. But graphene is just one example of a wider class of 2D materials. In this work we also studied for the first time the application of 2D materials beyond graphene as reconfigurable elements in THz devices. For this purpose, Molybdenum Disulfide ($MoS_2$) was employed as the reconfigurable element in cross-slot metamaterial amplitude modulators. Our results evidence that smaller insertion loss is possible when employing 2D materials with a bandgap, such as $MoS_2$, rather than a zero-gap material such as graphene. Furthermore, because of a stronger optical absorption active control of the metamaterial properties is possible by altering the intensity of an optical pump.

We later investigate and discuss transparent conductive oxides (TCOs), which constitute an interesting choice for developing visible-transparent THz-functional metamaterial devices for THz applications. These materials show a metallic THz response thus can substitute the metal patterns in metamaterial devices. In our particular studies we analyzed samples consisting of: (i) two-dimensional electron gases at the interface between polar/non-polar complex oxides having record-high electron density, and (ii) thin-films of La-doped $BaSnO_3$ having record-high conductivity in a TCO. These materials exhibit a flat THz conductivity across a broad terahertz frequency window. As a result of their metal-like broadband THz response, we demonstrate a visible-transparent THz-functional electromagnetic structure consisting of a wire-grid polarizer.

TABLE OF CONTENTS

ACKNOWLEDGMENT

This journey would not have been possible without the support of my family, professors, and friends. To my family, thank you for encouraging me in all of my pursuits and inspiring me to follow my dreams, although staying away from you was the hardest part of this journey. I am especially grateful to my beloved husband, who supported me to do what I love to and was my companion in this journey. To my mom and dad, I always knew that you believed in me and wanted the best for me. To my sisters and brother that I missed in all these years.

I would like to express the deepest appreciation to my advisor and my committee chair Professor Berardi Sensale-Rodriguez, he continually and convincingly inspires me in regard to research and critical thinking. Without his guidance and persistent help this dissertation would not have been possible. He also was a great and supportive friend of me and help me in all these years.

I would like to thank my committee members David Schurig, Ajay Nahata, Steve Blair and Ashoutosh Tiwari, for their support and useful discussion. Also I would like to thanks Bharat Jalan and his students who closely work with us and provide samples and useful discussions. Also I would like to thanks my colleagues Kai Yang, Hugo Condori, Prashanth Golpan, and Ashish Chanana, without their spirit of teamwork this dissertation would not be possible.

I would also like to thank to MRSEC for their financial support granted through my graduate study at University of Utah.

# CHAPTER 1

# INTRODUCTION

## 1.1 The THz spectral range

The THz spectral range, also referred as sub-millimeter and far-infrared range, consists of electromagnetic waves with frequencies ranging from 300 GHz to 3 THz. Wavelengths in this spectral range correspondingly span from 1000 μm to 100 μm with respective photon energies ranging from 1.24 meV to 12.4 meV. THz electromagnetic waves are emitted as a part of black-body radiation from any objects with temperature larger than 10K [1]. Historically, the THz frequency range has been referred to as the "THz gap" [2], owing to the fact that for a long-time exploration of this spectral region was scarce, and the interaction of THz electromagnetic waves with different materials was a mystery. Furthermore, during many decades there was an absence of efficient THz generation and detection techniques. The main motivations and driving forces for exploring this spectral range were astronomy and chemistry. But after the introduction of efficient room-temperature THz sources and detectors over the last three decades of the 20th century, new applications have emerged, such as spectroscopy, medical imaging, security, communications, etc., in which THz wavelengths can offer some remarkable advantages. In addition to its practical applications, the THz spectral range also appears as an interesting subject in fundamental science to study light-matter interaction. After developing functional sources and detectors for THz spectroscopy, physicists and materials scientist have also used THz spectroscopy as a non-contact, non-invasive method to reveal many important low-energy physical properties of condensed matter.

As mentioned, astronomy was one of the first motivations for researchers to develop THz systems. Interstellar dusts sizes are in the range of 100 μm to 1 mm, thus associated with THz wavelengths [1]. In addition, the black-body radiation from interstellar particles with temperatures between 10-20K (with the cosmic background noise of 2.7K) occurs in the THz frequency range. Therefore, THz detection will improve our understanding of cosmos. Studies of the spectral energy distribution in galaxies show that half of the total luminosity and 98% of the emitted photons since the big bang are located in the THz spectral range [3]. Consequently, for achieving a comprehensive understanding of any cosmological phenomena, we are in need of proper THz detection mechanisms. Furthermore, many abundant molecules like, oxygen, water, carbon monoxide and nitrogen, have absorption lines in the THz spectral region [1]. Probing these molecules in regions of the universe where star formation occurs is essential for observation of the star formation phenomena in early stages. In addition, with regards to our planet, THz probing can be used to observe thermal absorption of the tropospheric layer and study ozone destruction, global warming, and other atmospheric phenomena [4].

Spectroscopy has been another primary application field motivating the development of THz technology; this is a result of the strong emission and absorption features associated with vibrational, rotational, torsional, phonon, inter-molecular and intra-molecular modes present in this spectral range [5]. For example, DNA molecules show phonon absorption in this frequency range, which motivates THz studies on detection of DNA molecule signatures [6], [7]. It is worth mentioning that far-IR spectroscopy is a technique that has been well established for a long time, however THz spectroscopy is different from far-IR spectroscopy in which black-body radiation and photon detection are used as methods of generation and detection. In this aspect, it is to note that the signal to noise ratio in conventional far-IR spectroscopy will decrease at longer wavelengths. In THz spectroscopy, bright sources and different detection techniques are used which improves the signal to noise ratio when compared to far-IR spectroscopy in longer wavelengths. Time domain spectroscopy (TDS) is a technique that is widely used for spectroscopy in the THz region of the spectrum. Recent developments in this technique have led to commercially available systems. In addition, availably of compact and portable TDS systems have popularized the use of this technique for chemical analysis. Due to the low photon energy and non-ionizing nature of this probing method, it is a safe, non-invasive, non-destructive technique to characterize chemical compounds, which make it THz spectroscopy attractive in drug and food industry, as well as for security applications [8].

THz radiation can penetrate through normal clothing and usual packaging with tolerable attenuation, so it is a feasible choice for security purposes. In this regard, currently, the majority of security systems in airports are based on millimeter wave technologies. By integrating THz spectroscopy and imaging techniques, this higher frequency range can be utilized for more complex and functionally integrated security purposes. In addition, THz spectroscopy can be used to identify explosive materials as well as illegal drugs [9], [10].

Aforementioned, THz radiation is non-invasive and non-ionizing, which makes it a potential candidate for medical imaging. Due to the small penetration depth of THz beams in tissues and owed to its sensitivity to the water content on the tissue, only visible-accessible parts of the body are potential targets for THz imaging. In this regard, the ultimate goal of THz biomedical imaging is to replace invasive methods like biopsy with non-invasive techniques at least in some areas such as skin cancer detection [11], [12]. Dermatology is the first application that widely uses THz imaging since the skin in the most accessible organ for reflection measurements. THz imaging was demonstrated in wound healing monitoring measurements. Using THz imaging, scar tissue and healthy tissue can be differentiated, and since THz can probe without removing the wound dressing it has advantages over other alternative methods [13]. Also, THz can differentiate between benign and malignant lesions and measure the depth of the lesion in the skin. So THz imaging is a suitable technique for diagnosing of malignant cancers at its early stages [14], [15]. Another application of THz imaging in the dermatology is to measure the water content and skin hydration, which can be employed so to monitor and reveal the impact of cosmetic products [16]. Teeth are the next accessible choice in our body for THz imaging. In dentistry, imaging options are fewer than those in dermatology and radiography is widely used for diagnoses. But it is shown that THz imaging due to its non-ionizing nature can detect caries on primary stages, which is not possible with X-ray imaging [17], [18]. If caries can be detected on its primary stages, by early therapy it can be reversed and filling procedures could be prevented. It is worth mentioning that by using this technique, primary caries can be differentiated from low mineralization regions. In addition to these applications in dermatology and dentistry, there are also some initial works showing THz imaging of human breast cancer tissue [19]–[21]. As the interest in THz imaging techniques increases, more and more potential applications will emerge.

Wireless communications is perhaps one of the most important potential applications for THz technology. As based on Edholm’s law, the bandwidth demand in communication networks triples every eighteen months [22]; the prediction for 2020 is 100 Gbps. There are two approaches for addressing this growth in demand, the first approach is to use advanced modulation techniques or MIMO systems, and the second approach is to utilize higher frequency carriers. Since there is a fundamental upper limit for the achievable bandwidth using signal processing techniques [23], eventually, using higher frequency carriers will become inevitable. Right now, microwave (i.e. GHz) carriers are used for wireless communication links, but ultimately, the demand for bandwidth will force us to use millimeter-waves and THz frequencies for communications.

In order to enable the use of THz for communication systems, we need to develop different system components such as sources, detectors, modulators, switches, filters, amplifiers, etc. Most conventional materials and design approaches typically used in optics or at microwave frequencies are not applicable in the THz frequency range. Since most materials are lossy at these frequencies it is essential to find or develop new materials suitable for the THz range [24]. Therefore, researchers have been interested in finding novel naturally occurring materials, such as graphene, liquid metals, etc. as well as in designing artificial metamaterial structures for application in the THz range.

## 1.2 Visible-transparent conductive materials and devices

Transparent electronics have always been a part of the image that describes the future as usually seen in science-fiction movies. In 1997, for a first time, a transparent circuit was demonstrated based on transparent conductive oxides (TCOs), which was called the "invisible circuit" [25], [26]. Subsequently, transparent electronics does not sound so unreachable. Since then there has been an enormous effort for realizing such devices and for discovering new materials enabling transparent electronic applications. TCOs are materials which are electrically conductive and visually transparent; during World War II, these materials were used as heaters in aircraft windshields so to prevent icing [27]. These materials have vast applications as passive electrical and optical elements, for example, in solar cells, antistatic coating, and flat panel display. Since the introduction of the first circuits based on TCOs [28]–[30], several electronic devices were proposed and fabricated based on these family of materials.

The first transparent thin-film transistor (TTFT) was introduced and demonstrated in 2003 [28]–[30]. Although the quality of these transistors is not yet comparable with that of silicon

MOSFETs, there is an intense effort to improve the quality of these devices by introducing new materials and device designs. The market for transparent electronics is growing fast and is expected to be worth more than 2 billion dollars in 2023 [31]. Usually, it takes one or two decades before a device technology can be commercialized, so we expect to have commercial transparent devices around 2020.

One of the many challenges that this technology is facing is the small operational frequency of these devices, which is a bound by kHz to few MHz with existing devices [32]. This drawback is caused by the small mobility available in these transparent thin films and can restrict certain applications such as telecommunications. The materials science community is constantly in the search of new materials which can improve the mobility and therefore the operational frequency of transparent devices.

Conductive transparent materials, are an essential group of materials for the demonstration of transparent devices as well as several other applications like liquid-crystal displays (LCDs), solar cells, and LEDs. These materials can be classified in two large groups: the first group are TCOs which possesses high band-gap and either n-type or p-type conduction. Typical examples of these materials are Sn-doped indium oxide (ITO), doped zinc oxide (ZnO) and Cadmium oxide (CdO) [33]–[35]. Also, conductive polymers have been introduced, for example poly(3,4-ethylenedioxythiophene) doped with polystyrene sulfonic acid (PEDOT:PSS) [36], [37] which offers these properties. The major problem with this group of materials is their limited conductivity, the largest achieved conductivity value among these group of materials was found in ITO [38], with conductivity values $\sim 10^4\ S/cm$ at room temperature. This material is widely used in transparent electrodes for many applications. However, the conductivity found in ITO is still two orders of magnitude smaller than metal conductivities and the search for new transparent materials with higher conductivity never stops. The second group of materials are not necessarily transparent, but their dimensions are engineered so that they are conductive in one direction and effectively transparent in another, nanowires [39], [40] and 2D materials [41], [42] belong to this group of materials. These materials are conductive in-plane and relatively transparent out-of-plane as a result of their very small thickness. Graphene is one of the most prominent 2D materials, and is widely used as a transparent electrode since its inception.

In this dissertation, I will study materials from each of these two groups: TCOs and 2D materials, and use them to realize electromagnetic structures and reconfigurable devices for applications in the THz frequency range. As mentioned before, most of the conventional materials that are used in millimeter wave or infrared frequencies, are lossy in the THz range and do not respond to these frequencies in a controllable manner. For achieving efficient devices, we are constantly in the search of new materials. Two methods are commonly used for realizing functional electromagnetic structures in the THz frequency range, namely, (a) taking advantage of the intrinsic properties of materials, e.g. plasmonic properties, or (b) by engineering "artificial" structures by merging several materials at subwavelength scales, e.g. metamaterial approaches. Here, we are going to use both of these approaches to demonstrate functional devices in the THz frequency range. In this regard, graphene apart from being used as a transparent electrode in electronic applications, is currently one of the materials of greatest interest so to develop THz devices. Furthermore, TCOs can be also beneficial for THz technology, since they can enable devices that are functional at THz frequencies and transparent in the near-IR and visible ranges. In the next two sections of this introductory chapter I will describe graphene and some of its THz applications as well as introduce my work on graphene and TCOs.

## 1.3 Graphene

Graphene is a two-dimensional (2D) material with a hexagonal lattice which is formed from carbon atoms. Single-layer graphene was synthesized experimentally for the first time in 2004 [43], however, its existence was predicted by physicist long time ago [44]. In early efforts, mechanical exfoliation (using a scotch tape) was employed to peel-off graphene from graphite crystals. However, the mechanical exfoliation method is a random process and only provides small flakes of graphene in the order of several 10 µm or smaller. Later, several other methods such as thermal annealing of SiC [45], [46], chemical vapor deposition (CVD) [47], [48], and reducing graphene oxide [49] were developed so to synthesize large area graphene.

Thermal annealing of SiC single crystals at 1500˚C will result in evaporation of Si, which leaves the substrate, and thus the remaining carbon atoms will form graphene [50]. This method is very easy to perform, however the SiC crystals are expensive and due to formation of a buffer layer between graphene and the substrate the transfer process is very difficult. But graphene devices can be potentially fabricated directly on the SiC substrate. Reduction of graphene oxide is popular method for bulk application, like manufacturing composites, conductive inks, or conductive pastes. The complete reduction of the oxide to graphene is very difficult and this will lead to samples suffering from defects. CVD graphene synthesis uses hydrocarbon molecules in the presence of a metal catalyst at a high temperature of 900-1080˚C. The number of layers in this growth procedure depends on the metal catalyst. In the presence of Ni and Pd the final structure will be multi-layer and its thickness depends on the gas flow and other growth parameters [51], [52]. However, by using Cu as metal catalyst one can achieve single layer graphene since the synthesis is a self-limiting process [53]. The graphene synthesized using CVD process is easy to transfer to any arbitrary substrate which makes it an interesting choice for flexible electronics and many other applications. Achieving large area is also another advantage of this method and is relatively cheap. Dark-field transmission electron microscopy (DF-TEM) on CVD grown sample indicates that this growth process leads to polycrystalline samples with grain sizes in the order of 500 nm to several micrometers, and these grains are randomly rotated [54].

In addition, several transfer techniques were developed to transfer the synthesized graphene films into arbitrary substrates. Wet transfer using a supporting layer, usually PMMA is the most common method in which the graphene layer is spin coated with a PMMA supporting layer and then the film is released in a metal etchant solution after what is finally transferred to the desired substrate. There are also several alternative dry transfer methods such as roll-to-roll transfer in which very large samples (~30 inch) can be transferred [48], [55]. Also there are other transfer techniques proposed in the literature such as electrochemical and imprinting techniques [56]–[58].

Graphene proves to be an interesting material for developing reconfigurable and active devices due to its extraordinary electro-optic properties. Graphene is a material without a bandgap, which means that the valence band and conduction band meet in one point, called the Dirac point. Like in semiconductors, the optical conductivity in graphene is determined by the superposition of two different optical absorption mechanisms, namely intraband and interband transitions, which can be expressed as [59]:

$$\sigma(\omega) = \sigma_{\text{intra}}(\omega) + \sigma_{\text{inter}}(\omega)$$

$$\sigma_{\text{intra}}(\omega) = \frac{ie^2 E_f}{\pi\hbar(\omega + \frac{i}{\tau})}$$

$$\sigma_{\text{intra}}(\omega) = \frac{ie^2\omega}{\pi}\int_0^\infty \frac{f(\varepsilon - E_f) - f(-\varepsilon - E_f)}{(2\varepsilon)^2 - (\hbar\omega + i\Gamma)^2} d\varepsilon$$

where e is the electron charge, $E_f$ is the Fermi level, ħ is the reduced Planck constant, ω is the angular frequency, $\tau$ is the electron momentum relaxation time, f is the Fermi-Dirac distribution function, and $\Gamma$ is the phenomenological scattering rate. In the low-frequency regime, i.e. the THz frequency range, the intraband transition mechanism is dominant, whereas in the high-frequency regime, i.e. the visible spectrum, the interband transition mechanism is dominant. In Fig. 1 the conductivity of graphene is shown versus frequency for different Fermi levels. Figure 1 also shows that by changing the Fermi level, graphene's optical conductivity can be tuned. In the THz frequency range, the impact of the interband transitions is negligible, due to the low THz photon energy, which is on the order of a few meV, which makes intraband transitions to be the dominant absorption phenomenon. Therefore, graphene conductivity can be expressed using a Drude model at THz wavelengths, as:

$$\sigma_{THz} = \frac{1}{1 + j\omega\tau}$$

in which $\sigma_{DC}$ is the direct current (DC) electrical conductivity, and $\tau$ is electron momentum relaxation time. In conventional 2D semiconductors with parabolic band structure $\sigma_{DC} \propto E_f \propto \sqrt{n}$ but in graphene, due to its conical band structure $\sigma_{DC} \propto E_f \propto n$. Therefore, as in semiconductors the conductivity of graphene can be effectively tuned when $\omega < 1/\tau$. $\tau$ depends on graphene quality and $\omega$ typically is between 1-4THz.

In higher frequency regimes, interband transitions are the dominant term and at the limit of very high frequencies ($\omega \to \infty$), the optical conductivity of graphene will be $\frac{e^2}{4\hbar}$, which results in ~2.3% absorption per layer of graphene. As indicated in Fig.1 the optical conductivity is a constant value and independent of Fermi level in this region. Therefore, graphene in single layer form, has a negligible absorption in the visible frequency range and it is mainly transparent.

Graphene is also known for its large carrier mobility. As a result of the linear E-k dispersion in graphene, its charge carriers behave like massless Dirac fermions. In practice, these massless fermions lead to a very large mobility in graphene, e.g. mobility in the order of $10^6$ $cm^2/V.s$ has been reported in suspended graphene samples [60]. Overall graphene can provide greater conductivity per unit length than any other known semiconductor. These properties prove graphene to be one of the most promising reconfigurable materials for THz frequencies.

As stated before, the conductivity of graphene can be tuned in the THz frequency range by means of changing the Fermi level. To control the Fermi level there are several proposed methods. One of the methods proposed is to gate graphene using a semiconductor substrate, e.g. silicon (e.g. [61]). Another way introduced in the literature is to self-gate graphene (e.g. [62]). In this method two layers of graphene are separated with a dielectric layer; and, by applying a voltage between these layers, accumulation of different type of charge carriers takes place in opposite layers. By means of injecting and depleting carriers in these layers the Fermi levels are therefore tweaked. Moreover, there are several other methods discussed in the literature such as gating graphene using ion-gels [63] or doping graphene chemically [48]. In the chemical doping method, the graphene layer is typically doped using nitric acid, which is a known dopant of other carbon products such as carbon nanotubes [64]. The attainable doping level depends on the acid concentration and the duration of soaking. Also, this doping can be either time-variant (i.e. reversible) or time-invariant (i.e. irreversible) depending on the experimental conditions.

As mentioned before graphene is one of popular active material in THz frequency, however over large areas graphene in its single layer form can achieve limited performance, due to its practically achievable conductivity swing being limited as a result of fabrication issues [61], but by patterning graphene or by integrating it with metamaterial we can boost the performance of THz devices.

## 1.4 Metamaterials for THz applications

Metamaterials are artificially-made structures that are engineered so to achieve desirable electromagnetic properties, thus capable of achieving properties beyond what is possible in naturally occurring materials [65]. Usually, metamaterials are made of unit cells arranged in lattices, and the lattice dimensions are small relative to the operational wavelength, which causes these structures to be recognized as a homogenized effective medium in the operational frequency range. The properties of metamaterials are affected by the properties of the component materials as well as by the geometrical features of the unit cells. Metamaterials are widely used to manipulate electromagnetic waves and other physical phenomena. Using this concept, various exotic unnatural phenomena like negative refractive index and ultrahigh refractive index have been demonstrated. Historically, the theory and first implementation of metamaterials took place at microwave frequencies, but later this was applied to higher frequency ranges such as the THz and the visible.

In the THz range, since most natural materials are not able to effectively interact with these electromagnetic waves, metamaterials provide a promising method so to enhance the light matter interaction. Using metamaterials, unit cell designs can be engineered so to achieve a set of desired properties. In addition, in the THz frequency range, fabrication of subwavelength structures can be achieved using standard lithographic techniques. As a result, the metamaterial approach has become very popular at this frequency range. There are many studies proposing different metamaterial structures for THz applications, such as modulators and filters [66], [67]. These designs can be divided in to two major groups of passive and active metamaterials. Usually passive metamaterial designs are based on metallic films patterned in different geometries, arranged in an array (e.g. [68]). Active control of THz signals is one of the biggest challenge in THz science and technology field. In order to address this challenge, a popular approach is to use an active metamaterial in which the effective THz permittivity of the metamaterial structure should be tuned. For demonstrating this active control, usually, a metamaterial design is either made out of or integrated with an active medium like conventional semiconductors, two-dimensional electron gas (2DEG), phase transition materials, etc. In the past decade, several actuation approaches have been taken so to effectively control the THz permittivity of metamaterials, such as electrical, thermal, mechanical, optical, and plasmonic effects. Among all these methods, electrical tuning is the most attractive option due to its integrability. Typically, semiconductors are the suitable choice for electrically tuning permittivity, but, as mentioned, most of the conventional semiconductor materials lead to high insertion loss when used in the THz frequency range, and their response are weak. Therefore, we are in search of new materials which can efficiently manipulate THz frequencies and meanwhile introduce low insertion loss. From this perspective, graphene, as a reconfigurable semimetal, proves to be good candidate for metamaterial applications at THz frequencies.

## 1.5 Graphene based THz metamaterials

Based on the extraordinary electro-optical properties of graphene and the capability of tuning its optical conductivity in a wide range of frequencies, graphene attracted lots of attention from the THz research community. Several devices were designed and demonstrated based on graphene. The first of such devices based on graphene were realized actually in the infrared. However, in 2011, Ju et al, demonstrated the first reconfigurable plasmonic device based on graphene stripes which were gated using ion-gel [63]. Later in 2012, Yan et al, proposed a stack of graphene/insulator layers in disk shape array, where resonance tuning was achieved via chemical doping [69]. Early works on graphene in the THz frequency range were motivated by modulator applications. In 2012, the first graphene based THz device, a large area modulator, was demonstrated by Sensale-Rodriguez et al, in this configuration gating was achieved by applying a voltage between a graphene layer and a semiconductor substrate [61]. More recently a new configuration of gating, self-gating, which was also proposed by Sensale-Rodriguez et al in [61], was experimentally introduced by Gomez-Diaz et al, in which two layers of graphene are separated with a dielectric layer and by applying voltage between these layers the Fermi level can be tuned [70]. Later work by Lee et al in 2012, introduced the first metamaterial integrated with a graphene film, which operates as a switch for THz waves [71]. Furthermore, graphene integrated with split ring resonators was experimentally demonstrated for THz modulation by Valmorra et al [72] and Degl'Inocenti et al [73].

Several studies also investigated the fundamental limits in the performance of devices based on graphene. In 2014, Tamagnone et al showed by random simulation across the whole design space that there are fundamental limits between the insertion loss and modulation depth in designing graphene based modulators [74]. Later work by Yan et al., indicates that the separation between the active graphene layer and the metallic metamaterial is a key parameter in controlling the tradeoff between insertion loss and the modulation depth in these devices, in other words: the strength of the light-matter interaction needs to be carefully tailored so to obtain the best tradeoffs in device performance [75]. In addition, in a simulation study based on split ring resonators made only of graphene, Zouaghi et al, showed that the Drude scattering-time is an effective parameter in defining the device performance [76].

Even though most of the first applications of graphene at THz frequencies were on developing amplitude modulators, several other devices were also proposed based on graphene integrated metamaterials. In 2014, Yang et al demonstrated for the first time active THz filters based on graphene, in which the frequency of resonance was tuned by controlling the Fermi level on a graphene film [66]. With all the recent progress in developing THz elements for future compact low-cost terahertz systems, several needs are not properly addresses yet. A beam steerer is an essential part for MIMO communications, tunable flat lenses for terahertz cameras, etc., since all of these applications will demand components capable of achieving active beam-shaping at some degree. One of the issues we are trying to address in this work is on approaches towards active beam steering using reconfigurable terahertz metamaterials.

## 1.6 Structure of this dissertation

In the second chapter of this thesis, a graphene-based metamaterial phase modulator is proposed. In this device, graphene is integrated within a deep-subwavelength frequency selective surface so to form a reconfigurable metamaterial for THz phase modulation. This phase modulator is designed as a building block for future THz beam steerers. From this perspective, our goal was to improve the device efficiency in terms of attainable phase modulation per unit cell area when compared to that of existing metamaterial phase modulators. The proposed devices are deeply-scaled and have the smallest unit cell dimensions demonstrated in the literature. This can lead from a practical perspective in sharper phase gradients and larger directivity.

However, as unit cells are made smaller and smaller, while trying to keep the resonance frequency constant, the performance of these devices in terms of loss degrades. Therefore, in the third chapter of this dissertation, I discuss on the geometrical tradeoffs when designing deeply scaled metamaterial phase modulators. In this study two families of deep subwavelength structures, namely spiral resonators and split ring resonators are considered. Two figures of merit, related to device performance are defined, and the effect of the metamaterial geometry on these figures of merit is investigated. Our results show that there is an optimal unit cell size ~$\lambda/20$, which leads to the best performance.

After having discussed graphene-based metamaterials as phase modulators, we try to answer a more general research question: what is the best role that graphene can play when realizing terahertz devices? In other words, is graphene more attractive as a reconfigurable plasmonic material by itself? Or, as a reconfigurable media providing tunability to, otherwise passive, metallic structures? Our simulations and experimental results evidence that graphene/metal hybrid approaches can offer much stronger responses that graphene-only plasmonic structures, therefore are more suitable for practical terahertz applications. A discussion on this analysis is presented in Chapter 4.

In addition to graphene, we investigated other 2D materials as reconfigurable element in THz devices. Graphene is a material with zero bandgap, which results in a non-zero minimum conductivity; therefore, all the device designs based on graphene suffer from a finite insertion loss in the off state. However, there are other novel 2D materials with bandgap, which can be used so to achieve reconfigurability and lower loss. One of these materials is Molybdenum disulfide ($MoS_2$). In Chapter 5, we discuss our investigations on THz metamaterials based on 2D materials beyond graphene.

Furthermore, in Chapter 6, another 2D conductive structure is discussed and studied. The structure under study is not a particular material, as was the case in Chapters 2 to 5, but the 2DEG that is formed at the interface between two complex-oxides. These complex oxides are transparent to visible wavelengths as a result of their wide bandgap. The 2DEG under study, in the interface between NTO and STO, has a world-record 2D carrier concentration. By using THz spectroscopy we show that the conductivity thus mobility at the nanoscale is much larger than the one extracted from DC transport measurements. The large extracted nanoscale conductivity positions these materials at the same conductivity levels typically attainable in high-mobility semiconductors. From this perspective, these materials can be very strong candidates for transparent electronics applications.

We followed this study on highly conductive oxide interfaces, by studying a transparent conducting oxide thin film with record high conductivity levels: BaSnO3 (BSO). Chapter 7 discusses this material and presents our results on its characterization by means of THz spectroscopy. As a result of the high terahertz conductivity in this material, which takes place across a broad frequency window, a visible-transparent / terahertz functional electromagnetic structure is designed and demonstrated so to reveal the application potentials of BSO for visible-transparent terahertz applications.

Finally, this dissertation ends in Chapter 8, which discusses conclusions and future work.

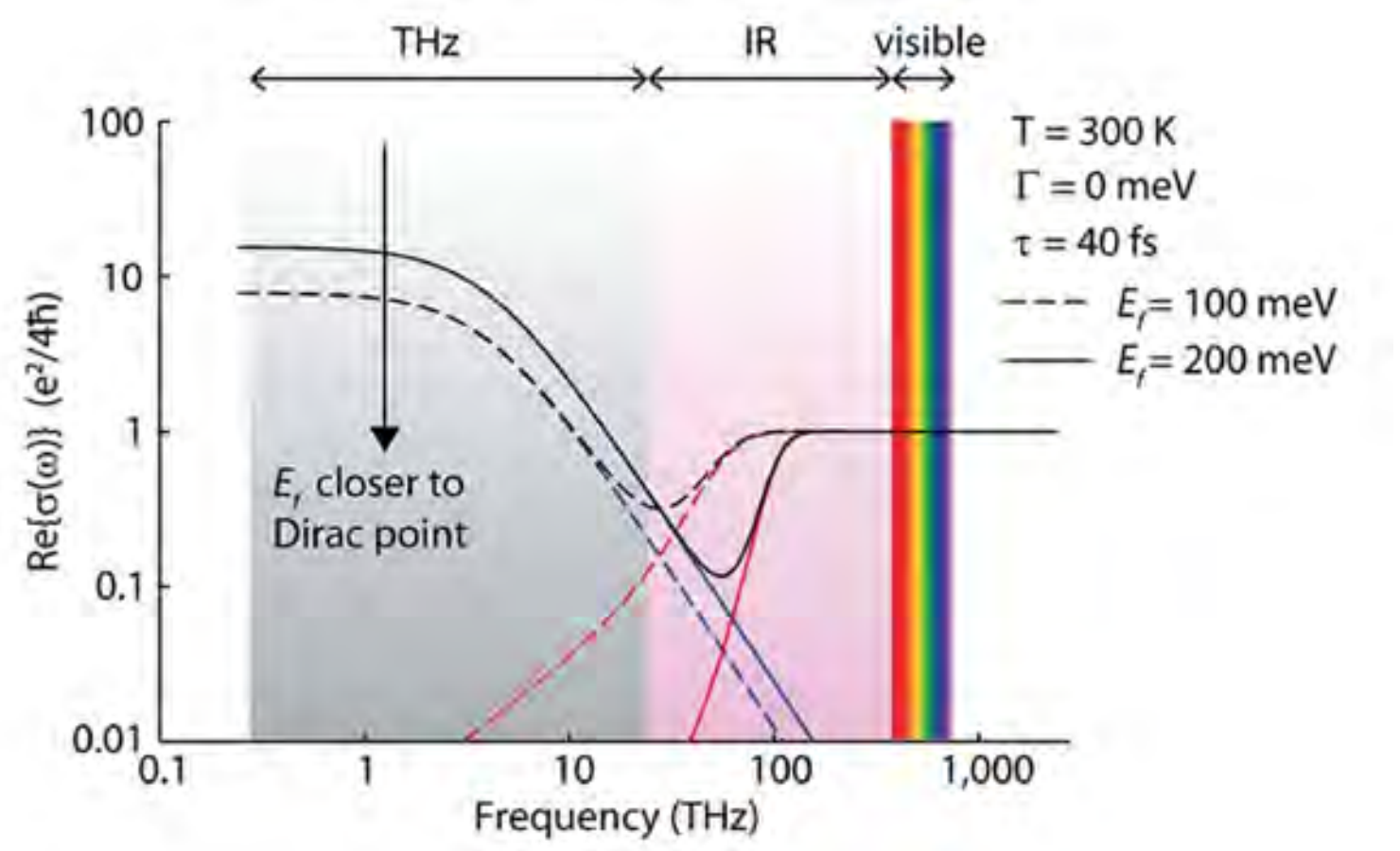

Figure 1: Graphene conductivity vs. frequency for different Fermi levels [59]

# CHAPTER 2

# GRAPHENE-BASED ELECTRICALLY RECONFIGURABLE DEEP-SUBWAVELENGTH METAMATERIALS FOR ACTIVE CONTROL OF THZ LIGHT PROPAGATION

CHAPTER 3

# GEOMETRICAL TRADEOFFS IN GRAPHENE-BASED DEEPLY-SCALED ELECTRICALLY RECONFIGURABLE METASURFACES

Abstract

In this work we study the terahertz light propagation through deeply-scaled graphene-based reconfigurable metasurfaces, i.e. metasurfaces with unit-cell dimensions much smaller than the terahertz wavelength. Two types of deeply-scaled metacell geometries are analyzed and compared, which consist of: (i) multi split ring resonators, and (ii) multi spiral resonators. Two figures of merit, related to: (a) the loss and (b) the degree of reconfigurability achievable by such metamaterials -when applied in beam shaping applications-, are introduced and discussed. Simulations of these two types of deep-subwavelength geometries, when changing the metal coverage-fraction, show that there is an optimal coverage-fraction that gives the best tradeoff in terms of loss versus degree of reconfigurability. For both types of geometries the best tradeoff occurs when the area covered by the metallic region is around 40% of the metacell total area. From this point of view, deeply-scaled metamaterials can indeed provide a superior performance for beam shaping applications when compared to not deeply-scaled ones; however, counterintuitively, employing very highly-packed structures might not be beneficial for such applications.

## 3.1 Introduction

Terahertz technology is a growing technological field, which in recent years has been finding multiple emerging applications in diverse areas including: medical imaging, biochemical sensing, security, wireless communications, and so on[1]. In this context, future compact low-cost terahertz systems, such as beam steerers for MIMO communications, tunable flat lenses for terahertz cameras, etc., will demand components capable of achieving active beam-shaping at some degree. Reconfigurable terahertz metamaterials[2] were shown capable of modulating the phase of an arbitrary terahertz beam[3], which is of special interest for beam-shaping applications. In this regard, via independently biasing each metacell, arbitrary phase gradients can be constructed[3], which in turn can shape the reflected or transmitted beams in accordance with the recently proposed generalized laws of reflection and refraction (generalized Snell's law)[4].

When a phase gradient is placed in the interface between two media of refractive index $n_t$ and $n_i$, Snell's law of transmission should be rephrased as the generalized law of reflection and refraction[4]:

$$\sin(\theta_t)\, n_t - \sin(\theta_i)\, n_i = \frac{\lambda_0}{2\pi}\frac{d\phi}{dx}\,, \qquad (1)$$

where $\theta_i$ and $\theta_t$ are the angle of incidence and the transmitted angle, respectively, $\lambda_0$ is the vacuum terahertz wavelength, and $d\phi/dx$ represents the phase gradient. Assuming normal incidence and $n_i = n_t = 1$, Eqn. (1) can be rewritten as:

$$\sin(\theta_t) = \frac{\lambda_0}{2\pi}\frac{d\phi}{dx}\,. \qquad (2)$$

Therefore, it can be easily seen that the shape of the transmitted beam can be arbitrarily controlled via designing an adequate phase gradient. For instance, assuming an incident collimated beam, a linear phase gradient can tilt the transmitted beam, whereas a parabolic phase gradient can focus it. It is therefore of interest the use of electrically-driven reconfigurable metamaterial phase modulators for constructing these arbitrary phase gradients. However, in order to enable the design of these arbitrary phase gradients, each metacell in the device should be able to provide: (a) the same transmission amplitude, and (b) 360° (2π) control over the transmitted phase. In this context, it can be observed, as discussed by Chen *et al*[3], that the terahertz transmission amplitude and phase through a metacell are not independent of each other, but they are related by Kramers-Kronig (KK) relations. Near frequencies where the amplitude has a strong dependence on the applied voltage bias, the phase experiences a maximum shift. In contrast at the frequencies where maximum amplitude modulation is achieved no phase modulation takes place. From this point of view (a) is guaranteed. Since terahertz metamaterial phase modulators proposed to date exhibit phase modulation much smaller than 360° (see Ref.[3, 5]), epitaxial stacking of multiple layers[3] is necessary in order to achieve (b), which in turn increases the loss in the device. Moreover, construction of arbitrary phase gradients is also limited by the geometrical length of each unit-cell. This is due to the fact that, when employing metamaterials, a continuous phase gradient is approximated by a discretely spatially-varying one. From this point of view, the smaller the unit-cell length (when compared to the target terahertz wavelength), the better one can approximate an arbitrary phase gradient, therefore the more functionality and better performance the metamaterial beam shaper might achieve. In this context, a problem of metamaterial structures proposed to-date as phase modulators is that the unit-cell to wavelength ratio is not small enough to provide good performance. For instance, in order to provide ~90° control over the transmission angle using a 10 element phase-gradient discretization, a unit-cell length < $\lambda_0/10$, thus a unit-cell to wavelength ratio <0.1, is required. Terahertz metamaterial phase shifters reported to-date have unit-cell to wavelength ratios in the order of ~0.15/~0.2 (see Ref.[3, 5]). From this point of view, *one of the main challenges of terahertz metamaterial phase modulators is: designing a metamaterial with small unit-cell to wavelength ratio, which has a large phase modulation and large transmission at the frequencies at which maximum phase modulation takes place.* In this work, the terahertz (THz) light propagation through deeply-scaled graphene-based reconfigurable metasurfaces is studied in the context of beam-shaping applications. Although graphene is used as an example reconfigurable semiconductor in these devices, the discussion presented here is general enough and the results are also valid if employing other semiconductor materials.

## 3.2 Results

Two types of deep-subwavelength metamaterial geometries are studied and compared. These consist of: (i) *multi spiral resonators* (MSRs), and (ii) *multi split ring resonators* (MSRRs), as depicted in Fig. 1(a) and Fig. 1(b), respectively. A sheet of graphene was considered as the tunable element to reconfigure the terahertz transmission properties of the metamaterial[6, 7], which was placed in some strategic regions of the device, as depicted in Fig. 1. The electromagnetic properties of this graphene layer, and therefore the effective properties of the metamaterial, can be adjusted via controlling the Fermi level of graphene therefore its density of states available for intra-band transitions and thus its optical conductivity[6]. Although graphene metamaterials have been widely employed in devices modulating the amplitude of a transmitted terahertz beam[8-11], to the author's knowledge, graphene-based terahertz metamaterials controlling phase, under normal incidence of a THz beam, have not yet been proposed to-date. Actuation over the graphene terahertz optical conductivity can be achieved electrostatically via either gating graphene with another graphene layer (self-gated structure)[12], or via employing ion-gel as the gating element[13].

Shown in Fig. 2 are the characteristic transmission and phase frequency responses as a function of graphene conductivity for one of these metamaterials (a MSR with 30% metal to unit-cell area coverage-fraction). Maximum phase modulation, 108°, was observed at 500GHz; at this frequency the transmittance was found to be 20%, independently of the graphene conductivity.

Metacells consisting of MSRRs and MSRs were numerically simulated. In order to extract useful information regarding the design tradeoffs in these structures, simulations were performed by changing the metal coverage-fraction in each of both geometries. The width of the metal rings/spirals was set to 2-µm, which is a dimension comparable with that of the minimum features achievable in optical lithography; the unit-cell edge-length was taken between 52-µm and 58-µm (depending on the particular metacell). Therefore, a larger coverage-fraction translates into: (a) a larger number of rings and smaller spacing between adjacent rings for the MSRR metacell geometries, or (b) a larger number of turns and smaller spacing in-between metals for the MSR metacell geometries. Shown in Fig. 3 are the sketches of the eight simulated devices (4 MSRR geometries and 4 MSR geometries, each of them having a different metal coverage-fraction); the results of these simulations are shown in Table I (where $f_p$ stands for the frequency at which maximum phase modulation takes place).

## 3.3 Discussion

As discussed in the introductory section, for beam shaping applications, an ideal metamaterial geometry should provide: (i) large phase modulation, (ii) large transmittance, and

(iii) small unit-cell to wavelength ratio. Arbitrary phase gradients need to be constructed when reconfiguring the phase-shift inserted by each metacell. Therefore, a full control of the transmitted phase, i.e. between 0 and 360°, is desirable in each unit-cell in order to achieve truly arbitrary designs. But the phase modulation achievable by each metacell is finite, e.g. prior metamaterial phase-modulator proposals[3, 5] show phase modulation < 50°; therefore, epitaxial stacking of layers is required in order to obtain a 360° control over phase in each metacell. When many layers are epitaxially stacked, although the phase shifts can be added[3], loss increases with number of layers, which is not desirable. From this point of view, the following figure of merit, related to loss, is defined: $FoM_1 = PM \times T / ([360°] \times [100\ \%])$, where $PM$ is the metacell maximum attainable phase modulation, and $T$ is the transmittance through the metacell at the frequency where maximum phase modulation takes place. For an ideal metacell geometry $FoM_1$ should approach unity (since $PM$ and $T$ are bounded by 360° and 100%, respectively); the larger the $FoM_1$ the most suitable a metamaterial geometry is for beam steering, i.e. the less loss the device will provide. But also, a small unit-cell to wavelength ratio is required in order to construct sharp phase gradients, which are needed, for instance, in order to achieve large swings in beam steering applications as discussed in the introductory section. From this point of view, a second figure of merit is defined: $FoM_2 = L / \lambda_P$ , where $L$ is the edge-length of the metacell and $\lambda_p$ is the wavelength associated with the frequency at which maximum phase modulation takes place. For an ideal metacell geometry $FoM_2$ should approach zero, the smaller the $FoM_2$ the most suitable a metamaterial geometry is for beam steering.

As depicted in Table I, it was observed that for MSRRs, the resonance always red-shifts as the metal coverage-fraction is increased. However, for MSRs, when the coverage-fraction is increased, the resonance frequency first starts red-shifting and then blue-shifts. Moreover, if the metal coverage-fraction is further increased (as depicted in Fig. 4), the response becomes even less monotonic. The first blue-shift is observed when the metal coverage-fraction is increased to larger values (i.e. from 50% to 64%). The trends observed in both structures can be qualitatively explained with an equivalent circuit model (the series of an equivalent inductance and an equivalent capacitance), see Ref.[14]. For instance, for the case of MSRRs, the equivalent inductance ($L_0$) is given by the average inductance of the rings. Therefore, the unit-cell size will be determinant in $L_0$; since for the simulated structures the unit-cell dimensions remain almost constant, the equivalent inductance can be considered as independent of the metal coverage-fraction. Shown in Fig. 5(a)-(b) are sketches of the equivalent circuit models for MSR and MSRR geometries, respectively. Here $C_0$ is the total capacitance (capacitance between two adjacent rings) and $L_0$ is the equivalent inductance; in the figure it is assumed that this capacitance, which is related to the spacing between rings, remains constant and its value is $C_0$. It is worth mentioning that effects such as the capacitance between non-adjacent rings and the resistances arising from losses in the metal and the dielectric will be neglected. As the metal coverage-fraction increases (by adding more turns to the structure) the total effective capacitance of the structure increases as depicted in Fig. 5(b). In practice, because of the smaller spacing when coverage-fraction is increased, $C_0$ also increases. This increase in effective total capacitance explains the observed red-shifting of the resonance frequency in MSRRs as the metal coverage-fraction is increased. The equivalent circuit model for a MSR geometry consisting of two spirals is shown in Fig. 5(a). As the metal coverage-fraction is increased, there are two competing effects taking place in this geometry: (i) as the number of turns increases, as depicted in Fig. 4(a), the equivalent capacitance first (at small number of turns) increases and then (at large number of turns) decreases. From this point of view, since indeed the number of turns is increased when the coverage-fraction is increased, the resonance is expected to blue-shift for very large metal coverage-fractions. In the other hand, (ii) as the coverage-fraction is increased, $C_0$ increases due to a smaller spacing. From this point of view, the resonance is expected to red-shift. These two effects, (i) and (ii), are competing and both are important when increasing the coverage-fraction. For small coverage-fractions the same resonance-frequency evolution trend as in the MSRR is observed. However, for large coverage-fractions (i) can become dominant, and the overall effect we observe might be a decrease in the equivalent circuit capacitance, and therefore a resonance frequency blue-shift. However, when the metal coverage fraction is increased even further, (ii) can become dominant again and cause the resonance frequency to again red-shift; this leads to a very non-monotonic characteristic at large coverage-fractions.

It can be also observed (Fig. 4) that when the resonance red-shifts its strength diminishes, until eventually it becomes so weak that it disappears (blue, red, and sky-blue data points which correspond to 30%, 42%, and 49% metal coverage-fractions, respectively). When the metal coverage-fraction is further increased then a former higher-order resonance becomes the first resonance (e.g. purple data point, which corresponds to 64% metal coverage-fraction), leading to a saw-tooth characteristic for resonance-frequency versus metal coverage-fraction as observed in Fig. 4. The resonance strengths for metal coverage-fractions above 80% become considerably weaker since the gaps between adjacent gold-stripes become much smaller than the width of the gold-stripes and thus the structure, effectively, is mostly covered with metal.

Shown in Fig. 6 are the plots of $FoM_1$ and $FoM_2$ versus metal coverage-fraction for the two metamaterial geometries that were studied. When analyzing $FoM_1$, it is observed that in MSRs the smallest coverage-fractions give the best tradeoffs. However, in MSRRs, decreasing the coverage-fraction below 30% significantly decreases transmission and phase modulation due to a very weak interaction between rings. Therefore, it can be noticed that moderate coverage-fractions (i.e. around 40%) give the best tradeoff.

When analyzing $FoM_2$, in MSRRs, it can be observed that large coverage-fractions give the best tradeoff. This is a result of the monotonically increasing dependence of effective capacitance with coverage-fraction in this geometry. However, in contrast, for MSRs a completely different trend is observed when the metal coverage-fraction is increased. There is an optimal coverage-fraction, which occurs around 40%, that gives the best tradeoff. This is a result of the non-monotonic dependence of equivalent capacitance with coverage-fraction in this geometry.

When (overall) considering all the above described trends, it can be concluded that for both types of geometries the best tradeoff between $FoM_1$ and $FoM_2$ occurs in the region where coverage-fraction is around 40%. *From this point of view, deeply-scaled metamaterials can indeed provide a better performance than traditional metamaterials in beam-shaping applications. However, counterintuitively, use of very highly packed structures can actually be not beneficial. There is an optimal metal coverage-fraction, ~40%, which offers the best tradeoff. Interestingly, this optimal coverage-fraction is the same for both types of metamaterial geometries, MSRRs and MSRs.*

## 3.4 Methods

*Numerical simulations and structural parameters*

In the analyzed metamaterials, gold was chosen as the material for the metallic layers, whereas $Al_2O_3$ was considered as the dielectric-in-between (see Fig. 1). These materials were set in top of a 2-µm thick polyimide layer, which has the role of a substrate; the thickness of the gold and $Al_2O_3$ layers was 1-µm. The metamaterials were numerically simulated employing high frequency structural simulator (HFSS). In these simulations graphene was taken as a finite-thickness material as discussed in Ref.[10, 15].

**Table I.** Simulation results for the geometries depicted in Fig. 3

| MSRR | | | | | MSR | | | | |
|---|---|---|---|---|---|---|---|---|---|
| Metal coverage fraction | $f_p$ (GHz) | T @ $f_p$ | PM @ $f_p$ | unit cell (μm) | Metal coverage fraction | $f_p$ (GHz) | T @ $f_p$ | PM @ $f_p$ | unit cell (μm) |
| 62% | 220 | 60% | 19° | (58) | 64% | 510 | 20% | 33° | (52) |
| 48% | 290 | 34% | 36° | (58) | 42% | 320 | 40% | 42° | (54) |
| 39% | 350 | 28% | 66° | (56) | 37% | 400 | 26% | 67° | (52) |
| 24% | 440 | 17% | 55° | (56) | 30% | 500 | 20% | 108° | (54) |

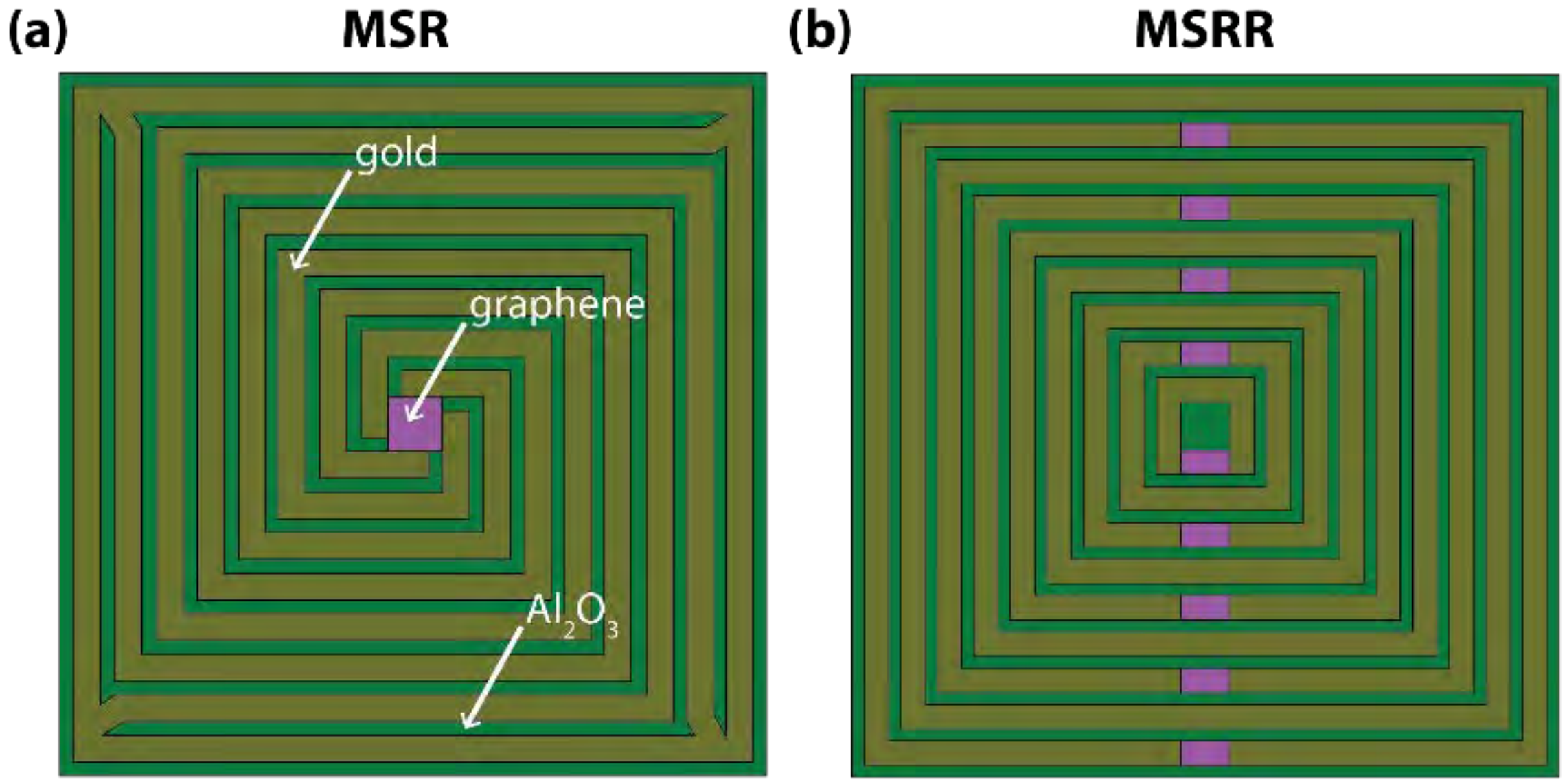

**Figure 1. Sketches of the analyzed metacell geometries.** (a) A multi spiral resonator (a), and (b) a multi split ring resonator.

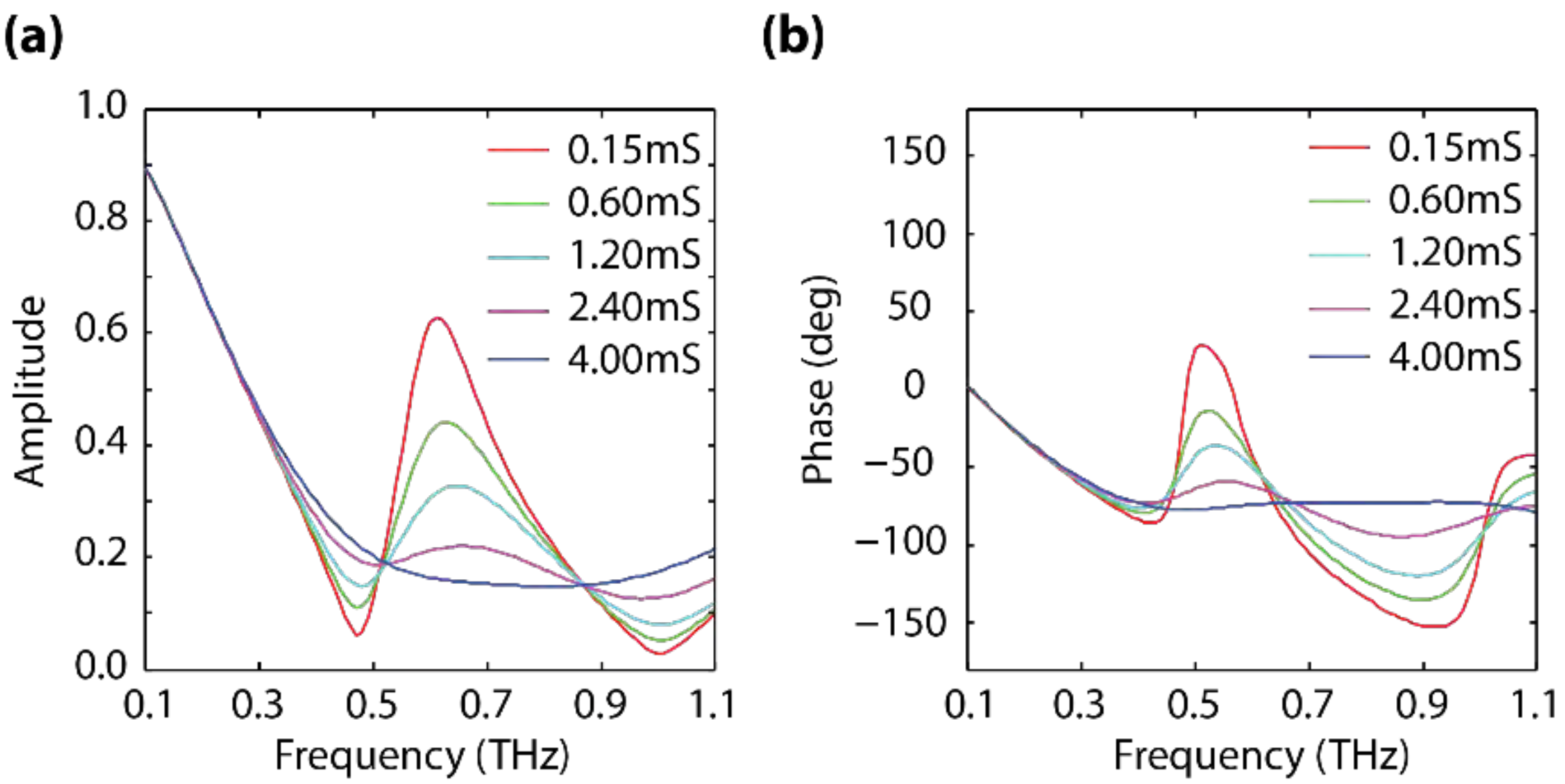

**Figure 2. Characteristic response of one of the analyzed metamaterials.** (a) Amplitude and (b) phase of the transmission as a function of frequency for different graphene conductivities -for a MSR with 30% metal coverage-fraction-. The conductivity of graphene is varied from 0.15mS to 4mS.

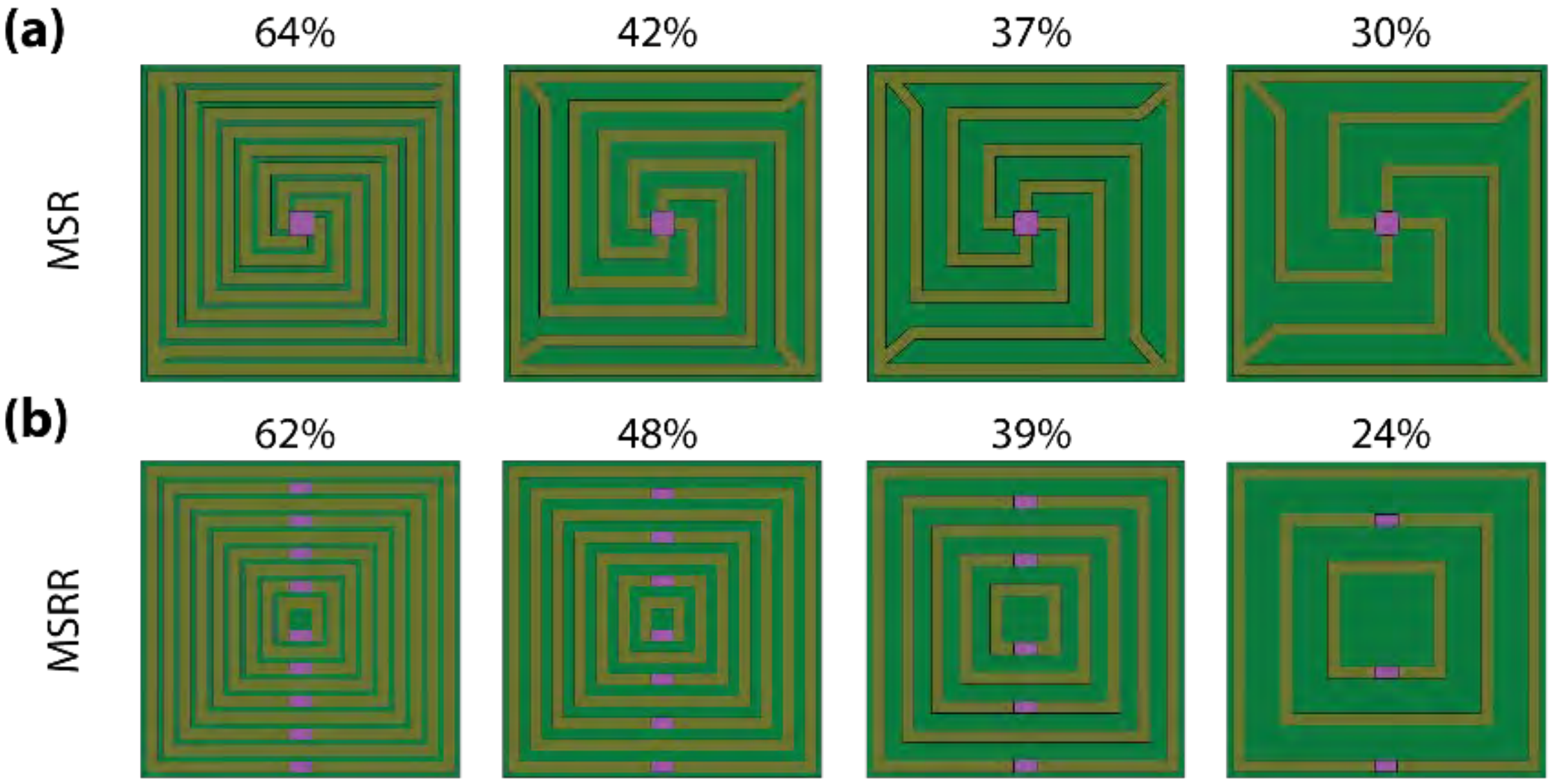

**Figure 3. Sketch of the simulated metamaterial geometries.** (a) Multi spiral resonators with metal coverage-fraction 64%, 42%, 37% and 30%, and (b) multi split ring resonators with metal coverage-fraction 62%, 48%, 39%, and 24%. The coverage-fraction is defined as the ratio between the area covered by metal and the total area of a metacell.

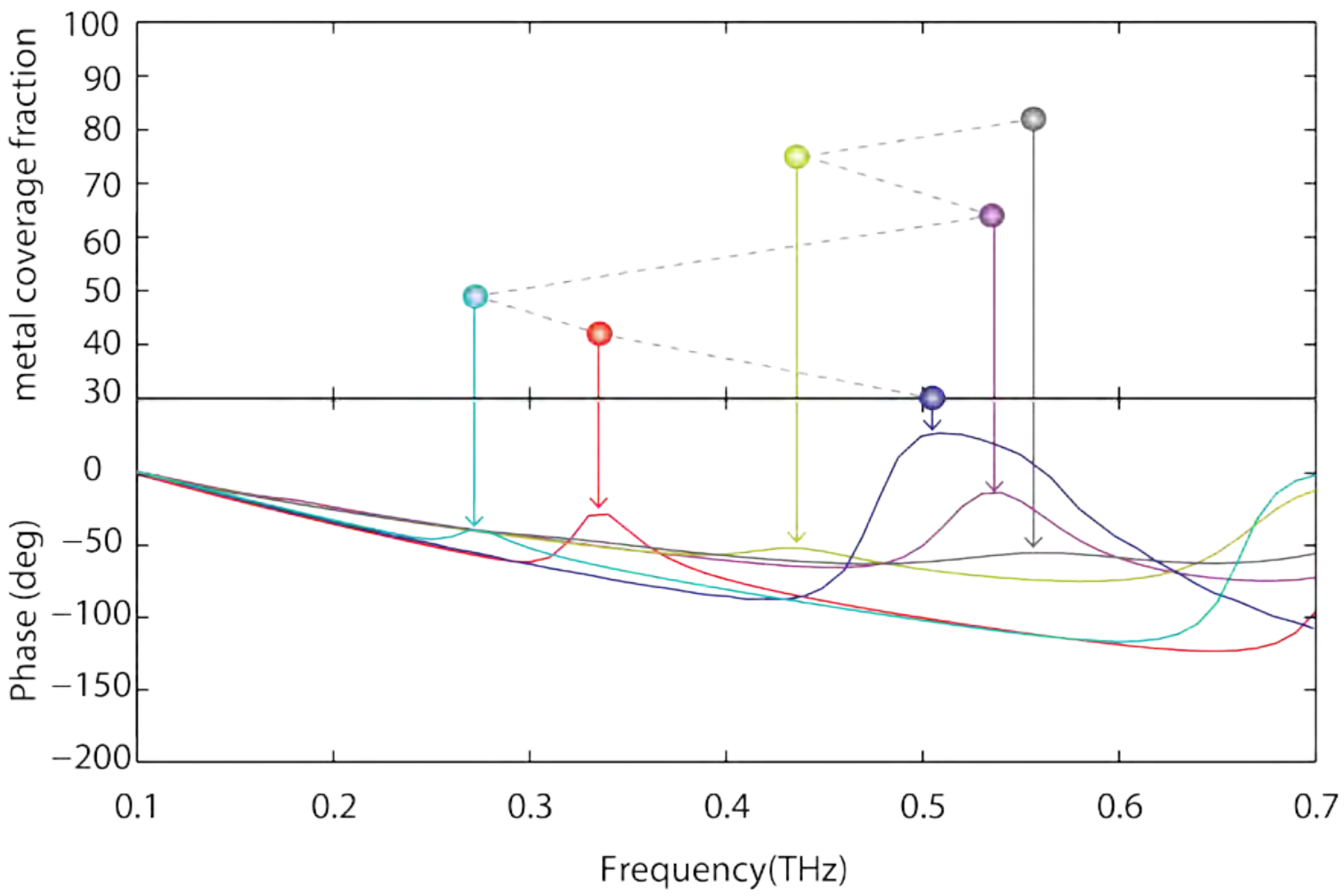

**Figure 4. Phase of transmission for different metal coverage fraction in MSRs.** The upper plot depicts the frequency at which maximum phase modulation takes place versus metal coverage fraction. The lower plot shows phase versus frequency for different metal coverage fractions.

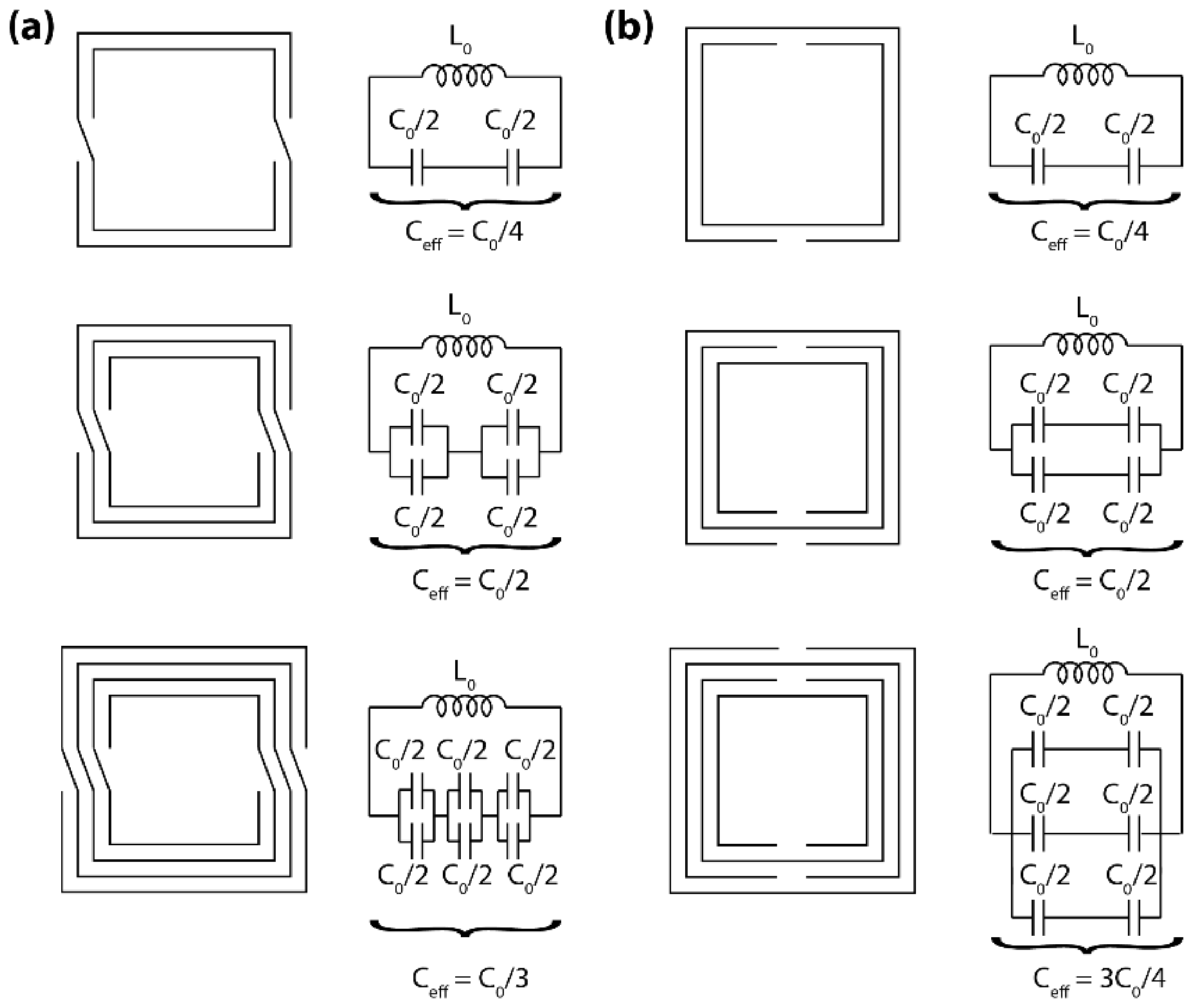

**Figure 5. Equivalent circuit models.** Different (a) MSR and (b) MSRR structures and their equivalent circuit model (following the discussion in Ref.[14]). For MSRs equivalent capacitance first (at small number of turns) increases and then (at large number of turns) decreases when the number of turns is increased, whereas for MSRRs the equivalent capacitance always increases. Here $C_0$ is the total capacitance between two adjacent rings, which is assumed to be the same for all the depicted structures.

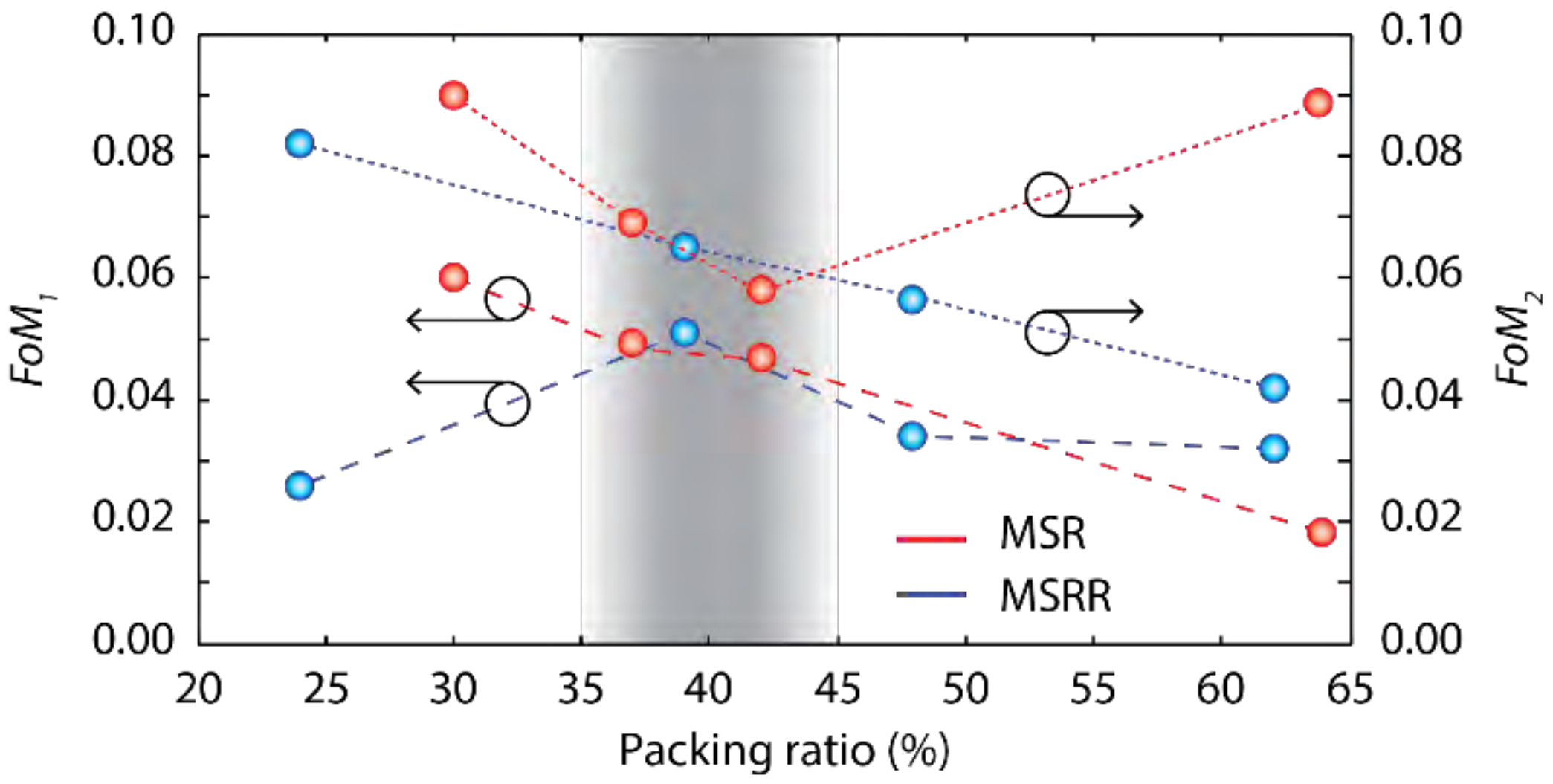

**Figure 6. Figures of merit (*FoM₁* and *FoM₂*) versus metal coverage-fraction for both analyzed metamaterial geometries.** In both cases, the best tradeoff occurs when the metal coverage-fraction is around 40% (gray shaded region).

ACKNOWLEDGMENTS

The authors thank David Schurig and Ajay Nahata for useful discussions. The authors acknowledge the support from the NSF MRSEC program at the University of Utah under grant # DMR 1121252 and from the NSF CAREER award #1351389 (monitored by Dimitris Pavlidis).

AUTHOR CONTRIBUTIONS: S. A. and B. S-R. carried out the numerical simulations, analyzed the data, and contributed to the preparation of the manuscript.

AUTHOR INFORMATION: The authors declare no competing financial interests. Correspondence and request for materials should be addressed to Berardi Sensale-Rodriguez (berardi.sensale@utah.edu).

CHAPTER 4

# GRAPHENE-BASED RECONFIGURABLE TERAHERTZ PLASMONICS AND METAMATERIALS

## Abstract

This work discusses and compares two proposed practical approaches for realizing graphene-based reconfigurable terahertz metamaterials, namely: graphene-only plasmonic structures, and graphene/metal hybrid structures. From rigorous theoretical analysis, full-wave electromagnetic numerical simulations, as well as supporting experiments, several reconfigurable structures are analyzed and compared in terms of their: (i) Quality-factor, (ii) Extinction-ratio, (iii) Unit-cell dimensions, and (iv) Resonance-frequency tunability-range. From this analysis it is observed that at terahertz frequencies, although typically possessing larger unit-cell dimensions and being limited by a restricted resonance-frequency tunability-range, reconfigurable metamaterials based on graphene/metal hybrid structures can provide much larger quality-factors, extinction levels, and, when reconfigured, smaller extinction-level degradation than graphene-only plasmonic structures. As a result, when analyzed in the context of reconfigurable terahertz metamaterials, graphene might result attractive as a reconfigurable media providing tunability to otherwise passive metallic structures rather than as a reconfigurable plasmonic material per-se.

## 4.1 Introduction

Over recent years, graphene has emerged as an attractive material for plasmonic and metamaterial applications [1-3]. Due to the possibility of efficiently tuning its electromagnetic properties via controlling its Fermi level [4-5], graphene is particularly appealing for applications demanding reconfiguration. Proposals and demonstrations of such reconfigurable structures have been reported in a wide range of frequencies, ranging from the microwaves to the near-infrared [6-9]. In particular, at terahertz frequencies, multiple device structures have been proposed for realizing graphene-based reconfigurable metamaterials. In this regard, one of the main advantages of graphene resides on the possibility of integrating it onto arbitrary substrates, i.e. low-loss, low-*k* substrates. Graphene-based metamaterials can be loosely classified in two categories: (*a*) *graphene-only plasmonic structures*, and (*b*) *graphene/metal hybrid metamaterial structures*. Whereas in the first type of structures graphene is used as both the plasmonic media as well as the reconfigurable media, in the second type graphene behaves solely as the reconfigurable media, i.e. the electromagnetic response of the structure is completely set by the metal geometry. At terahertz frequencies reconfigurable *graphene-only plasmonic structures* were first shown by Ju *et al* in 2011 employing arrays of graphene stripes gated by ion-gel [10]. Later work by Yan *et al* experimentally demonstrated tunable properties in plasmonic structures consisting or graphene/insulator stacks arranged in disk arrays, where tunability was achieved by means of chemical doping [11]. More recently, Gomez-Diaz *et al* showed that self-biased graphene structures, i.e. capacitively coupled graphene layers, could be used for such applications [12]. Moreover, Liu *et al* reported on electrically tunable graphene anti-dot array terahertz plasmonic crystals exhibiting multi-band resonances [13], and Zouaghi *et al,* on a simulation study, analyzed split-ring resonator arrays made completely of graphene, identifying the Drude scattering-time as an important parameter affecting the device performance [14]. Early work on reconfigurable *graphene/metal hybrid metamaterial structures*, was driven by modulator [15] applications. In this context, Lee *et al* showed that large-area un-patterned graphene could be employed as the tunable element in metamaterial switches consisting of graphene together with hexagonal and double split-ring-resonator metallic structures [16]. Later work by Yan *et al* discussed the geometrical tradeoffs in such structures for modulator applications, finding that the graphene-to-metal separation is a key parameter, which controls the strength of light-matter interaction in graphene [17-18]. Experimental demonstrations of terahertz modulators based on structures consisting of active graphene and passive metallic split-ring resonators have been also reported by Valmorra *et al* [19] and Degl'Inocenti *et al* [20]. More recently, Gao *et al* reported on graphene/metal hybrid structures using ring resonators with modulation depth approaching 50% and low loss [21], and Liu *et al* [22] demonstrated a tunable modulator by coupling graphene plasmons to a metamaterial. In terms of theoretical analysis, Tamagnone *et al* studied the fundamental limits of graphene-only and graphene/metal hybrid structures in terms of modulation performance via performing random simulations and exploring the geometrical parameter design space [23]. However, besides modulating the transmission amplitude at resonance, graphene/metal hybrid terahertz metamaterials have also been shown capable of actively tuning its resonance-frequency (e.g. work by Yang *et al* [24]). *In this context of recent advances in both graphene-only and graphene/metal hybrid structures, the aim of this paper is to rigorously discuss and compare both approaches for realizing graphene-based reconfigurable terahertz metamaterials*. In particular, structures enabling active tuning of their resonance-frequency will be discussed, which can find application as reconfigurable terahertz filters.

## 4.3 Discussion

In order to give physical insight into the properties of *graphene-only plasmonic structures*, let us consider an array of periodically patterned graphene disks, as studied in Ref. [11], and as depicted in **Fig. 1(a)**. Such structure is chosen as a well-known example, one of the few that has an analytical solution, so to address the advantages and limitation of graphene-only plasmonic structures at terahertz frequencies. The physical origin and the properties of plasmons in such structures can be readily derived starting from the Drude terahertz conductivity of graphene. By employing classical electromagnetic analysis based on Maxwell-Garnett theory, and neglecting disk-to-disk interactions, the following average effective conductivity is obtained (see [25]):

$$\sigma_{av,eff} = F\frac{\sigma_{graphene}(\omega)}{1+\frac{\pi\sigma_{graphene}(\omega)}{2d\varepsilon_0(1+\varepsilon_S)i\omega}} \ , \qquad (1)$$

where $F$ is the filling factor, i.e. fraction of the area covered by the disks, $d$ is the disk radius, $\varepsilon_0$ is the vacuum permittivity, $\varepsilon_s$ is the substrate relative permittivity, and $\omega = 2\pi f$ is the angular frequency. Moreover:

$$\sigma_{graphene}(\omega) = \frac{\sigma_{DC}}{1+i\omega\tau} \ , \qquad (2)$$

is the terahertz sheet conductivity of graphene, where $\sigma_{DC}$ represents its DC sheet conductivity and $\tau$ its electron momentum relaxation time. In the terahertz regime the conductivity of graphene can be modeled by a Drude model [26-29] since the contribution arising from interband transitions is negligible at this frequency range. At normal incidence, the transmission ($T$) through such structure -normalized to that of the substrate ($T_0$)-, is given by [11]:

$$\frac{T}{T_0} = \frac{1}{\left|1+\frac{Z_0\sigma_{av,eff}(\omega)}{1+\sqrt{\varepsilon_S}}\right|^2} \ , \tag{3}$$

where $Z_0 = 377\ \Omega$ is the vacuum impedance, and the substrate was assumed to be optically thick. By substituting Eqn. (1) and (2) into Eqn. (3), the following expressions can be derived:

$$\omega_p = \sqrt{\frac{\pi\sigma_{DC}}{2d\varepsilon_0(1+\varepsilon_S)\tau}} \ , \tag{4}$$

$$E = 1 - \frac{T}{T_0}\bigg|_{@\omega_p} = 1 - \frac{1}{\left|1+\frac{Z_0 F\sigma_{DC}}{1+\sqrt{\varepsilon_S}}\right|^2} \ , \tag{5}$$

$$Q = \omega_p\tau\sqrt{1-E} \ , \tag{6}$$

where $\omega_p$, and $Q$ represent the frequency and quality factor of the plasmonic resonance, respectively, and $E$ represents the extinction evaluated at the plasmonic resonance frequency. The detailed derivation of these equations is presented in the *Supplementary Information*. In this regard it is worth mentioning that Eqns. (5) and (6) are general enough to hold for any convex geometry, e.g. squares, rectangles, triangles, etc. From Eqn. (6) it results the following fundamental upper bound for the quality factor: $Q < \omega_p\tau$. From this perspective, *this simple example demonstrates that quality factors above unity are not possible in structures designed to exhibit resonances at frequencies such that* $\omega_p < 1/\tau$. In this context, due to typical wavelengths practically employed in terahertz technology being on the order of > 100 µm (where 100 µm corresponds to 3 THz) and therefore terahertz beam spot sizes being > 100 µm$^2$, single-layer graphene samples with area > 100 µm x 100 µm are usually required for terahertz applications. From this point of view, large-area chemical vapor deposition (CVD) grown graphene films are normally employed for terahertz plasmonic and metamaterial applications. In this regard, it is worth noticing that the electron momentum relaxation time in such samples is typically on the order of < 0.1 ps (e.g. Ref. [30], where $\tau$ < 0.1 ps was observed even in CVD graphene samples grown on single-crystal copper), and thus, in practice, $Q << 1$ in structures exhibiting plasmonic resonances up to 1.6 THz.

The DC sheet conductivity of graphene can be expressed as:

$$\sigma_{DC} = \sqrt{n_{2D}}\tau e^2 v_F / \sqrt{\pi}\hbar \,, \tag{7}$$

where $v_F$ is the Fermi velocity in graphene ($v_F \approx 10^6 m/s$), $n_{2D}$ is the graphene electron sheet density, $\hbar$ is the reduced Planck constant, and $e$ is the electron charge. It is worth mentioning that the charge density dependence in Eqn. (7) is different to what is seen in typical (parabolic-band) semiconductor materials, where a $n_{2D}$ dependence rather than $\sqrt{n_{2D}}$ is observed. In the limit when $E \rightarrow 1$, which is desired in terms of practical applications, it results:

$$\omega_p^2 = \frac{\sqrt{\pi} e^2 v_F}{2\hbar\varepsilon_0(1+\varepsilon_s)} \frac{\sqrt{n_{2D}}}{d} \,, \tag{8}$$

$$Q = \frac{1+\sqrt{\varepsilon_s}}{Z_0 F} \frac{\pi^{3/4}\sqrt{\hbar}}{e\sqrt{2\varepsilon_0(1+\varepsilon_s) v_F}} \frac{1}{d n_{2D}^{1/4}} \,, \tag{9}$$

where Eqn. (8) has been directly derived by substituting Eqn. (7) into Eqn. (4), and Eqn. (9) has been derived from Eqn. (6) via a Taylor expansion. From Eqns. (8) and (9) it is observed that when $E \rightarrow 1$, the relaxation time, which is related to the graphene quality, does not affect neither $\omega_p$ nor $Q$. It is worth noticing that for the extinction-ratio to approach unity, large filling-fraction, approaching the hexagonal packing density of $F = (\pi\sqrt{3})/6 \approx 0.91$ is desired, in addition to: small substrate refractive index, and large $\tau$, as well as large $n_{2D}$ in accordance with Eqns. (5) and (7). However, as can be noticed in Eqn. (8), terahertz plasmonic resonances require small $n_{2D}$ in addition to large geometric dimensions, i.e. large $d$. By inspecting Eqn. (9) it is observed that small $n_{2D}$ is also associated with large $Q$, yet, since $Q$ is inversely proportional on $d$, small dimensions are preferred from this point of view. In general, there are multiple tradeoffs between these parameters. Shown in **Fig. 1(b)** are the contour plots of $\omega_p$ and $Q$ versus $d$ and $n_{2D}$, in accordance with Eqns. (8) and (9), and assuming $\varepsilon_s = 1$ (so to maximize $E$). The curves for which $\omega_p = 0.5$ THz and $\omega_p = 1$ THz are depicted in orange-colored and red-colored solid-traces, respectively. Contour plots for $Q$ are filled in blue color-scale (light-blue representing $Q < 1$ and dark-blue representing $Q > 1$). It is observed that $Q > 1$ and $\omega_p = 1$ THz requires $n_{2D} < \sim 7\text{x}10^{12}$ cm$^{-2}$, whereas for $Q > 1$ and $\omega_p = 0.5$ THz, $n_{2D} < \sim 2\text{x}10^{12}$ cm$^{-2}$ is required.

In order to describe the most general behavior of the structure we lifted the assumption of $E \rightarrow 1$. Shown in the *Supplementary Information*, **Fig. S1**, are the general contour plots for $Q$, $E$, and $\omega_p$ versus $n_{2D}$ and $d$ for different values of $\varepsilon_s$ and $\tau$, in accordance with Eqns. (4)-(7). It is observed that for a fixed disk geometry (i.e. constant $d$): (*i*) for $\tau < 1$ ps, as $n_{2D}$ is increased $Q$ increases; (*ii*) for $\tau \sim 1$ ps, $Q$ depends very slightly on $n_{2D}$ (when $n_{2D}$ is varied), which is particularly evident on low permittivity environments (i.e. when $\varepsilon_s$ approaches unity); (*iii*) for $\tau > 1$ ps, as $n_{2D}$ is increased $Q$ decreases. Case (*iii*) typically corresponds to a situation where $E \rightarrow 1$, thus $Q$ follows the behavior described by Eqn. (9) and depicted in **Fig. 1(b)**. Moreover, it is also observed that for a given $\omega_p$ design-target, i.e. when looking at the curves representing constant $\omega_p$ in **Fig. S1**, if $\tau \sim 0.1$ - $0.2$ ps, $Q$ cannot be significantly altered by changing $d$ and $n_{2D}$. In these cases it is particularly difficult to achieve $Q > 1$. On the other hand, when $\tau > 1$ ps, as both $n_{2D}$ and $d$ are simultaneously varied (so that $\omega_p$ is fixed), it is observed that $E$ increases as the values of these parameters are increased, but at the expenses of a lower $Q$ (in agreement with the discussion in **Fig. 1(b)**). In synthesis, *at typical frequencies of interest of terahertz technology*, i.e. $f < 2$ THz, *it is extremely difficult to attain simultaneously high-Q* (i.e. $Q > 1$) *and large extinction* (i.e. $E > 0.8$), *unless employing graphene with very large* $\tau$ (i.e. $\tau$ exceeding 1 ps), *which from a practical perspective is extremely challenging in large-area CVD-grown graphene*; *the lower the* $\omega_p$ *design-target the more pronounced is this tradeoff.*

On the other hand, *graphene/metal hybrid metamaterials* can also provide for active tuning of resonance properties [24]. Depicted in **Fig. 2(a)** is the sketch of one such a structure composed of metallic split-ring resonators (SRRs) in conjunction with strategically-placed active graphene layers. The reason why tuning of the resonance-frequency is observed when altering the graphene conductivity, rather than tuning of the transmission amplitude at resonance, is a result of the strategic placement of the patterned graphene film, which is located solely inside the SRR gap. **Figure 2(b)** shows an equivalent circuit model for the structure. In this model, which is validated in the *Supplementary Information*, graphene is modelled as an impedance of value $Z_g = R_g + i\omega L_g$, where $R_g$ and $L_g$ represent its associated resistance and inductance, respectively. Moreover, the metallic pattern is represented by the series connection of a resistor ($R$), an inductor ($L$), and two capacitors ($C$, and $C_{gap}$) [31-34]. In this regard, it is worth noticing that $L$ represents the self-inductance of the metal loop and $R$ the metal losses. Furthermore, whereas the capacitor $C$ results from the separation between adjacent unit cells [34], the capacitor $C_{gap}$ has its geometric origin on the gaps inserted along the metal loops (see e.g. [35]). When the graphene conductivity is altered, therefore $R_g$ is varied, the structure switches between the two-resonant states as depicted on the center and right panels of **Fig. 2(b)**, corresponding to $R_g = 0$ and $R_g \rightarrow \infty$. These two resonant-states are uniquely set by the geometry of the metallic pattern. In this case the graphene quality ($\tau$) has a negligible role in the overall response of the structure.

## 4.4 Results

In order to provide a comprehensive study, *we prepared two sets of samples, consisting of: (Sample Set #1) graphene-only plasmonic structures made of arrays of periodically patterned graphene-disks which are set on an hexagonal lattice.* As a difference with previous works (e.g. Ref. [11]), our plasmonic structures were designed so to exhibit resonances at $f$ < 1 THz. Therefore, large disk radius are employed (radius > 10 µm). *(Sample Set #2) graphene/metal hybrid metamaterial structures consisting of split-ring resonator arrays, in which graphene is strategically-placed inside the SRR gap* (as proposed in Ref. [24]). Moreover, a third geometry is also analyzed, but solely via numerical simulations: *(Sample Set #3) high-Q metal/graphene hybrid metamaterials; where graphene is integrated as an active element in the high-Q metallic metamaterial geometry described in Ref.* [36]*, which consists of concentric ring-resonators with interdigitated-fingers (ICRs).* The fabricated samples (*Sample Set #1* and *Sample Set #2*) are schematically depicted in **Fig. 3(a)**. Each sample contains a control region from where the graphene properties are directly extracted by means of terahertz spectroscopy and FTIR measurements. Sketches of the unit-cells for all the analyzed structures, together with details of their relevant geometric dimensions, are pictured in **Fig. 3(b)**. For details about the experimental methods, sample fabrication, and full-wave electromagnetic numerical simulation procedures please refer to the *Methods Section*.

At this end we proceed to study the *graphene-only plasmonic structures* (*sample Set #1*). These samples consist of disk-patterned single layer CVD graphene on top of a 4 µm thick polyimide (PI) substrate. The graphene conductivity was altered by means of chemical doping (see *Methods Section*). Disks were arranged on a hexagonal lattice as depicted in **Fig. 1(a)**; the disk radius ($d$) was set to $d$ = 23 µm, whereas the separation between centers of adjacent disks was set to 50 µm. From terahertz and FTIR transmittance measurements through the control region of the sample after successive cycles of chemical doping, sheet conductivity / relaxation-time values of: (*i*) 1.4 mS / 50.5 fs, (*ii*) 1.9 mS / 63.4 fs, and (*iii*) 2.4 mS / 57.4 fs were extracted. Full-wave electromagnetic simulations were performed for the disk array employing these extracted graphene parameters. The simulated extinction is depicted in **Fig. 4(a)**; cases (*i*), (*ii*), and (*iii*) are depicted in red, green, and blue-colored curves, respectively. A decrease in $E$ is observed as the graphene conductivity, thus $\omega_p$, is decreased, in accordance with Eqn. (5). Quality factors, $Q$, of 0.16, 0.23, and 0.24 are observed for cases (*i*), (*ii*), and (*iii*), respectively, as depicted in **Fig. 4(b)**. In all cases $Q \ll 1$ in accordance with the previous theoretical discussion. Shown in **Fig. 4(c)** is the measured extinction through the fabricated samples after each cycle of chemical doping together with an optical image (inset in the figure) depicting a section of the array; a good qualitatively agreement is observed between measurements (**Fig. 4(c)**) and the full-wave electromagnetic simulations (**Fig. 4(b)**).

We also studied samples consisting of *graphene/metal hybrid structures (Sample Set #2).* These samples consist of SRRs, whose geometry was optimized following the discussion in Ref. [24], but for a target resonance-frequency swing between 0.4 THz (low-frequency resonant-state) and 0.7 THz (high-frequency resonant-state). These structures consist of a 110 nm thick metallic metamaterial structure laying on top of a 4 µm thick PI substrate, and contain wet-transferred single-layer CVD graphene whose conductivity is reconfigured by means of stacking different number of layers following the procedure in Ref. [18] (see *Methods Section*). Shown in **Fig. 5(a)** is an optical microscopy image of one of the fabricated samples. The inner and outer radius ($R_{in}$ and $R_{out}$) were set to 110 and 120 µm, respectively, the gap width ($g$) was set to $g$ = 2 µm, and the unit-cell (separation between centers of adjacent rings) was set to 300 µm. The measured and simulated (HFSS) extinctions through these structures are depicted in **Figs. 5(b)** and **5(c)**, respectively. The resonance-frequency red-shifts as the overall graphene conductivity is increased, i.e. when the number of transferred graphene layers is increased from zero to five as shown by the blue (zero layers), red (one layer), yellow (two layers), violet (three layers), and green curves (five layers), respectively. The observed resonance red-shift in structures with three and five graphene layers is larger than that in structures containing one and two layers, which is owed to the transmission through the structure being a nonlinear function of the graphene conductivity. In this regard, it is worth noticing that it exists a particular conductivity level that maximizes the sensitivity of extinction to graphene conductivity. Overall the resonance shifts from 0.65 to 0.45 THz. The frequency-tunability range of this reconfigurable structure is set by the geometry of the metallic SRRs; as the graphene conductivity is enhanced, the resonance approaches that of the complete rings (black colored curves). This can be physically understood on basis of the equivalent circuit model depicted in **Fig. 2(b)**. A fit of the data to the equivalent circuit model together with a discussion on the effect of momentum relaxation time on the electromagnetic response of the structure, which is found to be negligible, is discussed in the *Supplementary Information* (**Figs. S2, S3,** and **S4**). **Figure 5(d)** shows the quality-factor (extracted from the full-wave simulation data plotted in **Fig. 5(c)**) as a function of graphene conductivity; the quality factor shows a non-monotonic dependence on conductivity. $Q$ ranging between 2.2 and 1.0 is observed when the graphene sheet conductivity is varied from $\sigma_{DC}$ = 0.15 to 2.4 mS. In this regard is worth mentioning that the quality factor at the limit resonant-states are 3.0 (no-graphene) and 1.5 (complete ring), respectively.

Finally, we analyzed structures consisting of concentric ring resonators with interdigitated fingers (ICRs) (*Sample Set #3*). As discussed by Jensen *et al* [36], such structures can attain even larger quality factors at terahertz frequencies. In order to provide for reconfiguration, the structure is modified by inserting graphene on a section of the gap between the rings and the interdigitated fingers as depicted in **Fig. 3(b)**. A 4 µm thick PI substrate is assumed. The inner and outer radius ($R_{in}$ and $R_{out}$) were set to 35 and 90 µm, respectively; the width of each ring ($l$) was set to $l$ = 20 µm, thus leading to a distance between rings ($d + a$) of 15 µm. The finger length ($w$) was set to $w$ = 2 µm, whereas $d$ was set to 12 µm. The unit-cell was set to 300 µm. The length of the gap where graphene is placed ($g$) is varied in order to optimize the sensitivity of the structure to a typical graphene sheet conductivity swing; a swing between 0.15 mS and 2.4 mS was assumed in accordance with previous discussions. Depicted in **Figs. 6(a-c)** is the simulated (HFSS) extinction when the graphene sheet conductivity is varied. Three cases are analyzed, corresponding to $g$ = 0.25 um, $g$ = 0.75 um, and $g$ = 1.25 um, which are depicted in **Figs. 6(a)**, **6(b)**, and **6(c)**, respectively. In a similar way to what has been shown in *Sample Set #2*, as the graphene conductivity is increased, the resonance red-shifts; the structure again shifts between two resonant-states, which are uniquely set by the metal geometry. For smaller gaps (**Fig. 6(a)**), the resonances tend to cluster near the low-frequency resonant-state, whereas for large gaps (**Fig. 6(c)**), the resonances tend to cluster near the high-frequency state. An optimal gap-length ($g$ ~ 0.5 µm) exists. Depicted in **Fig. 6(d)** is the resonance-frequency versus graphene conductivity for the three analyzed gap-lengths showing the latter effect; moreover, it can be observed that for a fixed graphene conductivity as the gap-length is increased the resonance-frequency increases. This can be understood as a result of the associated gap-capacitance ($C_{gap}$) decreasing as the gap-length is increased. Therefore, as the gap-length is increased the overall capacitance associated with the structure decreases, which results in a resonance blue-shift. When the graphene conductivity is varied, the quality factor ranges between 3 and 5 (see **Fig. 6(e)**); this is ~ 3X larger than what we observed in *Sample Set #2* and > 10X larger than what was observed in Sample Set #1. However, the resonance-frequency tunability range attainable in *Sample Set #3* is just 40 GHz (~ 0.47 to ~0.51 THz), which is much smaller than what is achievable in *Sample Set #1* (unconstrained frequency tunability) and *Sample Set #2* (~200 GHz)

## 4.5 Methods

The fabrication procedure for all samples is analogous to the process we reported in Ref. [18]. It starts with coating a glass substrate with polyimide (PI-2600) and curing it by gradually increasing the temperature to 300°C. Fabrication for *Sample Set #1* is followed by transfer of a graphene layer. Single-layer graphene films (monolayer percentage >95%) grown by means of CVD on a copper foil were obtained from a commercial vendor (Bluestone). These films are transferred to the samples using PMMA and wet etch methods [37]; in this process, graphene grown on copper foil is first covered by means of spin-coating PMMA (950-PMMA-C2). Then, the graphene on the back side of the foil, i.e. the side not covered by PMMA, is etched employing reactive ion etching (RIE). Subsequently, the copper foil is etched in copper etchant (COPPER ETCHANT APS-100), and the suspended graphene-on-PMMA layer is moved to DI water. Finally, transfer into the desired substrate is carried out in the DI water. Depicted in the *Supplementary Information* (**Fig. S5**) are typical Raman spectra for our graphene films showing characteristic features of single layer graphene. The transferred films are then patterned by means of photolithography and RIE steps. Finally, the fabricated sample is peeled-off from the glass substrate, forming a flexible, free-standing, and ultrathin film. In *Sample Set #2*, after the PI film is coated and cured, we proceed to define the metal pattern. For this purpose, photolithography (Shipley 1813 photoresist process) and metal deposition processes are employed. Ti/Au with thickness of 10nm/100nm are deposited so to define the metallic structure. After this, one to five layers of graphene are transferred. Finally, we perform lithography and RIE steps so that graphene remains just in the gap region.

In *Sample Set#1*, the graphene conductivity is varied by means of chemical doping. In this regard, nitric acid ($HNO_3$) is well known to be a *p*-type dopant for graphitic materials such as carbon nanotubes and graphene [37-38]. Chemical doping was realized by immersing a single-layer graphene in a $HNO_3$ solution, followed by a dry and baking step at 85℃, as we previously reported in Ref. [18]. In these samples, no change in the doping level was observed over time. In *Sample Set #2,* we observed that the metallic film defining the metamaterial peels-off when immersing the sample in nitric acid solution. This is the reason why the conductivity in *Sample Set #2* was varied by means of stacking graphene layers rather than by means of chemical doping.

It is also important to mention that in order t*o account for the process induced variations on the graphene properties so to therefore improve the accuracy when comparing our experimental results with simulations and theory, each fabricated sample is composed of two regions: (i) a test region, consisting of the structure under test (left part of the sample), and (ii) a control region, consisting of un-patterned-graphene (right part of the sample)*, as shown in **Fig. 3(a)**. The graphene properties are directly extracted by means of terahertz spectroscopy and FTIR measurements over the control region of each sample. The optical conductivity of graphene is determined by fitting the transmission spectra through the control region to an analytical model. In this regard, a Drude model is employed to represent the frequency-dependent graphene optical conductivity, as dictated by Eqn. (3). The experimentally extracted relaxation time ($\tau$) and DC sheet conductivity ($\sigma_{DC}$) of graphene were then used to validate our experimental results with simulations. A discussion on the fitting and typical results obtained from this procedure are depicted in the *Supplementary Information* (**Fig. S6**) showing an excellent fit to a Drude model.

Full-wave 3-D electromagnetic numerical simulations were conducted employing High Frequency Structure Simulator (ANSYS HFSS) by considering graphene as a zero-thickness graphitic layer (layered impedance) with sheet conductivity set to the experimentally determined values. In this regard it is worth mentioning that results using a finite thickness model for graphene (e.g. [17]) match those obtained employing the layered impedance model as discussed in our early work (Ref. [18]). Depicted in the *Supplementary Information* (**Fig. S7**) is a detail of some of the simulated geometries as set in HFSS. It can be observed that the simulated geometries represent exactly the fabricated ones and fully take into account the effect of the substrate. In this regard, a relative permittivity of 3.24 was assumed for polyimide [39-40]. Periodic boundary conditions were set around the unit cells so to represent an infinite array and to account for interactions between adjacent elements.

For terahertz characterization, a CW THz spectrometer (Toptica Photonics) was employed, which is based on InGaAs photo-mixers at 1550 nm for both generation and coherent detection [41]. A sketch of the system, which can cover a spectral range from 0.1 to 2 THz, is depicted in the *Supplementary Information* (**Fig. S8**). A FTIR system (Bruker IFS-88) was employed for finding the spectra at shorter wavelengths (i.e. above 4 THz) thus extract the momentum relaxation time through the control regions of the samples by means of fitting the measured transmittance to an analytical model, as previously discussed. All the reported terahertz data is normalized to a reference polyimide substrate.

## 4.6 Conclusions

In this work we analyzed and compared the terahertz response of graphene-only plasmonic and graphene-metal hybrid metamaterial structures. In these structures, when the graphene conductivity is altered, e.g. by means of changing its Fermi level, the resonance is shifted. From theory, simulations, as well as supporting experiments, we find that graphene-metal hybrid structures can offer: (*a*) stronger resonances, (*b*) larger quality factors, and (*c*) less sensitivity to the graphene quality (given by momentum relaxation time) than the graphene-only plasmonic structures. However this occurs at the expenses of a limited resonance frequency-shift and increased unit-cell footprint. Overall, we conclude that at terahertz frequencies graphene might result more attractive as a reconfigurable media providing tunability to, otherwise passive, metallic structures rather than as a reconfigurable plasmonic material by itself.

**ACKNOWLEDGEMENT**

The authors acknowledge the support from the NSF MRSEC program at the University of Utah under grant no. DMR 1121252. B.S.-R. and H.C.Q. also acknowledge support from the NSF CAREER award no. 1351389 and the NSF award ECCS #1407959. Students N.R. and C.N. acknowledge support through the University of Utah MRSEC REU program.

(a)

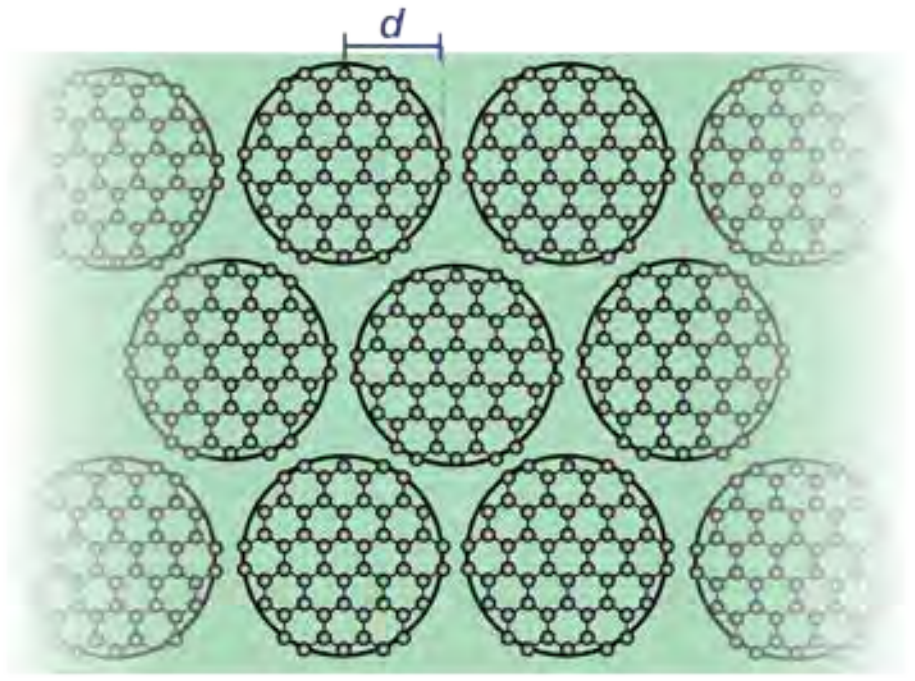

(b)

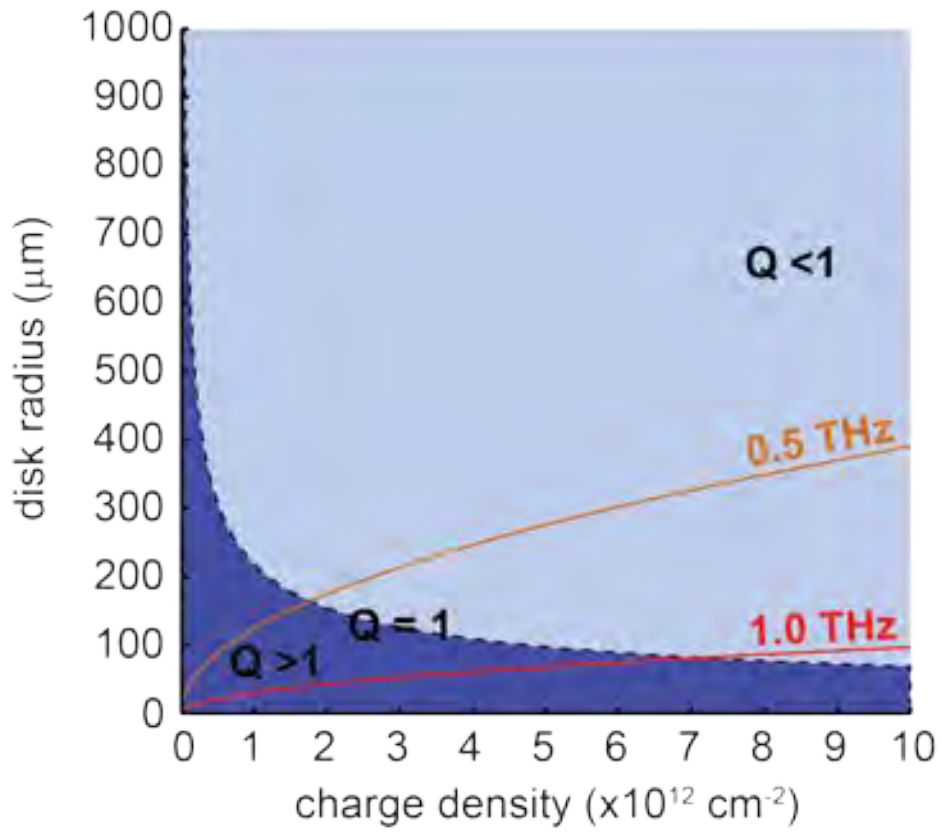

**Fig. 1. (a)** Sketch of a plasmonic structure consisting of an array of graphene disks. **(b)** Contour plots of quality factor (*Q*) and resonance-frequencies as a function of charge density and disk radius (in the limit when assuming the extinction to approach unity).

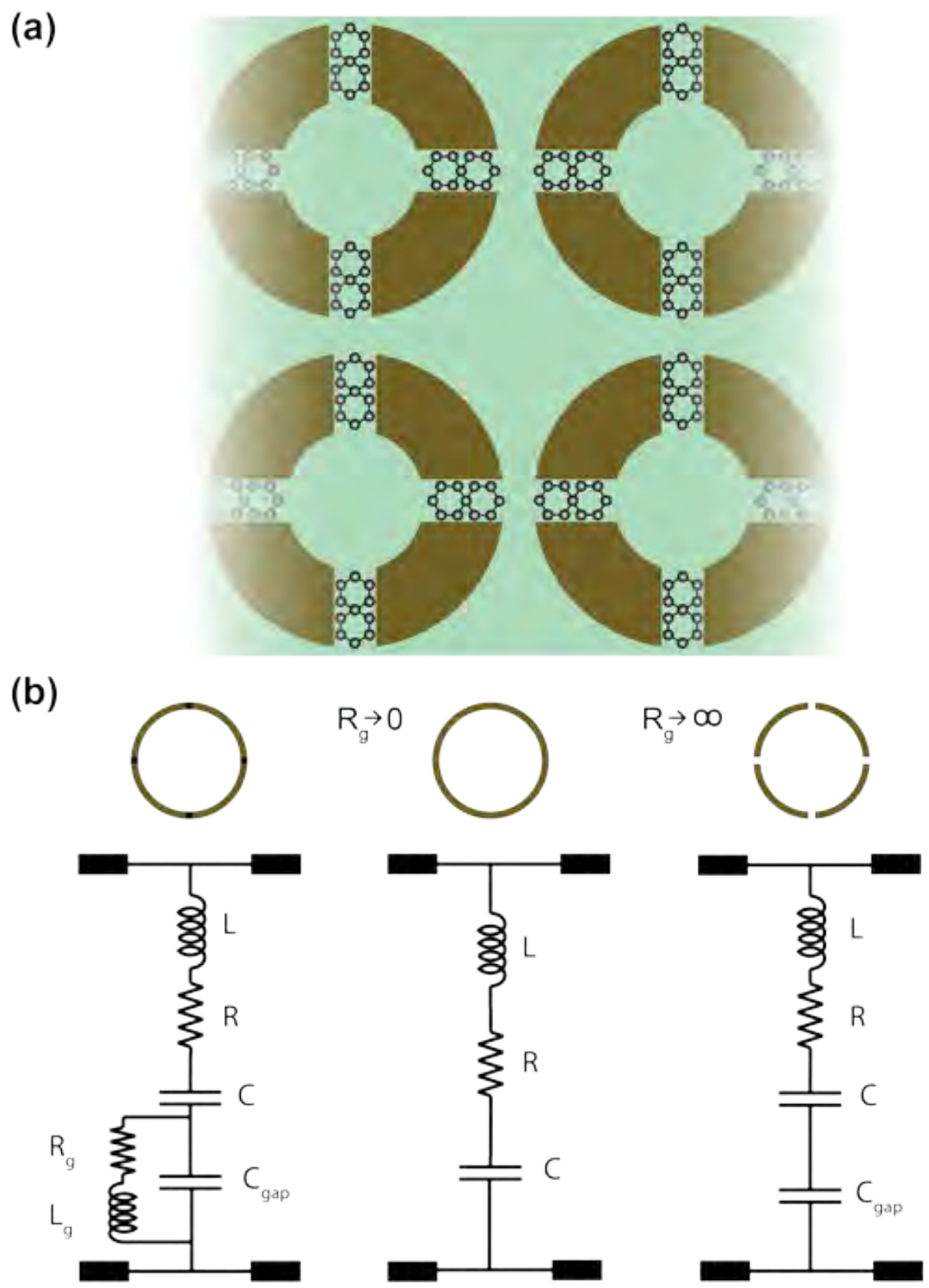

**Fig. 2.** **(a)** Sketch of a graphene/metal hybrid metamaterial structure consisting of SRRs with graphene placed inside the gaps. **(b)** Equivalent circuit model for the structure. Graphene is modelled as an impedance of value $Z_g = R_g + i\omega L_g$.

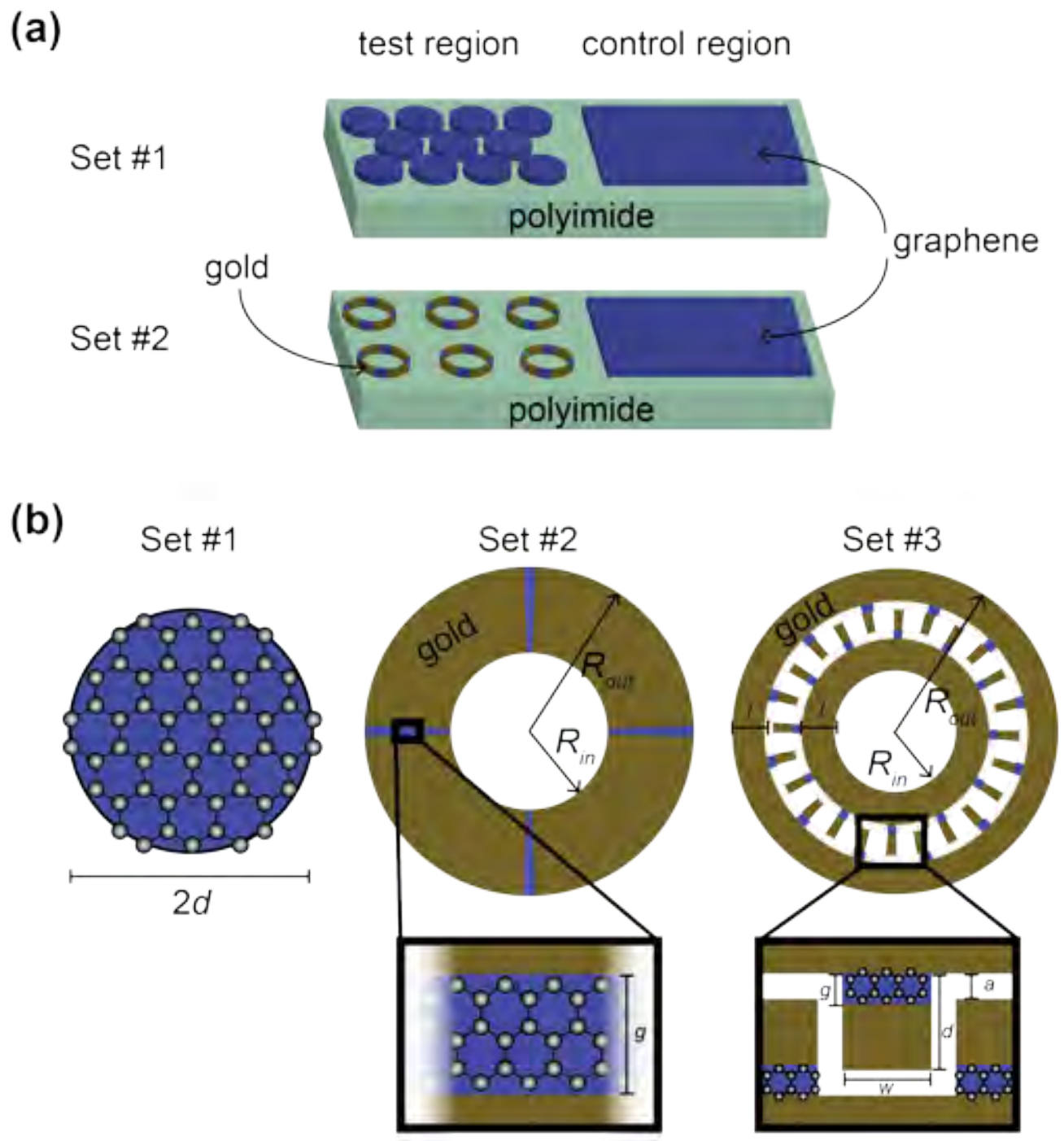

**Fig. 3. (a)** Schematic of the fabricated samples (*Sample Set #1* and *Sample Set #2*). The right half of the samples (control region), which contains un-patterned graphene in top of polyimide, is used to monitor the sheet conductivity of graphene. **(b)** Unit-cell detail and parameter definition for all the analyzed structures.

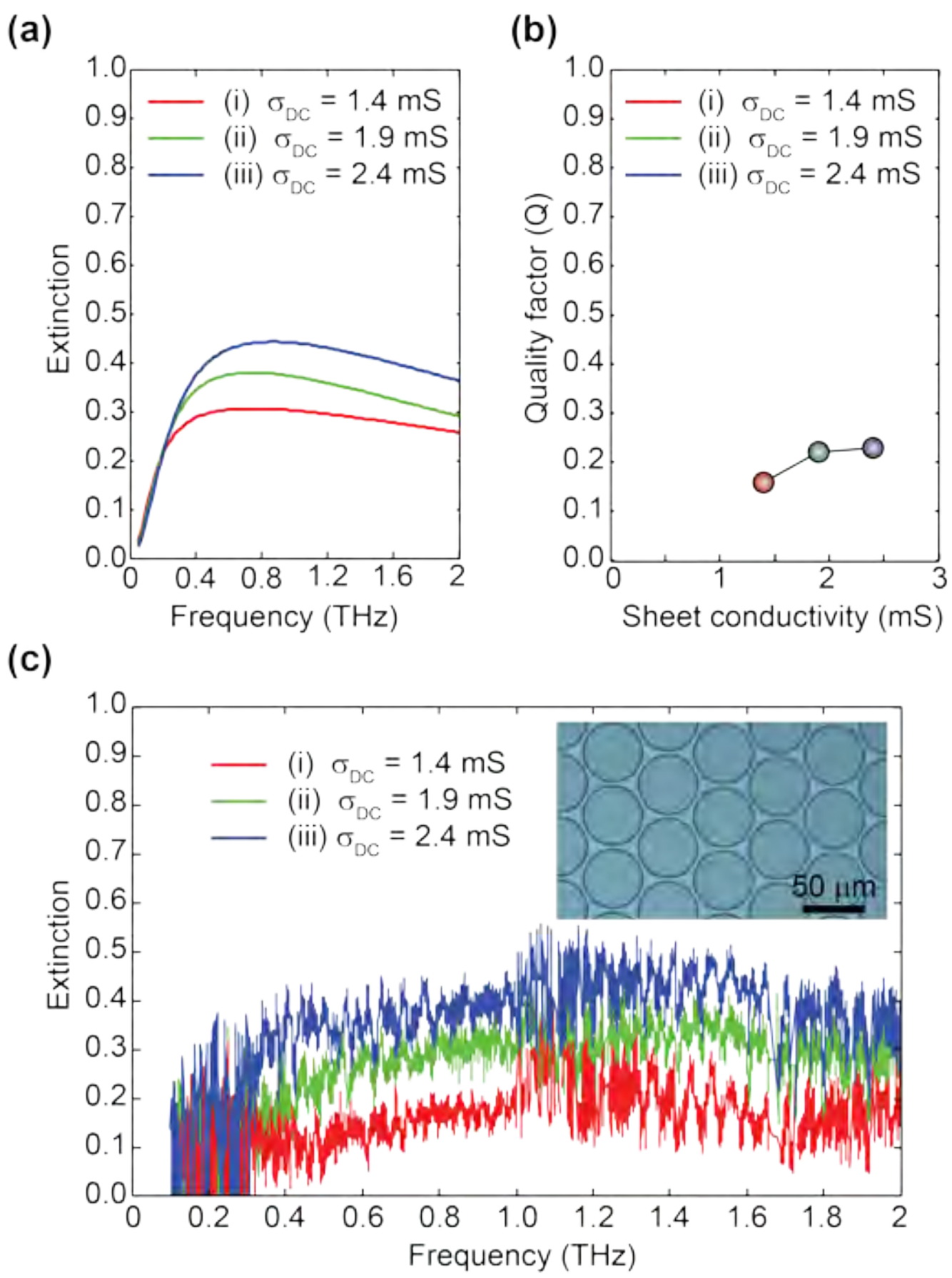

**Fig. 4. (a)** Simulated extinction versus frequency for Sample Set #1. **(b)** Quality factor as a function of sheet conductivity (extracted from the data in (a)). **(c)** Measured extinction versus frequency for Sample Set #1. The inset depicts an optical micrograph detail of a fabricated sample before removing the photoresist.

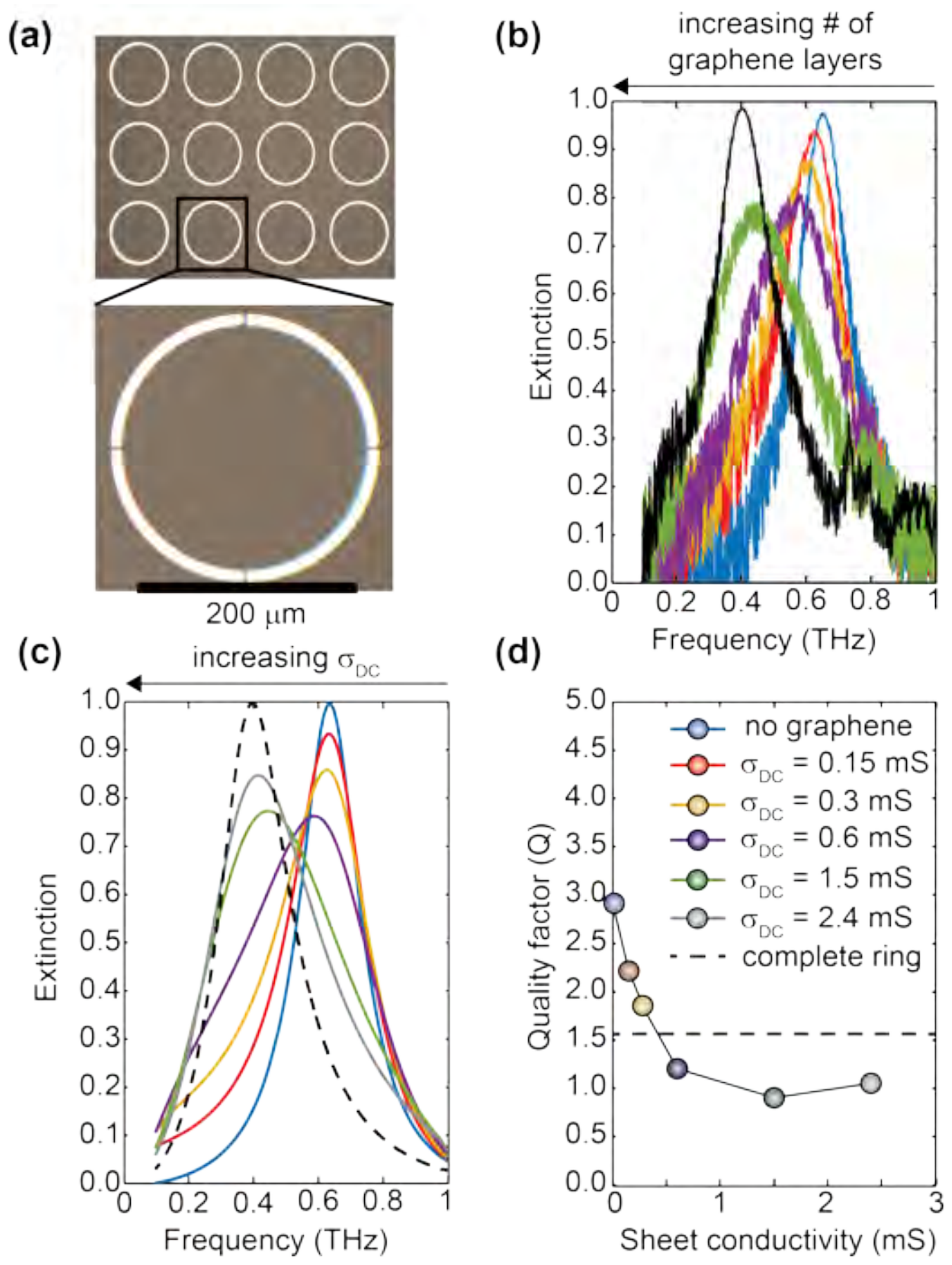

**Fig. 5. (a)** Optical image showing a detail of the graphene/metal hybrid metamaterial structure consisting of SRRs (*Sample Set #2*). **(b)** Measured extinction versus frequency for the sample depicted in (a) when varying the number of stacked graphene layers in the gap. **(c)** Simulated extinction versus frequency. **(d)** Extracted quality factor as a function of graphene sheet conductivity.

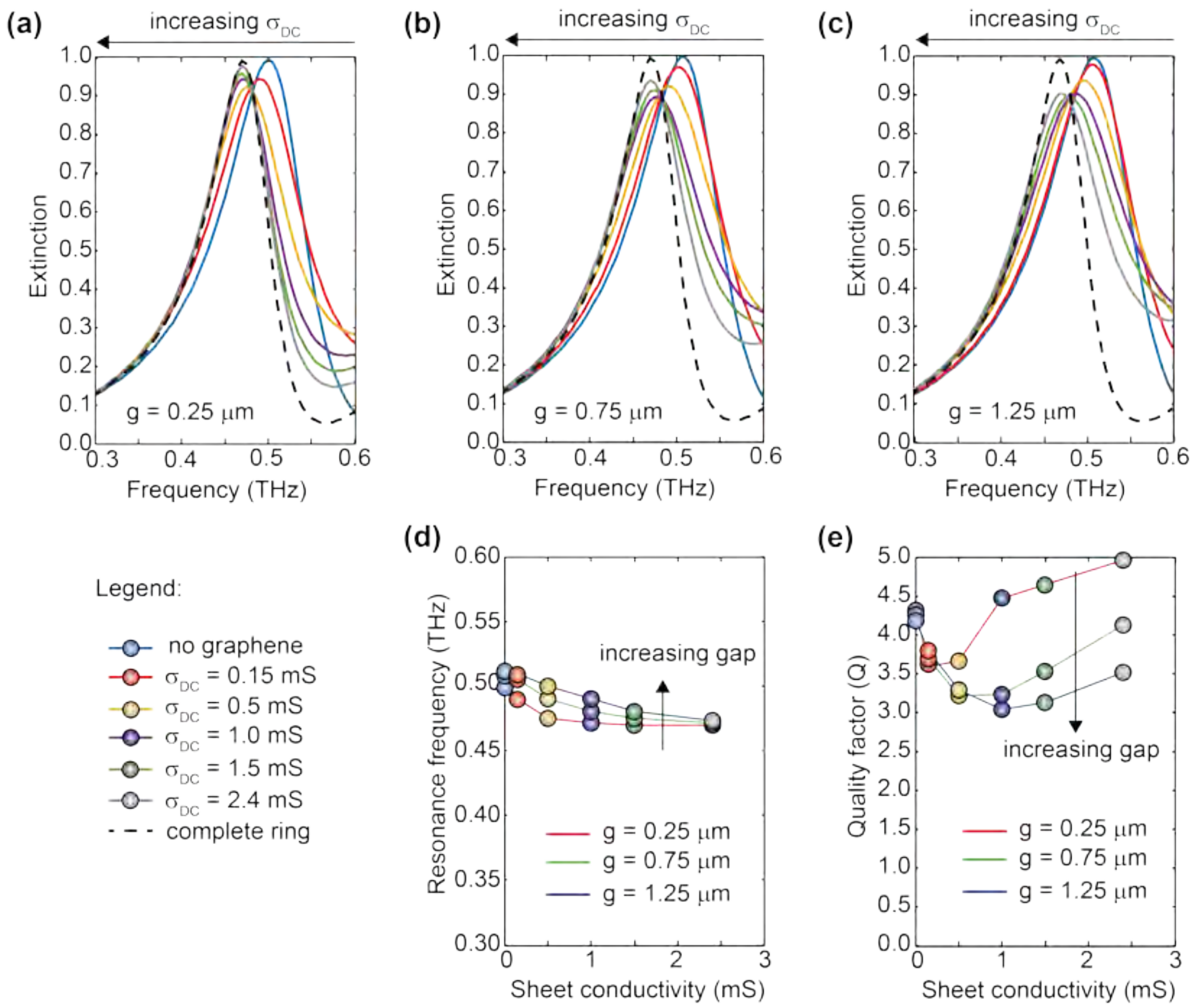

**Fig. 6. (a)-(c)** Simulated extinction versus frequency for *Sample Set #3*. Cases (a), (b), and (c) correspond to $g$ = 0.25, 0,75, and 1.25 μm, respectively. **(d)** Extracted resonance-frequency versus sheet conductivity for the cases depicted in (a)-(c). **(e)** Extracted quality factor versus sheet conductivity for the cases depicted in (a)-(c).

CHAPTER 5

# TUNABLE TERAHERTZ METAMATERIALS EMPLOYING LAYERED 2D-MATERIALS BEYOND GRAPHENE

Abstract

In this work we extend recent investigations on graphene/metal hybrid tunable terahertz metamaterials to other two-dimensional (2D) materials beyond graphene. For the first time, use of a non-graphitic 2D-material, molybdenum disulfide ($MoS_2$), is reported as the active medium on a terahertz metamaterial device. For this purpose, high-quality few-atomic layer $MoS_2$ films with controlled numbers of layers were deposited on host substrates by means of pulsed laser deposition (PLD) methods. The terahertz conductivity swing in those films is studied under optical excitation. Although no-appreciable conductivity modulation is observed in single-layer $MoS_2$ samples, a substantial conductivity swing, i.e. 0 to ~0.6 mS, is seen in samples with ~60 atomic layers. Therefore, although exhibiting a much smaller maximum terahertz conductivity than that in graphene, which is a consequence of a much smaller carrier mobility, $MoS_2$ can still be employed for terahertz applications by means of utilizing multilayer films. With this in mind, we design and demonstrate optically-actuated terahertz metamaterials that simultaneously exhibit a large modulation depth (i.e. >2X larger than the intrinsic modulation depth by a bare $MoS_2$ film) and low insertion loss (i.e. <3 dB). The advantages of using a 2D-material with a bandgap, such as $MoS_2$, rather than a gapless material, such as graphene, are: (a) a reduced insertion loss, which is owed to the possibility of achieving zero minimum conductivity, and (b) an enhanced modulation depth for a given maximum conductivity level, which is due to the possibility of placing the active material in a much closer proximity to the metallic frequency selective surface, thus allowing us to take full advantage of the near-field enhancement. These results indicate the promise of layered 2D-materials beyond graphene for terahertz applications.

*Index Terms*— Terahertz, Metamaterials, Optoelectronics, 2D materials, Graphene, $MoS_2$.

## 5.1 Introduction

The terahertz region of the spectrum comprises the broad set of frequencies lying between the microwave and the optical spectral ranges. Terahertz technology promises multiple diverse applications including imaging, communications, spectroscopy, security, astronomy, and so on [1-2]. However, the scarcity of active materials capable of efficiently responding and effectively manipulating terahertz waves currently limits the progress in developing many of these applications. In this regard, tunable terahertz metamaterials have been proposed as an efficient means for manipulating terahertz wave propagation. Significant research has been carried out over the last decade on tunable metamaterial terahertz modulators enabling amplitude [3] or phase modulation [4], and employing various actuation mechanisms, i.e. including: optical [5-6], electrical [3-4, 7-8], thermal [9], and mechanical [10] actuation. In this context, graphene has been proposed as a promising active material for reconfigurable terahertz applications because it can be integrated into a wide variety of substrates, its low cost, and its excellent terahertz response [11-18]. However, a key design factor (i.e. the possibility of tailoring the electric field enhancement in graphene) has been largely unexplored [19].

Graphene-based metamaterials demonstrated to-date usually comprise two elements: (*i*) a passive metallic frequency selective surface (FSS), and (*ii*) a single layer -or multiple layers- of graphene. Although in most reported metamaterials the graphene layers are placed in the same plane as the FSS, vertically stacking these layers and optimizing the separation between them can lead to better modulation performance. In this regard, Yan *et al.* showed both theoretically as well as experimentally that the field enhancement in graphene could be controlled via tailored placement of the graphene layers away from the plane of the FSS [20]. In general, the higher the electric-field enhancement in the plane of graphene, the more sensitive the terahertz transmission is to the graphene conductivity. However, close placement is often accompanied by a high insertion loss, which is owed to graphene having a finite minimum conductivity; i.e. if graphene is placed very close to the FSS the minimum conductivity of graphene will lead to a large loss in the terahertz transmission through the structure. Therefore, in graphene-based metamaterials, there exists a trade-off between modulation depth and insertion loss, which needs to be taken into account (and optimized) when designing the geometry of these structures [20]. Proof-of-principle demonstration of this phenomena was shown using graphene layers, which were separated by a polyimide (PI) spacer from a passive metallic FSS. By placing the active graphene layers at various distances from the FSS, it was demonstrated that: (*a*) it is possible to tailor the strength of the interaction between the terahertz waves and graphene, which is equivalent to arbitrarily enhancing the effective conductivity of graphene; and, (*b*) for a given conductivity swing in graphene, there exists an optimal placement leading to optimal modulation depth versus insertion loss tradeoff. *In this work we extend this hybrid terahertz metamaterial concept to other 2D-materials beyond graphene.*

Our discussion will be focused on metamaterials employing molybdenum disulfide ($MoS_2$). In recent years, a number of investigators have reported on time-resolved terahertz conductivity measurements in 2D-$MoS_2$ films. For instance, Strait *et al* performed low temperature mobility measurements in multilayer $MoS_2$ and reported acoustic phonon scattering as the mobility limiting mechanism [21]. Studies on exciton annihilation in optically excited monolayers of $MoS_2$ were performed by Sun *et al.* [22]. Furthermore, Docherty *et al.* demonstrated ultrafast photoconductive response (< 1ps) in a single layer of $MoS_2$ [23]. These early works shed light on the potential of molybdenum disulfide for use in terahertz applications. More recently, *Cao et al*. reported terahertz modulators based on large-area multilayer $MoS_2$ films in Si-substrates [24]. The enhancement in the terahertz modulation was achieved by *p*-type doping of the $MoS_2$ sample. Moreover, a recent study by Chen *et al.* [25], showed an optically pumped Si-$MoS_2$ terahertz modulator with a high modulation depth (~75%). However, (i) optical excitation in these experiments not only altered the terahertz optical properties of the $MoS_2$ films, but also altered the properties of the Si substrate, and (ii) large levels of modulation depth were observed in cases where the $MoS_2$ samples were intrinsically doped (i.e. there was an initial finite terahertz conductivity even in the absence of optical pump). Thus, these reported modulation depth levels were not only heavily dependent on the initial doping in $MoS_2$, but also, more importantly: *the fact that large modulation was observed could be attributed to an enhanced sensitivity of the terahertz transmission to the terahertz conductivity swing in the Si substrate under optical illumination, rather than by an enhanced terahertz absorption by* $MoS_2$*.*

## 5.2 Sample fabrication

The $MoS_2$ films employed in this work were fabricated using a pulsed laser deposition technique.

For this purpose, a pulsed KrF excimer laser with a wavelength of 248 nm and pulse width of 25 ns was used along with a polycrystalline $MoS_2$ target, which was made by pressing $MoS_2$ powder. The substrate, single crystal sapphire with (0001) orientation, was kept at 700˚C and at a vacuum pressure of $1\times10^{-6}$ torr during the deposition process. *The quantity of $MoS_2$ layers in these films was controlled by means of controlling the number of laser pulses*. Further details on the experimental growth procedures and on the quality of the films are given in [26].

*Three samples, with different number of $MoS_2$ layers are analyzed in this work. These samples contain 1, 4, and 60 atomic layers of $MoS_2$*. Figure 1 shows an optical image of the samples organized by number of layers (low-to-high from left-to-right). These $MoS_2$ samples were studied by means of terahertz time domain spectroscopy (with and without optical excitation) from where their dynamic terahertz response could be extracted (*Section III. A*). Following the metamaterial concept reported in Ref. [20], we then proceeded to fabricate $MoS_2$/metal hybrid structures. To fabricate these structures, we spin coated polyimide (PI-2600) on the $MoS_2$-on-sapphire samples. The polyimide was cured by gradually increasing temperature to 300˚C. The sample was held at this set point for 30 minutes and then gradually cooled down to room temperature. Samples with varied polyimide spacer thickness were fabricated by means of altering the spin casting conditions. Finally, lithography, metal deposition, and lift-off steps were performed to define the metallic FSS. In Fig. 2a, we show a sketch of the final metamaterial structure. Either Al (in structures with finite PI spacer thickness) or Ti/Au (in structures with no PI spacer) were employed as the FSS metals. The choice of metal arises from the fact that Al adheres well to PI but Ti/Au does not adhere; conversely, Al does not adhere to $MoS_2$ while Ti/Au does. The overall metal thickness was set to 100 nm. Figure 2b shows an optical microscope image of a section of the fabricated metamaterial device and a detail of a unit cell.

## 5.3 Results and discussion

### 5.3.1 terahertz conductivity of PLD grown MoS2 films.

Samples were measured using a conventional terahertz time domain spectroscopy (TDS) set up, where the terahertz beam was generated via optical rectification in a ZnTe crystal pumped using an 810 nm amplified Ti-Sapphire laser with a pulse width of 75 nm and repetition rate of 1 kHz. The generated terahertz beam was normally incident on the samples that were placed between a set of parabolic mirrors. The transmitted THz beam was measured using an optical probe beam employing a second ZnTe crystal using an electro-optic sampling technique [27]. The measured time-domain terahertz waveform was Fourier transformed to extract the frequency response of the sample.

To optically excite the $MoS_2$ samples, a broadband 250 W quartz tungsten halogen (QTH) lamp color temperature of 3400 K was employed. The lamp radiation was focused onto the sample yielding an optical intensity of 5 $W/cm^2$. A detailed diagram of the terahertz measurement setup is depicted in Fig. 3.

Each sample was measured with and without optical excitation. The extracted terahertz spectra was normalized to that of a reference sapphire substrate under the same illumination conditions. We found that samples measured in the absence of optical excitation (dark) showed unity transmittance, corresponding to zero terahertz conductivity, which is an evidence of the as-grown samples being intrinsically undoped. However, when they were optically excited using the halogen lamp we observed evidence of a finite terahertz conductivity that increased with number of layers.

The measured terahertz spectra for samples with 1, 4 and 60 atomic layers of $MoS_2$ (under lamp illumination) is depicted in Fig. 4, the extracted terahertz conductivities under these conditions were: 0.01 ± 0.07, 0.42 ± 0.06, and 0.65 ± 0.08 mS, respectively. The reason why the conductivity does not scale linearly with the number of layers is unclear at this point and will be the subject of future investigations. In the case of a single layer film, where the transmission is very close to 1.0, drift in the laser power can lead to an apparent transmission that is greater than unity at some frequencies.

### 5.3.2 $MoS_2$/metal hybrid metamaterials: numerical simulations.

In order to predict the behavior of $MoS_2$/metal hybrid metamaterials, numerical simulations were performed using ANSYS HFSS. The unit cell dimensions were taken as 265 x 265 $\mu m^2$, and the length and width of the cross-shaped apertures were set to 194 and 23 µm, respectively, as depicted in Fig. 2b. $MoS_2$ films were modeled as layered impedances due to their small thickness, i.e. a few atomic layers, which is much smaller than the relevant terahertz wavelengths. A relative permittivity of 10 was assumed for $MoS_2$ in accordance with Ref. [28]. No frequency dispersion was considered at this end. *In the first set of simulations, we varied the PI spacer thickness for a given $MoS_2$ conductivity*. Figure 5a depicts the simulated transmittance spectra for 0 mS and 0.65 mS conductivity levels, respectively, in agreement with the maximum conductivity swing observed in our films (see Fig. 4). In accordance with the discussion in Ref. [20], *it was observed that maximum modulation depth takes place in structures without PI spacer, which results from a stronger near-field light-matter interaction under this condition*.

We also analyzed the effect of electron relaxation time ($\tau$) on the terahertz transmittance. For this purpose, the $MoS_2$ films were modelled employing a Drude dispersion. Assuming an effective mass of $0.45m_0$ [29-30] and mobility ranging between 2 and 100 $cm^2/V.s$ [31], which correspond to typical reported values in $MoS_2$, it can be observed that a common range of values for its electron momentum relaxation time should be between ~1 and ~25 fs. The Drude model used in our simulations is therefore of the form:

$$\sigma = \sigma_{DC}/(1 + \omega^2 \tau^2); \quad \epsilon = \epsilon_r - (\sigma_{DC}\tau/\epsilon_0)/(1 + \omega^2 \tau^2),$$

where $\varepsilon_r = 9$ is employed in agreement with the results in Ref. [32] for multilayer $MoS_2$. In the above expression $\sigma_{DC}$ corresponds to the DC conductivity of the $MoS_2$ film (in [S/m] as per the HFSS definition) and $\varepsilon_0$ is the vacuum permittivity. A thickness of 5 nm was assumed during the constitutive parameter definition, however results are independent of the assumed thickness due to the material being effectively modelled in HFSS as a zero-thickness layered impedance. It is also worth mentioning that the transmission is dominated by the components arising from $\sigma_{DC}$, thus the results do not either depend on the assumed value for $\varepsilon_r$. In Fig. 5b, we show the simulated transmittance versus frequency for metamaterial samples with 2 μm PI spacer thickness and a $MoS_2$ conductivity level of 0.65 mS for various relaxation times (0, 1, 10, and 25 fs). Aside from a small red-shift in resonance, which is accompanied by a small increase in transmittance, no substantial differences are observed between all these four cases. *From this point of view, employing a frequency independent conductivity model can be assumed as a good (first order) approximation when modeling $MoS_2$ films*. This is consistent with our previous observations in graphene reported in Ref. [20].

### 5.3.3 experimental results on optically-actuated $MoS_2$/metal hybrid metamaterials.

We experimentally analyzed the terahertz transmission, with and without illumination, through multiple fabricated metamaterial samples. The samples were illuminated from the substrate side so that similar amount of light flux reaches the $MoS_2$ layers upon illumination as in the case without FSS. As predicted by our simulations, the resonance in these structures occurs in the 0.2 to 0.4 THz frequency interval. In Fig. 6a, we show the measured transmittance through structures with varied PI spacer of $d = 0$, 2, and 6 μm, with and without illumination. We find that the closer the $MoS_2$ film is to the FSS, the largest the modulation depth, in accordance with the discussion in [20] and with the simulation results discussed in the previous section. This observation is a result of the strong light-matter interaction in the near field (see Fig. 5(c)). Moreover, it is noticed that the resonance blue-shifts as the PI spacer thickness is increased; the reasons behind this trend will be discussed in *Section III. C*. In order to evaluate the effect of conductivity swing, or more precisely, maximum attainable conductivity, on the modulation performance, we measured the transmittance through samples with different number of $MoS_2$ layers (4 and 60 layers) and no polyimide spacer. The results from these measurements are shown in Fig. 6b. Not surprisingly, and in agreement with the simulation results depicted in Fig. 4, we find that samples having larger number of $MoS_2$ layers provide a larger modulation depth. *The maximum experimentally achieved modulation depth in these $MoS_2$/metal hybrid metamaterial structures is ~20%, i.e. >2X larger than the intrinsic modulation by the same $MoS_2$ films when on the bare sapphire substrate.*

Finally, it is worth mentioning that the fact that the experimentally observed modulation is much smaller than what our simulations predict, as depicted in Fig. 5a, and also smaller than what we observed in our previous work utilizing graphene [20], is not because of the maximum conductivity attainable in $MoS_2$ being small, but rather a result attributable to the following two mechanisms: (*i*) since the $MoS_2$ films are grown on top of a sapphire substrate ($n = 3.3$), they are immersed in larger surrounding refractive index than the ones in our previous work, in which polyimide ($n$ = 1.7) was the substrate. Overall, an environment with larger refractive index reduces the sensitivity of the terahertz transmission to the $MoS_2$ conductivity. (*ii*) In our samples, the $MoS_2$ films are placed in close proximity to the metallic FSS, which alters the optical field at the $MoS_2$ plane. Consequently, most of the optical excitation power could be reflected by the overall structure rather than being absorbed by the $MoS_2$ film as in the case without FSS. As a result, the effective optical power in the $MoS_2$, which determines the terahertz conductivity level in the material, can be much smaller than that in structures without an FSS. From this point of view, when placed in metamaterial structures and under optical illumination, the effective conductivity swing in $MoS_2$ might dramatically reduce from that under the experimental conditions in Fig. 4.

Finally, it is worth mentioning that in terms of operation speed, the transient evolution of photo-excited carriers in $MoS_2$ is much slower than that in graphene. In this regard, decay times on the order of nano-seconds have been observed [33-34], which is longer than the pico-second times typically observed in graphene [35]. From this perspective, the maximum attainable operation speeds would be 3 orders of magnitude slower in $MoS_2$ structures than in graphene structures. Moreover, if employing electrical actuation so to reconfigure the $MoS_2$ conductivity, we will also expect a much slower response due to the smaller carrier mobility in $MoS_2$ [31] than in graphene.

### 5.3.4 Equivalent circuit model

*The trends observed from our experimental and simulation results can be analytically described by representing the metamaterial structure via employing an equivalent circuit model*, as discussed in Ref. [20]. In this model, the $MoS_2$ film can be modelled as a parallel impedance of value $1/\sigma_{MoS2}$, where $\sigma_{MoS2}$ corresponds to an “effective” terahertz conductivity, which can account for the near-

field enhancement effects [20]. In the absence of optical excitation, $\sigma_{MoS2}$ is zero, and the simulated transmission spectra can be described by:

$$\frac{T}{T_0} = \left|\frac{2}{2+j\omega C Z_o + Z_o/(j\omega L + R)}\right|^2. \quad (1)$$

The parameters R, L and C, derived from the equivalent circuit model depicted in Fig. 8a, can be extracted by fitting of the numerically simulated transmission spectra to Eqn. (1). In order to estimate these parameters for our particular FSS geometry and dielectric environment, we fitted our numerical simulation results for a device geometry with no PI spacer to the above formula, from where it was determined that: R = 0.35 Ω, L = 9.8 pH and C = 32 fF. A very good agreement is observed between the model and the simulation data, as shown in Fig. 7b.

In addition, by analyzing Eqn. (1), we see that the resonance frequency in the metamaterial is given by:

$$\omega_0 = \sqrt{\frac{1}{LC} - \frac{R^2}{L^2}}. \quad (2)$$

Therefore, the fact that the resonance blue-shifts as *d* is increased can be understood from a decrease in the equivalent circuit capacitance, which is owed to a diminished effective permittivity seen by the FSS as the PI spacer thickness is increased. This trend is opposite to the tendency observed in our previous work [20]. The reason behind this difference is that whereas in our previous work increasing PI thickness increases the effective capacitance due to the PI layer being placed on top of the FSS, in our current work, the PI spacer is placed below the FSS rather than on top. Thus, since the refractive index of sapphire is larger than that of PI, it follows that the effective capacitance should decrease as *d* is increased. Moreover, although the FSS dimensions in this work are the same as those in our previous work, we notice that the equivalent circuit capacitance is ~3X larger, which can also be explained from the larger terahertz permittivity of sapphire with respect to that in polyimide.

Furthermore, when analyzing the transmittance at resonance, it can be shown that:

$$\frac{T}{T_0} = \left| \frac{1}{1+\frac{Z_0 RC}{2L}} \right|^2 . \tag{3}$$

From Eqn. (3), two observations can be made: (a) the fact that the transmittance levels under no illumination through the metamaterial structures studied in this work, where sapphire is the substrate, is on the order of ~70%, whereas in our prior work, where polyimide substrates were employed, transmittance was >80%, is a result of having a larger equivalent circuit capacitance; and (b) since as C decreases the transmittance increases; therefore, the transmittance peak does not only blue-shifts as *d* is increased, but it also increases in magnitude as clearly depicted in Fig. 5.

## 5.4 Conclusion

In conclusion, we have analyzed 2D-material/metal hybrid tunable terahertz metamaterials. We showed for the first time that a layered 2D material other than graphene could be used. Through optical-actuation in few-atomic layer $MoS_2$ films, we showed that terahertz metamaterials simultaneously exhibiting modulation depth levels >2X larger than the intrinsic modulation depth by the $MoS_2$ film and low insertion loss (<3dB) can be realizable. The trends observed in our experimental measurements as well as in our simulation results were explained by an equivalent circuit model. Although $MoS_2$ was chosen as an example 2-D material, the discussion in this work is general enough and can be extended to any other 2D-material. Overall our results indicate the promise of 2D-materials beyond graphene for terahertz applications. However, these materials need to be employed in multi-layer form, and the operation speed is expected to be much smaller than that in graphene-based devices.

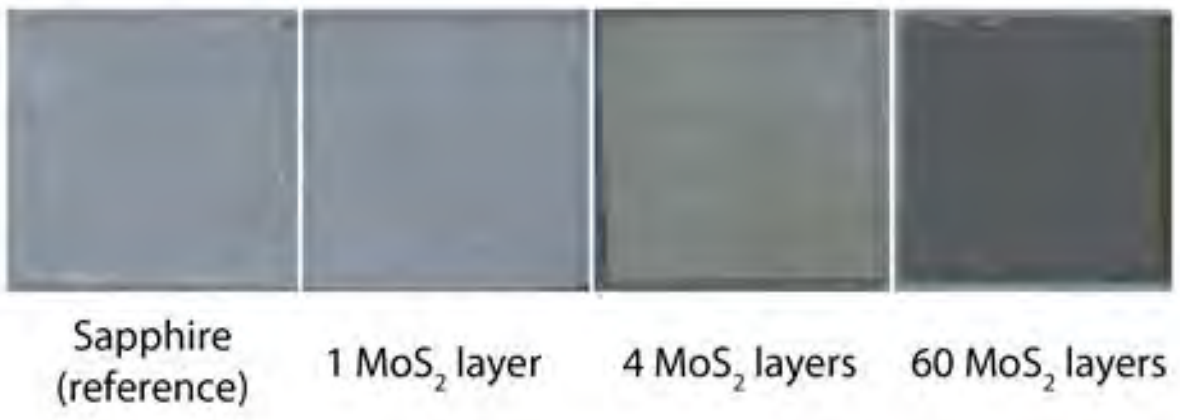

Fig. 1. Optical image of various $MoS_2$ films grown in sapphire substrates.

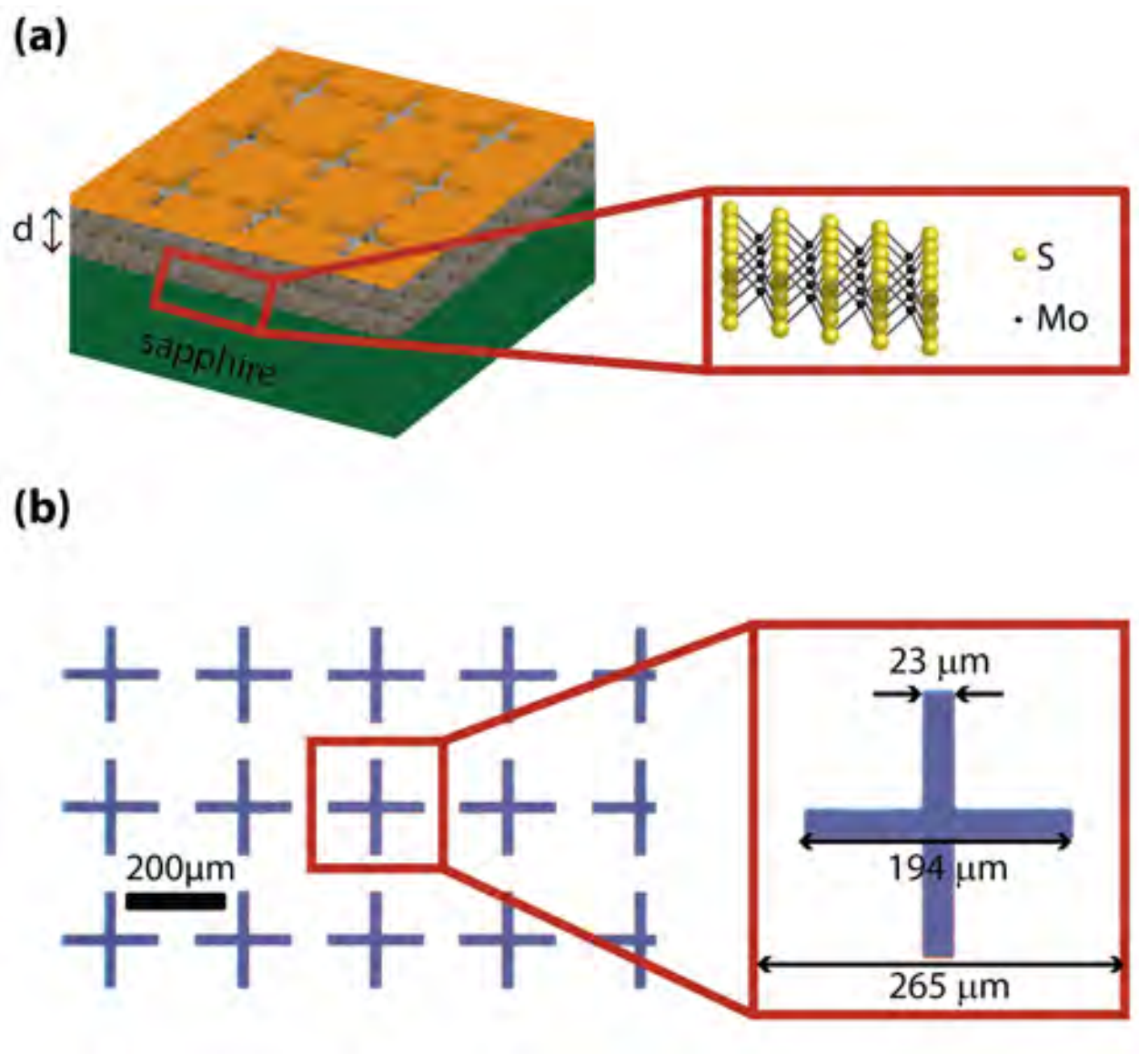

Fig. 2. (a) Schematic of a $MoS_2$/metal hybrid metamaterial structure. The polyimide spacer thickness, *d*, is varied in our experiments and simulations. (b) Optical image of a section of the metamaterial and detail of a unit cell.

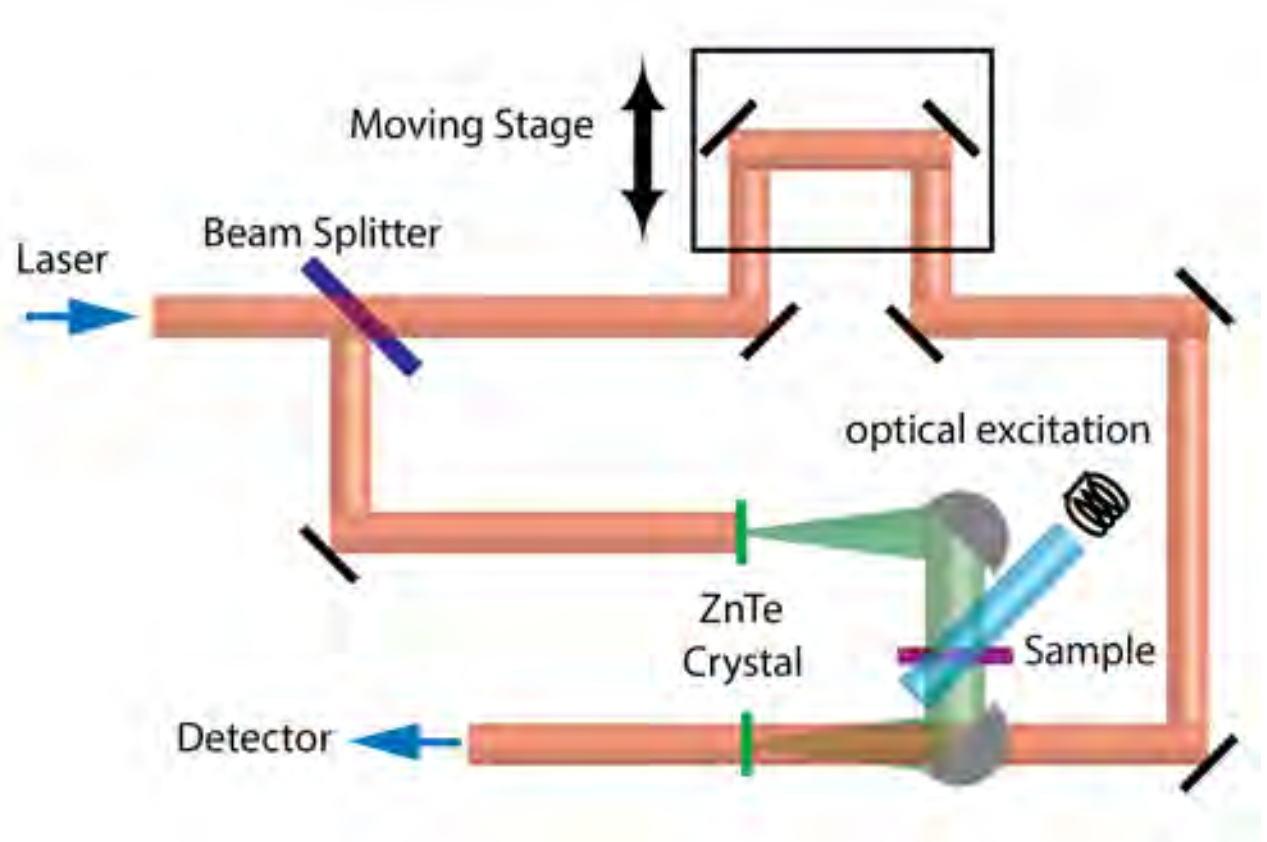

Fig. 3. Diagram of the terahertz time domain spectroscopy (TDS) setup. Actuation over $MoS_2$ is obtained via optical illumination employing a quartz tungsten halogen (QTH) lamp. The samples were illuminated through the substrate side.

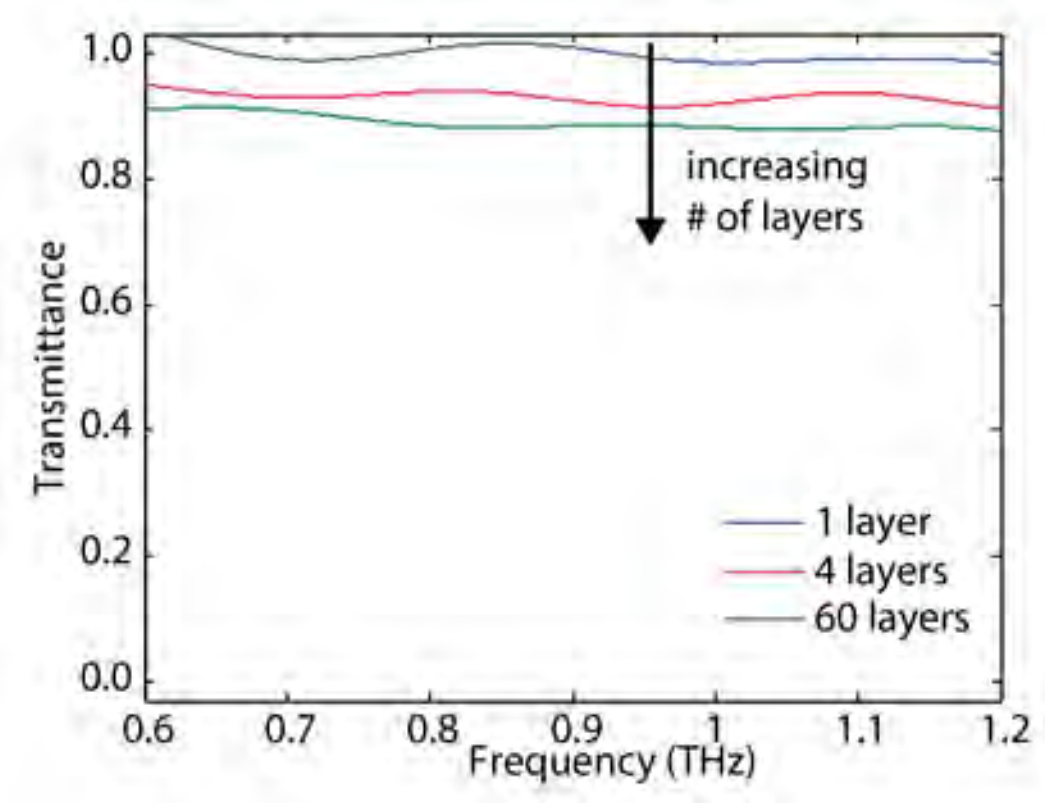

Fig. 4. Terahertz transmittance versus frequency for samples with varied number of $MoS_2$ layers under optical excitation. The measured transmittance under no optical excitation (not shown) is unity for all the analyzed samples, which is an evidence of the as-grown material being intrinsically undoped.

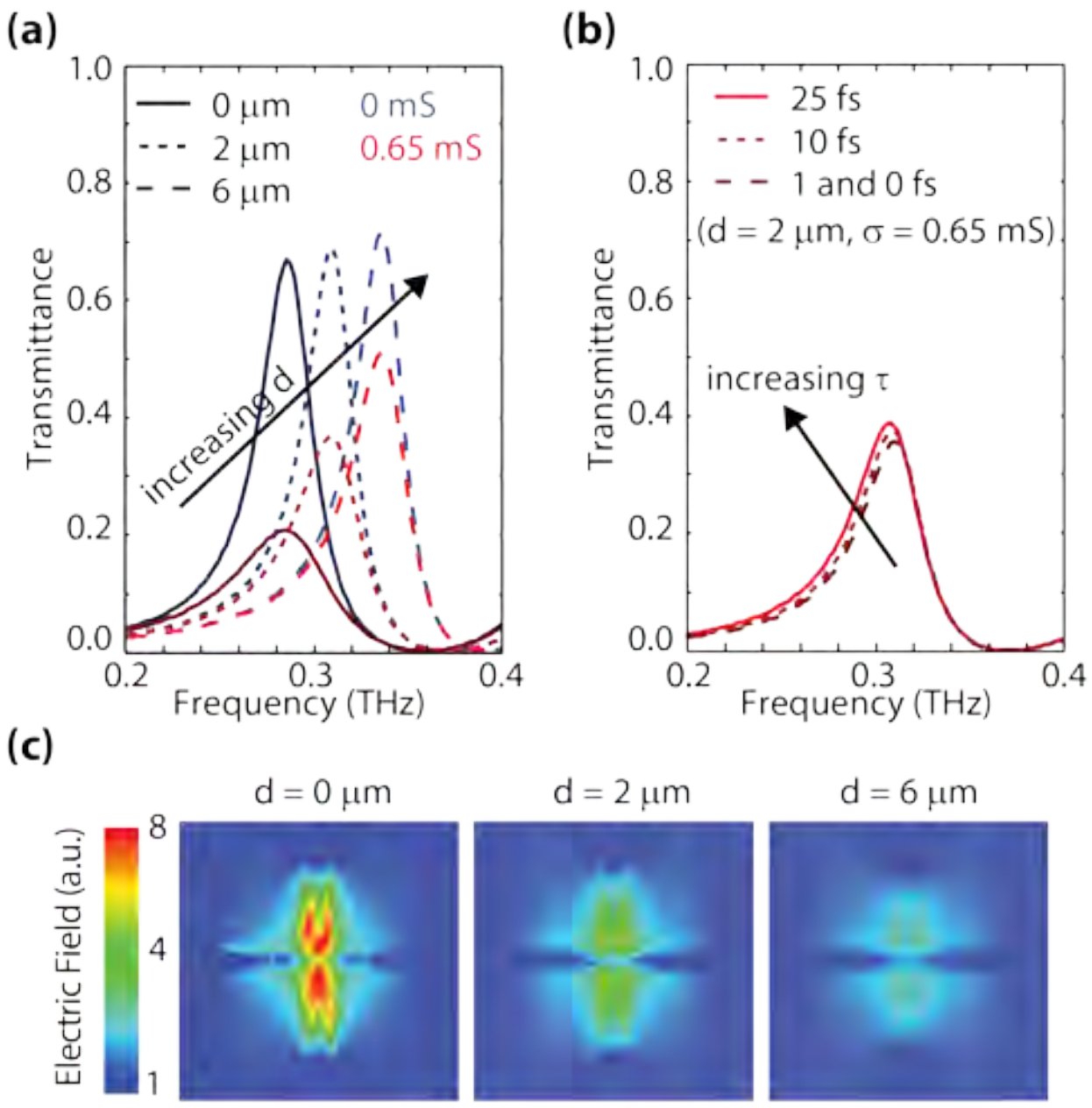

Fig. 5. (a) Simulated terahertz transmittance for metamaterial structures with varied spacer thickness (*d* = 0, 2, and 6 μm) and $MoS_2$ conductivity levels of 0 and 0.65 mS. (b) Simulated terahertz transmittance for metamaterial structures when employing a Drude model for $MoS_2$ and varying the electron momentum relaxation time (τ = 0, 1, 10, and 25 fs). The PI spacer thickness and $MoS_2$ conductivity were set to 2 μm and 65 mS, respectively. (c) Simulated electric field at resonance at the position where the $MoS_2$ film is places for different PI thicknesses.

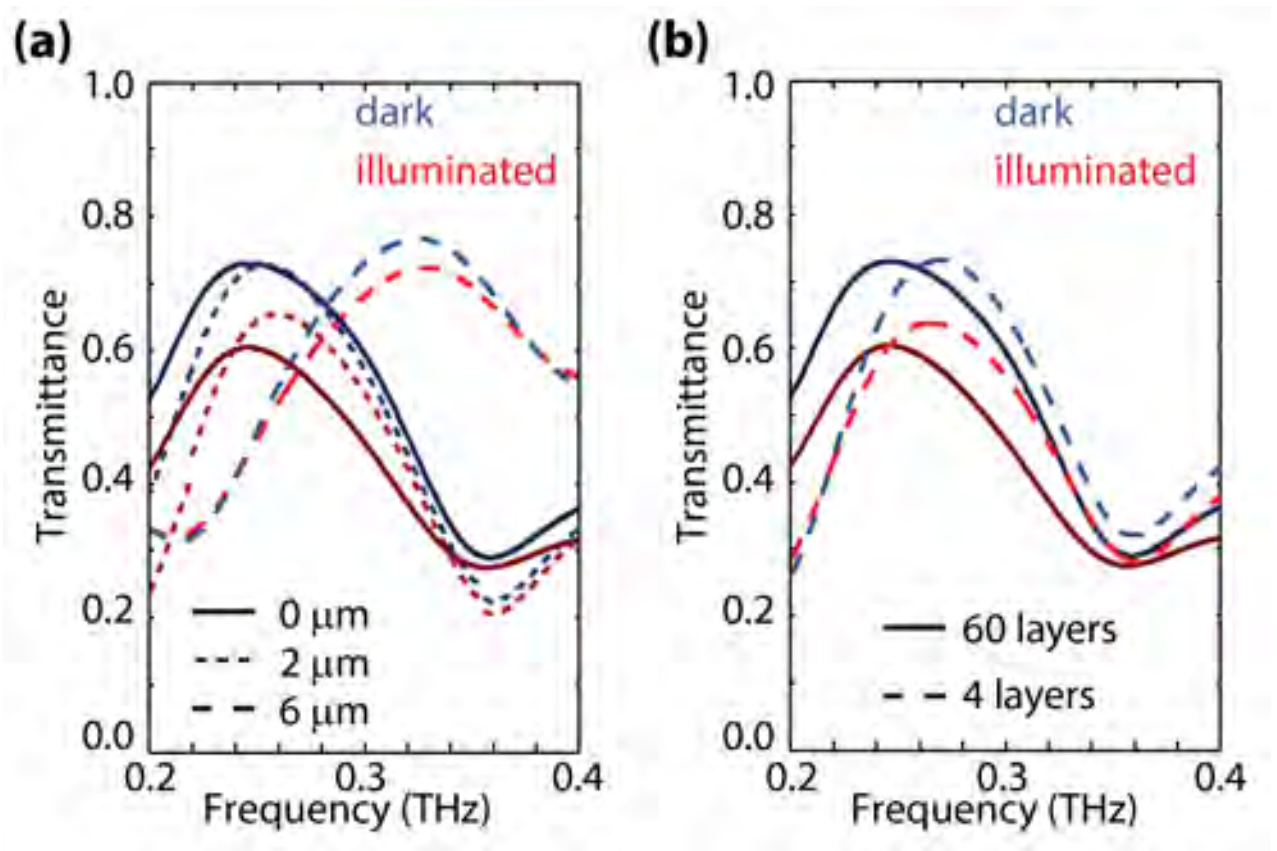

Fig. 6. (a) Measured terahertz transmittance (with and without illumination) for metamaterial structures with different spacer thickness ($d$ = 0, 2, and 6 μm) for a sample with 60 atomic layers of $MoS_2$. (b) Measured terahertz transmittance (with and without illumination) for samples with no PI spacer and 4 and 60 $MoS_2$ layers. In all cases the insertion loss is <3 dB.

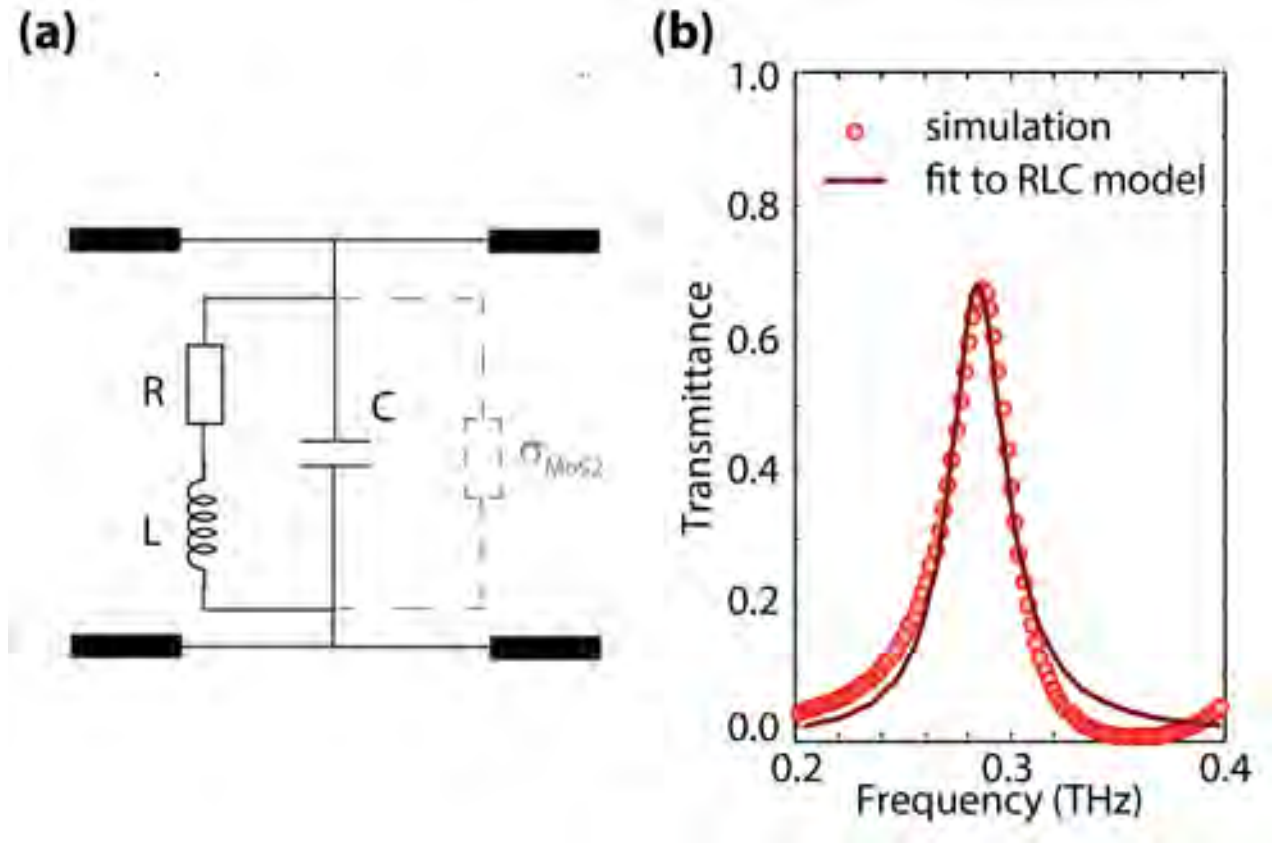

Fig. 7. (a) Equivalent circuit model. (b) Simulated and fitted transmission spectra for a structure containing the FSS in top of the sapphire substrate.

# CHAPTER 6

# TERAHERTZ CONDUCTIVITY OF ULTRA HIGH ELECTRON CONCENTRATION 2DEGS IN NTO/STO HETEROSTRUCTURES

## Abstract

We analyze the terahertz properties of complex oxide hetero-structures with record-high carrier concentration approaching $10^{15}$ $cm^{-2}$. Our results evidence a large room temperature terahertz conductivity, which corresponds to 3X to 6X larger nanoscale mobility than what is extracted from electrical measurements. That is, in spite of a relatively lower mobility, when taking into account its ultra-large carrier concentration, the 2DEG in complex oxide hetero-structures can still attain a large terahertz conductivity, which is comparable with that in traditional high-mobility semiconductors or large-area CVD graphene films.

The presence of a two-dimensional electron gas (2DEG) at the hetero-interface between, otherwise insulating, complex oxides has recently captured considerable attention [1-6]. This highly conductive 2DEG exists as a result of polarity discontinuities at the interface of the perovskite oxide crystals resulting from valence mismatch due to the presence of dangling bonds. Moreover, these hetero-structures exhibit intriguing physical phenomena and hold potential for novel electronic devices with promising properties like a high breakdown voltage and an ultra-high 2D electron density [7-8]. In addition, other interesting properties such as ferro-electricity [9-11], superconductivity [12-15], and negative magnetoresistance [16-17], have also been reported in these 2DEGs. In this context, complex oxide 2DEGs have been shown to support electron densities that are ~2 orders of magnitude larger than those in traditional semiconductor hetero-structures. Whereas for 2DEGs in III-V semiconductor hetero-structures typical maximum electron densities are in the order of ~$10^{13}$ $cm^{-2}$ [18-19], two-dimensional electron densities approaching ~$10^{15}$ $cm^{-2}$, thus half an electron per unit cell, have been reported in complex oxide hetero-structures [20]. Based on the above properties, complex oxide 2DEGs hold promise as suitable candidates for the next generation of electronic and terahertz devices.

In recent work, via band engineering, Xu *et al*. [20] discovered and demonstrated the possibility of attaining a 2DEG with ultra-high electron density at the interface of $NdTiO_3$ (NTO) and $SrTiO_3$ (STO). Although neither oxide conducts electricity as a pure material, such a 2DEG can be formed at the interface between polar NTO and nonpolar STO layers, and is confined within the STO side of the interface. Moreover, electronic reconstruction behavior above a certain critical thickness in these layers leads to a further enhancement in the electron density; hence, a record-high electron density of ~$3 \times 10^{15}$ $cm^{-2}$ was reported in these 2DEGs. The room-temperature electron mobility from Hall measurements under a Van der Paw configuration, was reported to be in the order of 1 to 10 $cm^2$/V.s. In this regard, it is worth noticing that the length-scale at which the 2DEG properties are characterized plays an important role on the measured values. In (DC) electrical measurements the characteristic length-scales at which electron transport is probed are set by the separation between contacts and thus are in the order of micrometers/millimeters. Therefore, the extracted electron transport properties can be affected by extended effects over >µm areas. At such large scales, these properties can be significantly affected by scattering by dislocations, defects, and stacking faults. However, these extended effects can become considerably less pronounced if probing with high-frequency excitations in a contactless, quasi-optical manner. For instance, if probing with a quasi-optical excitation associated with frequencies in the terahertz range, i.e. by means of THz spectroscopy, one can probe for a spatially-averaged nanoscale sheet conductivity due to the very high frequency alternating current (AC) electric fields associated with the THz radiation [21]. In this case, typical characteristic length-scales for semiconductors as well as for metals are in the order of a few nanometers [22-23], thus the extent at which such effects will degrade the extracted transport properties will be greatly diminished. In this context, *the aim of this paper is to analyze and compare the 2DEG transport properties extracted from DC measurements with the nanoscale properties extracted from THz spectroscopy in NTO/STO samples*. *For the first time, to the best of our knowledge, characterization of the electronic properties of such ultra-high electron concentration 2DEGs in NTO/STO hetero-structures is performed via a non-contact method using THz spectroscopy. Our results evidence terahertz sheet conductivity >1 mS in all the analyzed samples at room-temperature. These extracted values correspond to 3X to 6X larger conductivity than what is extracted from DC measurements, which is an indication of a much larger nanoscale transport mobility than what is evidenced from the electrical measurements. Our results overall indicate that the 2DEG in these complex oxide hetero-structures can possess large THz conductivity, comparable with those in the 2DEGs of traditional high-mobility semiconductors or in typical-quality large-area CVD graphene films, and, therefore, despite a still lower mobility, might result attractive for multiple terahertz as well as electronic applications.*

Three different samples were analyzed: Sample #1 and #2 correspond to NTO/STO structures grown under different conditions, whereas Sample #3 corresponds to a STO/NTO/STO hetero-structure containing two 2DEGs, one at each interface. Depicted in Fig. 1a are sketches of the structures as well as a typical TEM detail of the grown films and interfaces. All the samples were grown by means of molecular beam epitaxy (MBE) employing LSAT ($[La_{0.3}Sr_{0.7}][Al_{0.65}Ta_{0.35}]O_3$)

substrates. Details about the growth conditions for these samples are discussed in the *Experimental Section*. The electron concentration and mobility were obtained from Hall measurements; Table I depicts the extracted electrical transport properties as well as the thicknesses of the films forming the hetero-structures for each sample. The 2DEG in these hetero-structures shows indeed an ultra-high carrier concentration ranging from ~0.3 to ~$0.7\times10^{15}$ $cm^{-2}$; that is two orders of magnitude larger than what is typical in traditional semiconductor 2DEGs. The carrier mobility ranged from ~2 to ~4 $cm^2$/V.s at room temperature. The extracted DC sheet conductivity was 0.47 mS, 0.20 mS, and 0.19 mS, for Samples #1, #2 and #3, respectively.

Terahertz transmission measurements were performed for all samples employing a TOPTICA CW THz spectrometer [24]. The frequency sweep was limited to the range between 0.4 and 0.7 THz in such measurements in order to use a long averaging whereas not compromising on total scan time. Moreover, at this frequency range the terahertz properties of the LSAT substrate show very small frequency dependence [25]. Shown in Fig. 1b is the measured transmittance versus frequency for Sample #1, Sample #2, as well as a reference LSAT substrate. In order to extract the sheet conductivity of the 2DEG from the THz transmission spectra, the experimental data is fitted to an analytical model. In this regard, the effect of the 2DEG on THz transmittance can be evaluated using the scattering matrix formalism [26]. The air/NTO-STO hetero-structure/substrate region is modelled with Fresnel coefficients (S) by neglecting the optical thicknesses of the NTO and STO layers, which is a reasonable assumption since these layers are in the order of a few nanometers thus much smaller than the relevant THz wavelengths [26-27]. The associated S-matrix representing this region is given by:

$$S_{air/NTO-STO\ /\ substrate} = \begin{bmatrix} \frac{2n_{air}}{n_{air}+n_{substrate}+Z_0\sigma_{2DEG}} & \frac{n_{substrate}-n_{air}-Z_0\sigma_{2DEG}}{n_{air}+n_{substrate}+Z_0\sigma_{2DEG}} \\ \frac{n_{air}-n_{substrate}-Z_0\sigma_{2DEG}}{n_{air}+n_{substrate}+Z_0\sigma_{2DEG}} & \frac{2n_{substrate}}{n_{air}+n_{substrate}+Z_0\sigma_{2DEG}} \end{bmatrix}, \quad (1)$$

where $Z_0$ = 376.73 Ω is the vacuum impedance, $\sigma_{2DEG}$ is the 2DEG sheet conductivity, $n_{air}$ and $n_{substrate}$ are the refractive index of air ($n_{air}$ =1) and the substrate (LSAT), respectively. For each of the other interfaces and for the propagation in the substrate, classical formulas for the Fresnel coefficients can be employed:

$$S_{substrate\ /\ air} = \begin{bmatrix} \frac{2n_{substrate}}{n_{air}+n_{substrate}} & \frac{n_{air}-n_{substrate}}{n_{air}+n_{substrate}} \\ \frac{n_{substrate}-n_{air}}{n_{air}+n_{substrate}} & \frac{2n_{air}}{n_{air}+n_{substrate}} \end{bmatrix}, \quad (2)$$

$$S_{propagation\ substrate} = \begin{bmatrix} e^{-jn_{substrate}k_0 d} & 0 \\ 0 & e^{-jn_{substrate}k_0 d} \end{bmatrix}, \quad (3)$$

where $d$ is the substrate thickness, and $k_0 = 2\pi/\lambda_0$. With $\lambda_0=c/f$, where $c$ is the speed of light in vacuum, and $f$ is frequency. Since we are working in the 0.3 to 0.7 THz frequency interval, where the LSAT properties do not show a significant frequency dispersion [25], the substrate will be modeled in Eqn. (1)-(3) using a frequency independent complex index of refraction. Depicted in Fig. 1c are the calculated transmission spectra that best fit to the experimental data in Fig. 1b employing the model in Eqn. (1)-(3); the real and imaginary part of the substrate permittivity as well as the substrate thickness are set as fitting parameters thus extracted from the THz measurements. The substrate thickness was found to be non-uniform between samples, which explains the different positions of the Fabry-Perot peaks observed in the transmission spectra. The extracted 2DEG sheet conductivities obtained from fitting the experimental spectra to the model were 1.44 mS and 1.13 mS, for Samples # 1 and #2, respectively. This is 3.1X and 5.6X larger than the correspondent values extracted from DC measurements, respectively.

At this end, we proceed to study Sample #3. In contrast with Samples #1 and #2, which are grown in a 5 mm x 5 mm LSAT substrates, Sample #3 was grown in a 10 mm x 10 mm substrate, which besides being a suitable dimension for analysis in our CW THz system, where the spot size at the plane of the sample is in the order of 2 mm, it also constitutes a suitable dimension for analysis in our THz time-domain spectroscopy (TDS) system, where the spot size is in the order of 5 mm and the setup is more sensitive to sample misalignments. By selecting an appropriate time-window, so to subtract for contributions from multiple reflections, the 2DEG sheet conductivity can be experimentally determined by fitting the TDS transmission spectra to the following model, where a Drude dispersion was assumed for the 2DEG conductivity [28]:

$$\frac{T(\omega)}{T_0(\omega)} = \left|\frac{1}{1+Z_0\sigma_{2DEG}/(1+n_{substrate})(1+\omega^2\tau^2)}\right|^2 , \qquad (4)$$

where $T(\omega)$ and $T_0(\omega)$ are the measured (power) transmissions through the NTO/STO sample under test and through a LSAT reference substrate, respectively. Here, $\tau$ is the electron momentum relaxation time in the 2DEG, and $\omega = 2\pi f$ is the angular frequency. In the 0.4 to 1.2 THz frequency range the real part of the LSAT refractive index slightly increases from 4.8 to 4.9, whereas its imaginary part ($\kappa$) is in the order of 0.02 [25]. For the purpose of Eqn. (4), we will neglect the contribution of $\kappa$ and assume $n_{substrate}$ = 4.8 as the refractive index of the LSAT substrate. The maximum error introduced by these assumptions was evaluated to be <1%. The relaxation time $\tau$ and the conductivity $\sigma_{2DEG}$ are unknowns in Eqn. (4) thus are found from fitting of the measured TDS transmittance spectra to Eqn. (4). Since our THz TDS setup is integrated with a cryostat, and so to obtain a further insight into the electronic properties of this ultra-high carrier density 2DEG, we studied the temperature dependence of the extracted electrical transport properties in the temperature range from 77K to room-temperature. The measured transmission spectra as well as its fit to Eqn. (4), for various temperatures, are depicted in Fig. 2a. We observe a decrease in transmission as the temperature decreases, a signature of an increased sheet conductivity, and, therefore, increased mobility at low temperature. Moreover, the Drude-like frequency dependence becomes more clearly apparent as temperature is decreased, which is a signature of a longer momentum relaxation time at low temperature. From these measurements, sheet conductivity and momentum relaxation time were extracted as a function of temperature by fitting of the experimental data to Eqn. (4). Depicted in Table II is a summary of the extracted parameters from these TDS measurements at various temperatures. In this regard, it is worth noticing that at room-temperature an excellent agreement is observed between the extracted conductivities from TDS and CW measurements as can be observed by scrutinizing Table I and Table II. Moreover, a >2X increase in momentum relaxation time is observed when temperature is decreased from room-temperature to 77K, which correlates well with the observed increase in conductivity. Sample # 3 was also electrically characterized as a function of temperature, from where its temperature dependent DC conductivity was extracted. Shown in Fig. 2b are the extracted conductivities as a function of temperature from TDS as well as electrical measurements. On this regard, we notice that the >2X increase in the terahertz-extracted conductivity is accompanied with a similar increase in the DC-extracted conductivity when temperature is decreased from room-temperature to 77K. A strong correlation is observed between the temperature trends for conductivity when comparing the results from THz and DC measurements. The fact that the ratio between these two weakly depends on temperature could be interpreted as a signal of these methods probing transport at different scales, and, therefore, the differences in the extracted conductivity levels being mainly related to structural effects in the sample, which are temperature independent.

Overall, across all three studied samples, and across all the analyzed temperature range, the conductivity extracted from THz measurements (either CW or TDS) is consistently 3X to 6X larger than that extracted from DC measurements. At THz frequencies, the conductivity is determined by an effective characteristic length scale that follows a Debye model [29]. Assuming a diffusive electron transport under THz excitation [29-31], the characteristic length ($L$) follows:

$$L = \frac{D}{2\pi f}, \tag{5}$$

where:

$$D = \frac{\sigma_{DC} K_B T}{n_s e}. \tag{6}$$

With $D$ being the diffusion constant, $f$ the frequency, $n_s$ the carrier density, $e$ the elementary charge, $\sigma_{DC}$ the sheet conductivity, $K_B$ Boltzmann's constant, and T temperature. For the range of 2DEG electron densities and sheet conductivities extracted in our samples, we find that at THz frequencies, i.e. $f$ > 300GHz, the characteristic length is always < 10 nm. The scale of this characteristic length indicates that the sheet conductivity extracted from THz measurements is, in fact, a spatially-averaged nanoscale conductivity and is thus less affected by microscopic scale effects [30]. In contrast, the effects of defects and dislocations introduced during growth, can compromise the values of sheet conductivity extracted from electrical measurements, where transport is probed at much larger length scales. Thus, *measurement using a non-contact method and gauging the nanoscale properties, such as THz spectroscopy, provides a relevant estimate of the intrinsic electrical properties of the 2DEG*. The behavior of the 2DEG as a less-defective layer in the nanoscale transport lengths probed by THz spectroscopy in contrast with its behavior as a highly-defective sheet in the microscopic transport length scales probed by electrical measurements is an explanation for the discrepancy between the THz-extracted and electrical-extracted sheet conductivity values as well as for the much larger THz-extracted levels. The THz-extracted sheet conductivity seems more pertinent when projecting the use of such 2DEGs in optoelectronic or nano-electronic devices, where the relevant dimensions for electron transport, as dictated for instance by the gate-length in nano-scale field effect transistors, are also in the order of nanometers.

In conclusion, we discussed on the transport properties of ultra-high electron concentration 2DEGs in complex oxides hetero-interfaces. Samples with record-high 2DEG electron density in NTO/STO were analyzed. We showed that contactless THz spectroscopy can probe the spatially-averaged nanoscale transport properties in these 2DEGs, whereas electrical measurements are highly sensitive to microscale effects, thus the THz-extracted properties constitute a better estimate of the intrinsic 2DEG electronic properties. Our analysis reveals that the THz-extracted average sheet conductivity, thus nanoscale mobility in these 2DEGs, is 3-6X larger than what is extracted from DC measurements, which probe transport properties at much longer characteristic lengths. This can be attributed to defects, dislocation, and surface discontinuities which dominate on the microscale response. Our study suggests that NTO/STO hetero-structures can be attractive for electronic devices, and terahertz applications (e.g. in modulators), since the maximum attainable conductivity levels in its 2DEG are similar than those in III-V hetero-structures, such as AlGaN/GaN, or typical quality large-area CVD graphene.

The epitaxial oxide layers with cube-on-cube relationship with the substrates were grown on commercial, high structure quality (001) LSAT substrates (Crystech GmbH, Germany) using an hybrid MBE technique [MBE system: EVO 50, Omicron Nanotechnology, Germany] at a base pressure of $10^{-10}$ torr, as explained in Ref. [32]. The NTO/STO heterostructure were sequentially grown employing a metal-organic precursor of titanium tetraisopropoxide (TTIP) (99.999% from Sigma-Aldrich, USA) as a Ti source, which provides oxygen for the process. First a 3 nm stoichiometric STO layer was grown using an effusion cell as Sr source (99.99%, Sigma Aldrich APL, Milwaukee,Wisconsin) at 970º C; next the NTO layer was grown at substrate temperature of 900º C employing an effusion cell for Nd (99.99% from Ames Lab, USA). Post-growth, the structure stoichiometry was characterized with X-ray diffraction and thin film diffractometric techniques [33].

The macroscopic electronic properties of the heterostructures were measured by a four-point Van der Pauw method [20]. For this purpose, wire-bonded contacts were applied at the corners of the samples, connecting to the STO interface as well as to the NTO interface. For sample # 1 and #2, 300 nm of Au on top of 50 nm of Ti were sputtered as ohmic contacts for the NTO/STO/LSAT heterostructures, while 300 nm of Au on 20 nm Ni on top of 40 nm of Al were sputtered as ohmic contacts for the STO/NTO/STO/LSAT sample (Sample #3). The contacts were strategically placed so that they make a direct contact with the 2DEGs at the interfaces following the methods in [20].

For THz characterization two different setups were used, namely, a Continuous Wave (CW) THz spectrometer and a THz time-domain spectroscopy setup (THz-TDS). The CW THz spectrometer is a commercial setup from TOPTICA photonics, using InGaAs photo-mixers at 1550nm [24]. In the THz-TDS setup, a broadband THz pulse was generated by optical rectification of optical pulse in 1 mm thick (110) ZnTe crystal. The generated THz radiation is focused on to the sample and the response was modulated on the optical pulse using electro-optic sampling on the similar ZnTe crystal at the detector end [34]. The transmission is Fourier transformed to obtain the frequency response of the signal, which when normalized to the response of the LSAT reference provided for the properties of the complex oxide hetero-structures under test.

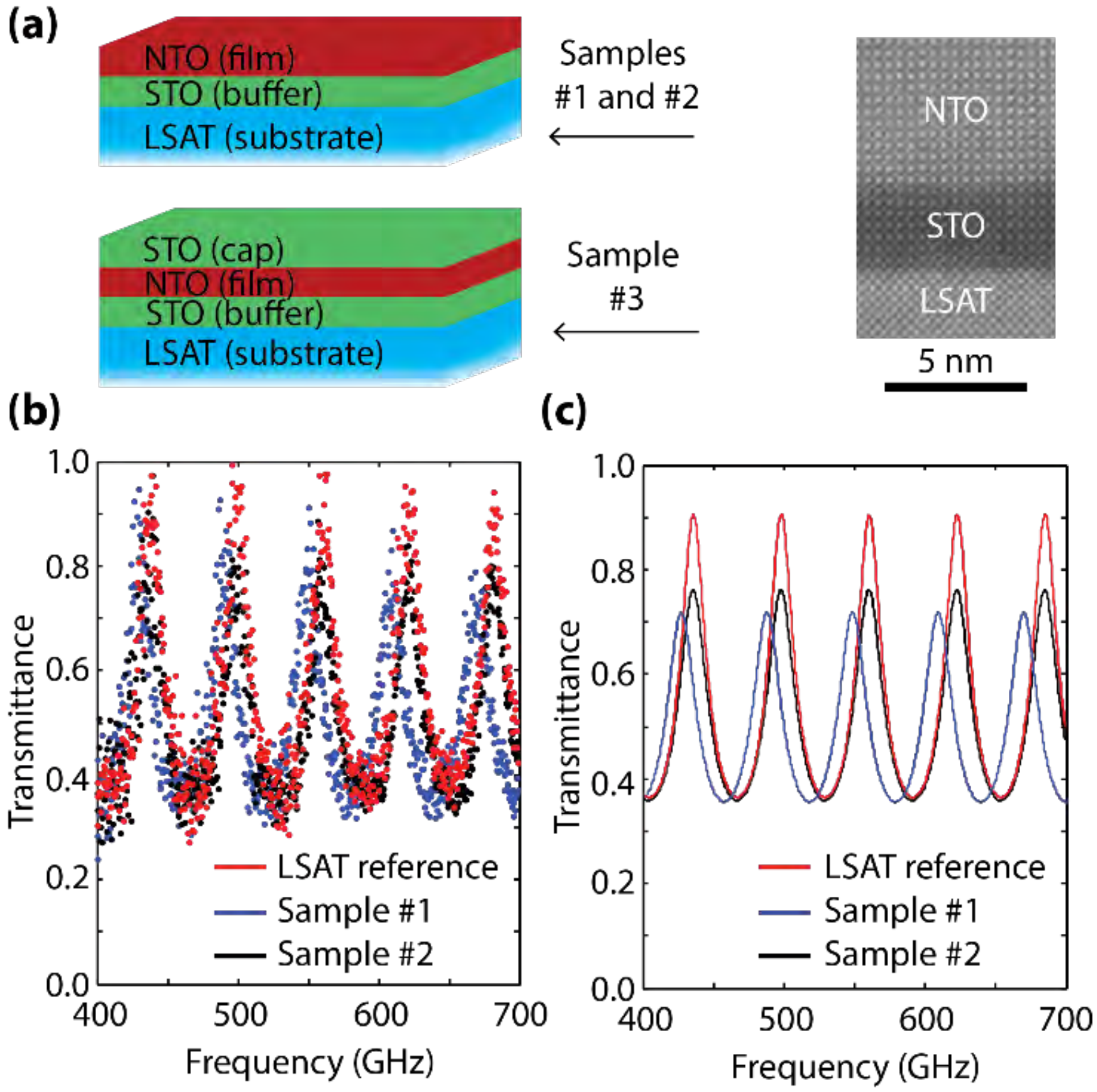

**Figure 1. (a)** Structure of the analyzed samples and TEM detail of the NTO/STO/LSAT interfaces. **(b)** Measured transmittance versus frequency for Samples #1, #2, and a LSAT reference substrate. **(c)** Best fit using the model in Eqns. (1)-(3) to the data in (b). Measurements were performed employing a CW THz spectrometer.

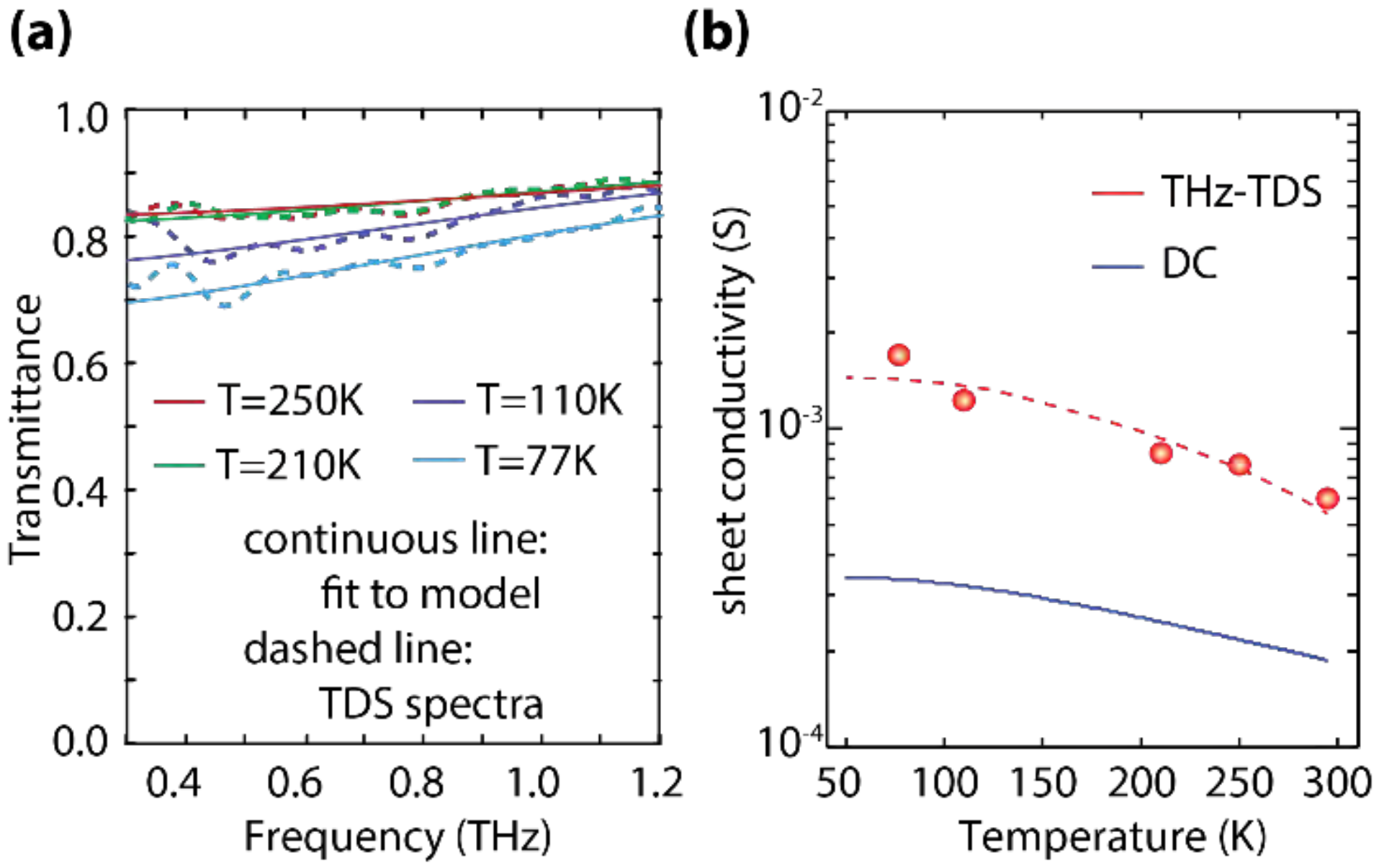

**Figure 2. (a)** Measured (dashed) and fitted (continuous) transmittance versus frequency for Sample #3 at various temperatures. Measurements were performed employing a THz-TDS setup, results were fitted to the model in Eqn. (4). **(b)** Extracted sheet conductivity versus temperature from THz-TDS measurements (red) as well as electrical measurements (blue).

**Table I.** (Sample properties and parameters extracted from electrical as well as CW-THz measurements)

| Sample | Structure | $n_{2DEG}$ ($cm^{-2}$) | $\mu_{HALL}$ ($cm^2/V.s$) | $\sigma_{extracted, DC}$ (mS) | $\sigma_{extracted, CW\text{-}THz}$ (mS) | $\mu_{nanoscale}/\mu_{HALL}$ |
|---|---|---|---|---|---|---|
| #1 | NTO/STO | $7.16x10^{14}$ | 4.11 | 0.47 | 1.44 | 3.1X |
| #2 | NTO/STO | $3.64x10^{14}$ | 3.36 | 0.20 | 1.13 | 5.6X |
| #3 [(a)] | STO/NTO/STO | $5.20x10^{14}$ | 2.33 | 0.19 | 0.65 (x2) | 3.4X |

[(a)] Sample #3 contains two 2DEGs; the values listed in the table correspond to those in a single 2DEG, for this purpose the THz-extracted conductivity was divided by a factor of 2 for the purpose of listing in this table.

**Table II.** (Extracted sheet conductivity and relaxation time for various temperatures in Sample #3 from TDS measurements)

| Temperature (K) | $\sigma_{extracted, TDS}$ (mS) | $\tau_{extracted, TDS}$ (fs) |
|---|---|---|
| 300 | 1.25 ± 0.02 | <65 |
| 250 | 1.55 ± 0.02 | 93 ± 2.1 |
| 210 | 1.68 ± 0.02 | 111 ± 1.9 |
| 110 | 2.47 ± 0.04 | 140 ± 3.15 |
| 77 | 3.41 ± 0.04 | 147 ± 2.1 |

# CHAPTER 7

# THZ CHARACTERIZATION AND DEMONSTRATION OF VISIBLE-TRANSPARENT / TERAHERTZ-FUNCTIONAL ELECTROMAGNETIC STRUCTURES IN ULTRA-CONDUCTIVE LA-DOPED BASNO3 FILMS

## Abstract

We report on terahertz characterization of La-doped $BaSnO_3$ (BSO) thin-films. BSO is a transparent complex oxide material, which has attracted substantial interest due to its large electrical conductivity and wide bandgap. The complex refractive index of these films is extracted in the 0.3 to 1.5 THz frequency range, which shows a metal-like response across this broad frequency window. The large optical conductivity found in these films at terahertz wavelengths makes them an interesting platform for developing electromagnetic structures having a strong response at terahertz wavelengths, i.e. terahertz-functional, whereas being transparent at visible and near-IR wavelengths. As an example of such application, we demonstrate a visible-transparent terahertz polarizer.

## 7.1 Introduction

Transparent electronics have garnered significant attention since the introduction of the first transparent thin film transistor [1]–[3]. In terms of commercial products, the market has been steadily flourishing, with demand being mostly led by the photovoltaic and the optoelectronic industries. Furthermore, these materials are expected to be convenient alternatives to conventional metals as well as doped-semiconductors so to develop fully transparent devices. Several types of transparent conductive oxides (TCOs) have been introduced and characterized for different applications in plasmonics, photovoltaics, and so on, namely tin-doped indium oxide (ITO), gallium-doped zinc oxide (GZO), and aluminum-doped zinc oxide (AZO), among others [4]. Also, because of their large transmission in the near-IR and visible spectral ranges, TCOs could be employed as contacts in terahertz optoelectronic devices, terahertz photoconductive antennas, photo-active metamaterial structures, plasmon-assisted spectroscopy of materials, etc. [5]–[7]; applications wherein the underlying material needs to be accessed optically. Moreover, being visible-transparent, these materials could be used to design structures whose optical response is encrypted at terahertz frequencies. On the other hand, spectroscopic characterization of these materials at terahertz frequencies reveals its high-frequency AC electronic transport properties, which could enrich the understanding of carrier dynamics in these materials and hence, help develop prospective devices with wider ranges of applications [8]–[10].

With rapid proliferation of optoelectronic devices and need for improved performance, demand for highly conductive and cost-effective TCOs has been on the rise. Among the TCOs demonstrated thus far, ITO has been shown to possess the largest conductivity of ~$10^4$ S/cm, where Indium oxide is degenerately alloyed with Tin ($n \sim 10^{20}$ $cm^{-3}$) [11], [12]. However, ITO might face in the future a major bottleneck due to the potential scarcity of Indium, which could translate into higher production costs. In this regard, doped Zinc-oxide compounds have been proposed as potential replacements for ITO; however, these materials suffer from a much lower electrical conductivity [13]. Recently, thin films of ultra-conductive $BaSnO_3$ (BSO) doped with Lanthanum were demonstrated and characterized [14], showing room-temperature conductivity ~$10^4$ S/cm; and high mobility of 120 $cm^2$/V.s at carrier concentrations of ~ $3x10^{20}$ $cm^{-3}$. It is to be noted here that these mobility values are exceptionally high for such high carrier concentrations. This has been made possible due to a combination of low electron effective mass and weaker phonon scattering. A wide band-gap of 3 eV, together with, ultra-high conductivity and relatively high mobility in thin films, makes BSO a strong candidate as a transparent conducting oxide. Further characterization is required to reveal the true potentials of this material for different applications, including terahertz applications. Terahertz frequencies occupy a crucial region of the electromagnetic spectrum; introduction of ultra-high conductive TCOs as a building block for visible-transparent terahertz devices would enrich the emerging technologies in this field.

Here, we present a comprehensive study on terahertz characterization of ultra-conductive La-doped BSO films grown on $(La_{0.3}Sr_{0.7})(Al_{0.65}Ta_{0.35})O_3$ LSAT(001). Details of film growth, structural and electronic characterizations can be found elsewhere [14]–[16]. Three sets of samples have been analyzed, with different levels of conductivity. The optical-transparency of the films at visible wavelengths was studied using visible spectroscopy while the terahertz optical properties and high-frequency carrier transport properties were characterized using terahertz spectroscopy. By fitting the results to proper models, the optical properties of this material were extracted. Finally, we demonstrate a terahertz polarizer as a proof-of-concept visible-transparent terahertz-functional electromagnetic structure. The extracted results are compared with those reported in the literature from terahertz studies in ITO. Our results reveal BSO as an efficient visible-transparent terahertz-functional material in addition to establishing new avenues of research directions in this novel wide band-gap oxide films as potential TCO applications.

## 7.2 Experimental

In this study, BSO/LSAT(001) samples were prepared using a hybrid molecular beam epitaxy

(MBE) technique [15], [16]. The films were grown epitaxially on 5mm x 5mm, 0.5 mm thick LSAT (001) substrates, where 45 nm thick undoped BSO was grown as a buffer layer. Next, an active layer of conductive La-doped BSO, $Ba_{1-x}La_xSnO_3$ was grown; where the value of *x* and therefore, the carrier concentration *n* can be varied by changing the La-cell temperature during growth. In this study, three samples with different dopant densities are analyzed. For *Sample #1*, 49 nm BSO film with nominally stoichiometric composition was grown. For *Sample #2*, a 46 nm film was grown intentionally with Barium deficiency whereas, *Sample #3*, a 48 nm film was grown with Tin deficiency. La-concentration was kept fixed by keeping La-cell temperature constant. ***Table 1*** summarizes the DC transport measurements for the three samples, which were obtained from Hall measurements. Rapid thermal annealing at 800$^o$C for two minutes was performed on the samples prior to the transport measurements. As shown in ***Table 1***, stoichiometric La-doped BSO (*Sample #1*) showed the highest electron mobility, 86 cm$^2$/V.s, the highest conductivity, and the highest carrier concentration. *Samples #2 and #3* with Barium and Tin deficiencies showed mobilities of 71 and 17 cm$^2$/V.s, respectively, consistent with higher disorder due to non-stoichiometry [16]. In order to gain a better understanding on the effect of crystal defects on our characterization method we also analyze *Samples* #2 and #3, which were intentionally grown non-stoichiometric.

For the purpose of terahertz spectroscopy, samples were grown on LSAT substrates. LSAT(001) is used instead of STO(001) substrate – a typical substrate- due to the strong phonon absorption by STO in the terahertz frequency range. In view of this fact, the LSAT substrate, apart from being transparent at visible wavelength, showed negligible absorption in the terahertz frequency range [12, 13]. To this end, samples grown on STO substrate were characterized using visible light spectroscopy in order to determine the transparency of the processed thin films. ***Figure 1*** shows the transmission spectrum for (*i*) a La-doped BSO-film grown on STO, and (*ii*) a bare STO substrate. In the sample under study, the thickness of the La-doped BSO film is 132 nm and the conductivity extracted from DC transport measurement is $6.85 \times 10^5$ S/m, which is comparable to the conductivity values observed in samples grown on LSAT substrates. Measurements were normalized to transmission through air. *The results indicate that the transmission for the BSO sample on STO is larger than that through the STO substrate. This can be understood as the BSO film acting as an antireflection coating*. The insets in ***Fig. 1*** depict an optical image of the analyzed La-doped BSO sample and the extracted refractive indices for BSO and the STO substrate by fitting of the transmission data in the visible spectrum to a proper model [19]. Also, depicted in ***Fig. 1*** is the fitting of the measured data to this model (dashed curves), showing an excellent agreement with the experiments as well as explaining the anti-reflection coating effect. The above results confirm that the La-doped BSO films are transparent across the entire visible range. The refractive index was found to be ~2 across a broad wavelength range from 400 to 800 nm, in agreement with previous reports [15, 16].

## 7.3 Results and discussion

For the purpose of terahertz characterization, we employed the terahertz time-domain spectroscopy (THz-TDS) technique as our primary characterization tool. Using a conventional THz-TDS setup, see the *Methods* section, the complex-transmission through the BSO/LSAT samples was determined; the measured spectral response was normalized to that of the substrate so to independently resolve for the effect of the conductive La-doped BSO film. The normalized transmission data was used to directly extract the complex refractive index of the conductive film (as described in the *Methods* section).

***Figure 2*** depicts the normalized transmission for *Sample #1* at different temperatures (77K, 120K, 170K, 220K and 295K) in the 0.3 to 1.5 THz frequency range. The inset in ***Fig. 2*** shows the time domain pulses measured for the reference LSAT substrate and the attenuated transmitted signal through the BSO/LSAT sample (at room temperature and 77K). Transmission in frequency is extracted from the Fourier transform of the received pulses and normalized to the transmission through the substrate and it is related to the conductivity value of the thin film inversely. As ***Fig. 2*** depicts, the transmission levels decrease as temperature is decreased. This observation indicates an increase in the film conductivity as temperature is decreased. Furthermore, it can also be noticed that the transmission is flat in this frequency range. This observation indicates a short carrier momentum relaxation time in this material; even with an increase in mobility at low temperatures, the momentum relaxation time is still relatively short and does not lead to spectral signatures in the analyzed frequency range.

From the complex transmission data, we extracted the refractive index and absorption coefficient following the procedure discussed in the *Methods* section. ***Figure 3*** shows these extracted values. These values were found to be relatively high when compared to those reported from THz spectroscopy of typical ITO samples, e.g. [9]. The behavior of these parameters, together with the flat spectral response obtained in the THz-TDS measurements, see ***Fig. 2,*** is an indication of metal-like behavior from these degenerate-doped semiconductors. From the complex-transmission spectra, it is possible to also obtain a THz-extracted AC conductivity for the film [22]. ***Figure 4*** depicts the THz-extracted AC conductivity for *Sample #1* as a function of temperature. Each transmission measurement was normalized to the measured transmission through the LSAT substrate at corresponding temperatures. The increase in conductivity with the decrease in temperature can be attributed to relaxation of phonon modes in these sample, hence curtailing the effect of electron-phonon scattering. The THz-extracted AC conductivity values at each temperature shown in ***Fig. 4*** are statistically consistent with those observed in DC measurements. This differs with our previous observations in complex oxide 2DEGs [23], where much larger conductivity levels were extracted in the THz-measurements than in DC. Furthermore, whereas in our previous work [23], largest differences between THz-extracted and DC-extracted conductivities were observed in non-stoichiometrically-grown samples, across all the analyzed samples in this work, i.e. *Samples #1* to *#3*, no-statistical differences were observed between DC-extracted and THz-extracted conductivity levels.

In our previous work [23], we reported on terahertz characterization of the two-dimensional electron gas (2DEG) formed at the interface between polar/non-polar complex oxides (e.g. NTO/STO). By contrasting results obtained from terahertz and DC characterization techniques, we showed that the electrical properties of the 2DEG are significantly affected by defects and scattering at the interface. Our measurements demonstrated a 3-6 X enhancement in the terahertz-extracted zero-frequency AC conductivity in these samples w.r.t. their DC conductivity [23]. Such enhancement was attributed to the characteristic length at which transport is probed in terahertz spectroscopy being much smaller than the mean free path between defect-induced scattering events [23]. Based on the extracted mobility levels, we estimated this characteristic length to be on the order of 10’s of nm[23]. However, in the BSO thin-films analyzed in the current work, the mobilities extracted from Hall-effect measurements are larger -by more than one order of magnitude- than those observed in 2DEGs in NTO/STO. From this perspective, we expect the terahertz-probe length to be on the order of 100’s of nm in the BSO samples under current study. Thus, as a result of (*i*) a longer characteristic length at which terahertz transport is probed in these samples having larger mobility, and (*ii*) a longer mean free-path between scattering events that leads to a larger mobility in these samples, no substantial differences between DC-extracted and THz-extracted conductivity levels are expected in the BSO films. Another important factor to be considered here is that free carriers in the NTO/STO heterostructure were confined to a quasi-two-dimensional interface; however, here carriers are present in a three-dimensional BSO layer. The (*iii*) additional degree of freedom for carrier transport as a result of this increased dimensionality would limit the impact of scattering by structural defects in these samples. Therefore, the close agreement between DC-extracted and THz-extracted conductivity across all samples can be understood on basis of a larger mean free-path between scattering events as well as on a higher carrier confinement dimensionality.

In order to further validate our results and observations, we performed continuous-wave (CW) terahertz spectroscopy. In the CW-THz setup, data was obtained in the 0.3 to 0.7 THz frequency range. Depicted in ***Fig. 5*** is the measured transmission spectra for Samples #1, #2, #3, as well as for a bare LSAT substrate; each transmission measurement was referenced to that in free-space. In this measurement since the transmission data is normalized to that in free-space, Fabry-Perot resonances resulting from the substrate optical thickness are present. Once the transmission data was taken, the measured response was modelled employing a constant conductivity model so to represent the conductive La-doped BSO film. Furthermore, the substrate was modelled with appropriate thickness and refractive index so to accurately represent the Fabry-Perot resonances associated with the finite thickness of the samples. By fitting the measured data to this model, see *Methods* section, THz-extracted AC conductivities of 7.9±0.2, 4.0±0.6, and 0.4±0.05 $\times 10^5$ S/m were obtained for Samples #1, #2, and #3, respectively. Overall, across all the analyzed samples a close agreement was observed between extracted conductivity levels from THz-TDS, CW-THz, and DC measurements.

## 7.4 Applications

High carrier concentration, hence, high-conductivity on the analyzed La-doped BSO films leads to a metal-like response from these degenerate-semiconductors at terahertz frequencies. The fact that these TCOs showed a very large terahertz-conductivity, implies that these films could be employed to design terahertz-functional electromagnetic structures that are transparent to near-IR and visible wavelengths. As a proof-of-concept, we designed a simple grid-polarizer by patterning of the La-doped BSO films. A polarizer is a structure that lets electromagnetic waves of a specific polarization pass and blocks waves with perpendicular polarizations. In the case of a perfect polarizer the co-polarized transmission should be 100% and the cross-polarized transmission should be zero. The design geometry for the polarizer was determined with the help of full wave simulations using ANSYS HFSS. We chose a design having a wire width of 17.2 µm and 20 µm periodicity. For this purpose, *Sample #1*, i.e. the sample with the highest conductivity, was patterned using standard lithography and etching techniques (Ar ion milling). The response of the patterned-film was measured using our THz-TDS system as a function of incident-beam polarization. Depicted in ***Fig. 6*** are the results from these measurements. For cross-polarized excitation, the transmission was 14%, while for co-polarized excitation the transmission was 88%. These experimental results were largely consistent with our simulated response, as also shown in ***Fig. 6***. In order to give further insight into the scaling of the polarizer performance with BSO-film thickness, we performed simulations for polarizers in 200 nm thick films by using the same terahertz-conductivity levels as experimentally determined in *Sample #1*. Furthermore, in order to provide a qualitative comparison, we also performed simulations for these polarizers using 50 nm thick gold as the conductive material. Although in both cases, the gold-based polarizer shows a better performance, in terms of co- and cross-polarized transmission levels, the BSO structures are transparent at near-IR and visible wavelengths, which might be attractive for many terahertz system level applications. Furthermore, by scaling the film thickness, better polarizer response can be obtained. It is to be noted here that, being a doped, wide band-gap semiconductor, our films demonstrate one of the highest reported conductivity levels from a TCO to-date; which is on-par of the best reported in ITO [24]. Furthermore, contrary to other highly conductive "effectively transparent" semiconductors, such as graphene, in which the high-conductivity rises from a very large electron mobility, the high conductivity in BSO is the result of an ultra-large charge density, in spite of a moderate mobility. As a result, highly conductive TCOs such as BSO have a metallic-like optical response across a wide range of terahertz wavelengths. In contrast, in high-electron mobility "transparent" semiconductors, such as high-quality graphene, the terahertz optical conductivity drops with frequency as a result of a long momentum relaxation time. From this perspective, terahertz-active and visible-transparent electromagnetic structures in BSO can provide for a broader terahertz frequency window of operation. From all these perspectives, ultra-conductive La-doped BSO holds bright prospects as a transparent material for the development of future terahertz devices.

## 7.5 Conclusions

In conclusion, we have reported on terahertz characterization of BSO thin-films. Our BSO films show one of the largest electrical conductivities in any TCO demonstrated to date, which is on par with the best reports in ITO. Our terahertz measurements evidence a metallic response of these films in a broad frequency range from 0.3 to 1.5 THz. Furthermore, visible spectroscopy shows that BSO is transparent in the 400 to 800 nm wavelength window. As a result of its large broadband terahertz conductivity, and large visible transmission, BSO constitutes an interesting material for the design of electromagnetic structures, such as polarizers, that are functional at terahertz frequencies but transparent in the near-IR to visible range. From this perspective BSO is a strong candidate for future transparent devices for terahertz applications.

## 7.6 Methods

*Film synthesis*

Thin films of doped $BaSnO_3$ were grown on LSAT (001) substrates using a hybrid molecular beam epitaxy approach. The method utilizes a metal-organic precursor (hexamethylditin) for Tin [15], a solid source for barium provided through an effusion cell, and an RF plasma source for oxygen. Oxygen plasma was operated at 250 W at a pressure of $5\times10^{-6}$ Torr. Barium beam equivalent pressure (BEP) was kept constant at $5\times10^{-8}$ Torr. Lanthanum (La) was used as an n-type dopant to make the films conducting. The dopant density was kept nominally fixed by keeping the La effusion cell temperature at 1230 °C. A substrate temperature of 900 °C was used for growing 45-50 nm thick of La-doped $BaSnO_3$ layer on top 45-50 nm of undoped $BaSnO_3$ buffer. Films were grown with three different tin precursor flux of $1.0\times10^{-6}$ Torr (Sn-deficient), $1.4\times10^{-6}$ Torr (stoichiometric) and $1.5\times10^{-6}$ Torr (Ba-deficient). Structural characterization was done using in situ Reflection High Energy Electron Diffraction (RHEED) and X-ray Diffraction. DC transport measurements were performed in a conventional van der Pauw configuration using a Physical Property Measurement System (Dynacool). Indium was used as Ohmic contacts.

*Ion Milling*

BSO films were dry etched using Argon plasma in an Intlvac Ion Mill. Argon plasma was operated at 75 W with a gas flowrate of 25 sccm. Substrate was tilted at 75° with respect to the normal and cooled down to 6°C to avoid overheating during the process. Etching was performed down to the substrate to form a polarizer structure as shown in the inset of figure 6.

*Terahertz spectroscopy*

THz-TDS: the terahertz signal was generated by optical rectification in a ZnTe crystal pumped by an 810 nm amplified Ti-Sapphire laser with pulse width of 75 nm and repetition rate of 1 kHz. The terahertz signal was then focused on the samples, which were placed inside a cryostat chamber for the purpose of cryogenic measurements, using two parabolic mirrors. Then, the response signal is focused and sampled on the optical probe beam by using another ZnTe crystal by electro-optic sampling technique. The obtained time domain terahertz waveform is Fourier transformed to extract the frequency response of the sample.

THz-CW: the samples were also tested in a diode-laser-driven photomixing spectrometer. The spectrometer is a commercial system from Toptica photonics. In this system, two lasers with center wavelength of 1500 nm are photo mixed for generation and detection of the terahertz signal. The sample is placed in the collimated region between parabolic mirrors.

*Visible spectroscopy*

Visible spectroscopy was performed using a Perkin-Elmer LAMBDA 950 UV-Vis-NIR Spectrophotometer with a 150 mm PbS integrating sphere in the wavelength range from 390 to 800 nm with steps of 1 nm.

*Modelling*

(a) THz-TDS: direct extraction of the complex refractive index. The transmission through a thin optical film (medium #2) sandwiched between two thick optical materials (mediums #1 and #3) normalized to the transmission from medium #1 to #3, can be modelled by [9,] [25]:

$$T_{sample}(\omega) = \frac{E_{sample}(\omega)}{E_{substrate}(\omega)} = \frac{t_{12}t_{23}e^{\frac{i(n_2-1)d\omega}{c}}}{t_{13}}FP(\omega)$$

Where $n_2$ is the refractive index of medium #2, $d$ is its thickness, $c$ is the speed of light, $\omega$ is angular frequency, $t_{ij}$ is transmission from medium i to medium j and is defined as $t_{ij} = \frac{2n_i}{n_i+n_j}$, and FP(ω) is a Fabry-Perot term set by the multiple reflections inside the optically thin slab and defined as:

$$FP(\omega) = \frac{1}{1 - r_{21}r_{23}e^{j2n_2d\omega/c}}$$

in which, $r_{ij}$ is the reflection from the interface of medium i and j and defined as $r_{ij} = \frac{r_i - r_j}{r_i + r_j}$. Here, medium #1 is air and medium #3 is the LSAT substrate. By fitting the visible transmission data to this model, the refractive index of the BSO layer can be extracted at each frequency.

(b) THz-TDS: extraction of conductivity from transmission spectra. The optical conductivity of the BSO film was extracted from the transmission data by fitting to [22]:

$$1 - \left|\frac{T}{T_0}\right|^2 = 1 - \frac{1}{\left|1 + \frac{\sigma(\omega)Z_0}{n_s + 1}\right|^2}$$

Where, $Z_0$ is the characteristic impedance of free space, $n_s$ is the refractive index of the substrate, and σ(ω) is an effective 2D conductivity for the La-doped BSO layer, i.e. conductivity [S/m] multiplied by thickness of the film. The above formula is valid for situations where the effective optical thickness of the film is much smaller than the relevant terahertz wavelengths, which is the case in our study. Since the measured transmission did not exhibit significant frequency variations, the measured data was fitted to the above formula assuming a constant conductivity.

(c) CW-THz spectroscopy: extraction of conductivity from transmission spectra. In CW measurements, multiple reflections from the substrate need to be considered, therefore a different theoretical framework needs to be employed so to model the data. For this purpose, the ABCD matrix formalism is employed. By fitting of the experimental data to this analytical model the conductivity of the film is extracted. It is worth mentioning that since the BSO film is very conductive, we are not capable to determine its real part of permittivity employing this approach. The LSAT substrate is modelled by [26]:

$$\begin{pmatrix} A & B \\ C & D \end{pmatrix}_{LSAT} = \begin{pmatrix} \cos(\Phi_{LSAT}) & \mathrm{j}Z_{LSAT}\sin(\Phi_{LSAT}) \\ \mathrm{j}\frac{1}{Z_{LSAT}}\sin(\Phi_{LSAT}) & \cos(\Phi_{LSAT}) \end{pmatrix}$$

Where: $\Phi_{LSAT} = \frac{n_{LSAT}\omega d_{LSAT}}{c}$ , $n_{LSAT}$ is the refractive index of the LSAT substrate, $\omega$ in angular frequency, $d_{LSAT}$ is the thickness of the LSAT substrate and $c$ is the speed of light. Furthermore, $Z_{LSAT} = \frac{Z_0}{n_{LSAT}}$ , where $Z_0$ is the vacuum impedance. The transmission through the LSAT substrate can be extracted using the following formula [26]:

$$T = \frac{2}{A + \frac{B}{Z_0} + CZ_0 + D}$$

The doped BSO film can also be modelled by an ABCD matrix:

$$\begin{pmatrix} A & B \\ C & D \end{pmatrix}_{BSO} = \begin{pmatrix} \cos(\Phi_{BSO}) & \mathrm{j}Z_{BSO}\sin(\Phi_{BSO}) \\ \mathrm{j}\frac{1}{Z_{BSO}}\sin(\Phi_{BSO}) & \cos(\Phi_{BSO}) \end{pmatrix}$$

Where: $\Phi_{BSO} = \frac{n_{BSO}\omega d_{BSO}}{c}$, $n_{BSO}$ is the complex refractive index of the BSO layer, and $d_{BSO}$ is its thickness. The complex refractive index can be defined as:

$$n_{BSO} = \sqrt{\varepsilon_r - j\frac{\sigma_{BSO}}{\omega\varepsilon_0}} \approx \frac{(1+j)}{\sqrt{2}}\sqrt{\frac{\sigma_{BSO}}{\omega\varepsilon_0}}$$

Here: $\varepsilon_0$ is the vacuum permittivity, and $\sigma_{BSO}$ is the conductivity of the BSO film. In this case, the ABCD matrix representing the total structure can be found from the matrix multiplication of $(ABCD)_{total} = (ABCD)_{LSAT}.(ABCD)_{BSO}$. It is worth mentioning again that in the analyzed frequency ranges, the transmission was found to be constant over frequency, therefore the terahertz optical conductivity was modeled as a constant across this frequency range.

**Acknowledgement**

The work involving terahertz characterization of the samples was supported primarily by the NSF MRSEC program at the University of Utah under grant DMR #1121252. We also acknowledge support from the NSF awards ECCS #1407959 and #1351389 (CAREER). The work involving thin films growth and DC transport measurements at the University of Minnesota was supported primarily by the AFOSR Young Investigator Program (FA9550-16-1-0205). The work also acknowledges partial support from National Science Foundation through DMR-1741801. We also acknowledge the use of facilities at the UMN Characterization Facility and the Nanofabrication Center, which receives partial support from the NSF through the MRSEC program at the University of Minnesota.

***Table 1.*** DC extracted parameters for the analyzed BSO samples

| | Thickness (nm) | Carrier concentration ($cm^{-3}$) | Mobility ($cm^2$/V.s) | Resistivity (Ω.cm) |
|---|---|---|---|---|
| Sample #1 | 49.3 | $6.75\times10^{20}$ | 86 | $1.06\times10^{-4}$ |
| Sample #2 | 45.8 | $3.73\times10^{20}$ | 71 | $2.36\times10^{-4}$ |
| Sample #3 | 48.0 | $1.27\times10^{20}$ | 17 | $2.93\times10^{-4}$ |

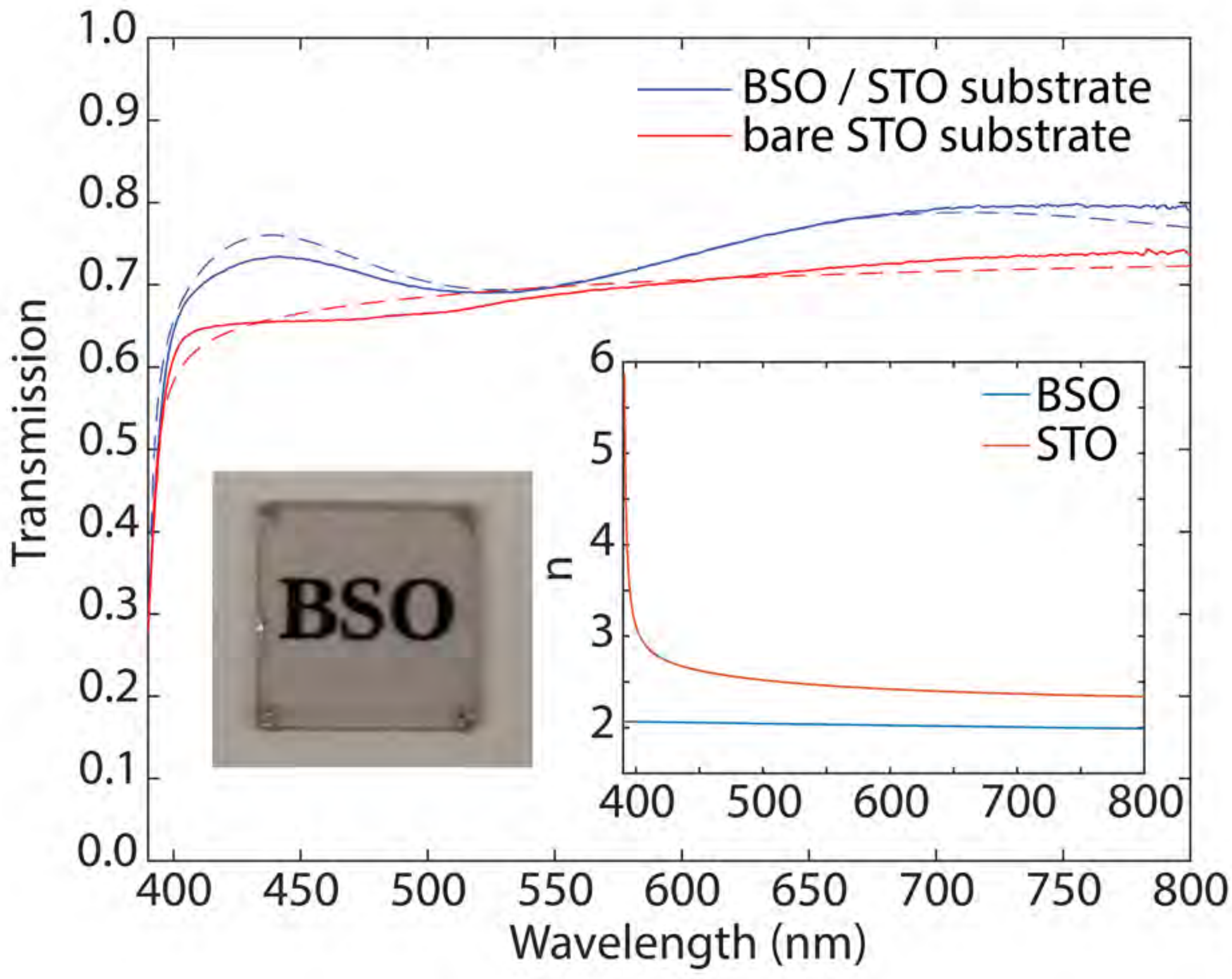

***Figure 1.*** Measured optical transmission through a 132-nm film of La-doped BSO grown on a STO substrate (blue trace) as well as optical transmission through a STO bare substrate (red trace) in the visible spectrum. Fitted curves for the measured data to the model described in the *Methods section* are indicated by dashed lines. The right inset shows the extracted refractive indices for BSO and the STO substrate extracted from this fitting. The left inset shows a picture of one of the analyzed samples showing its transparency.

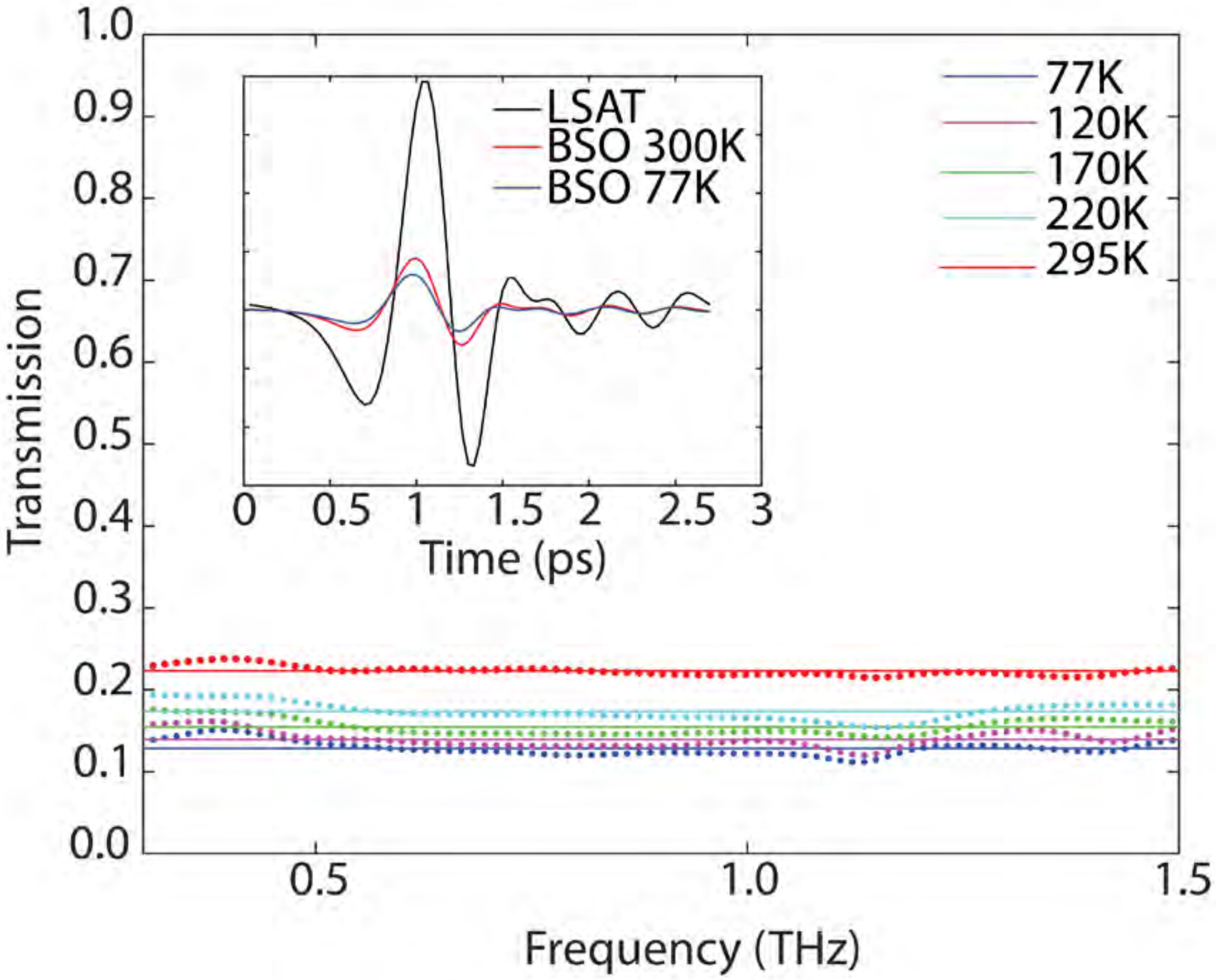

***Figure 2.*** Measured terahertz transmission through Sample #1. The transmission is normalized to that of a bare substrate at each temperature. The experimental data points are indicated with dots, whereas fits to the model discussed in methods section are depicted with solid lines. The inset shows the measured terahertz pulse through the LSAT substrate at room-temperature as well as the pulses through Sample #1 at room-temperature and 77K.

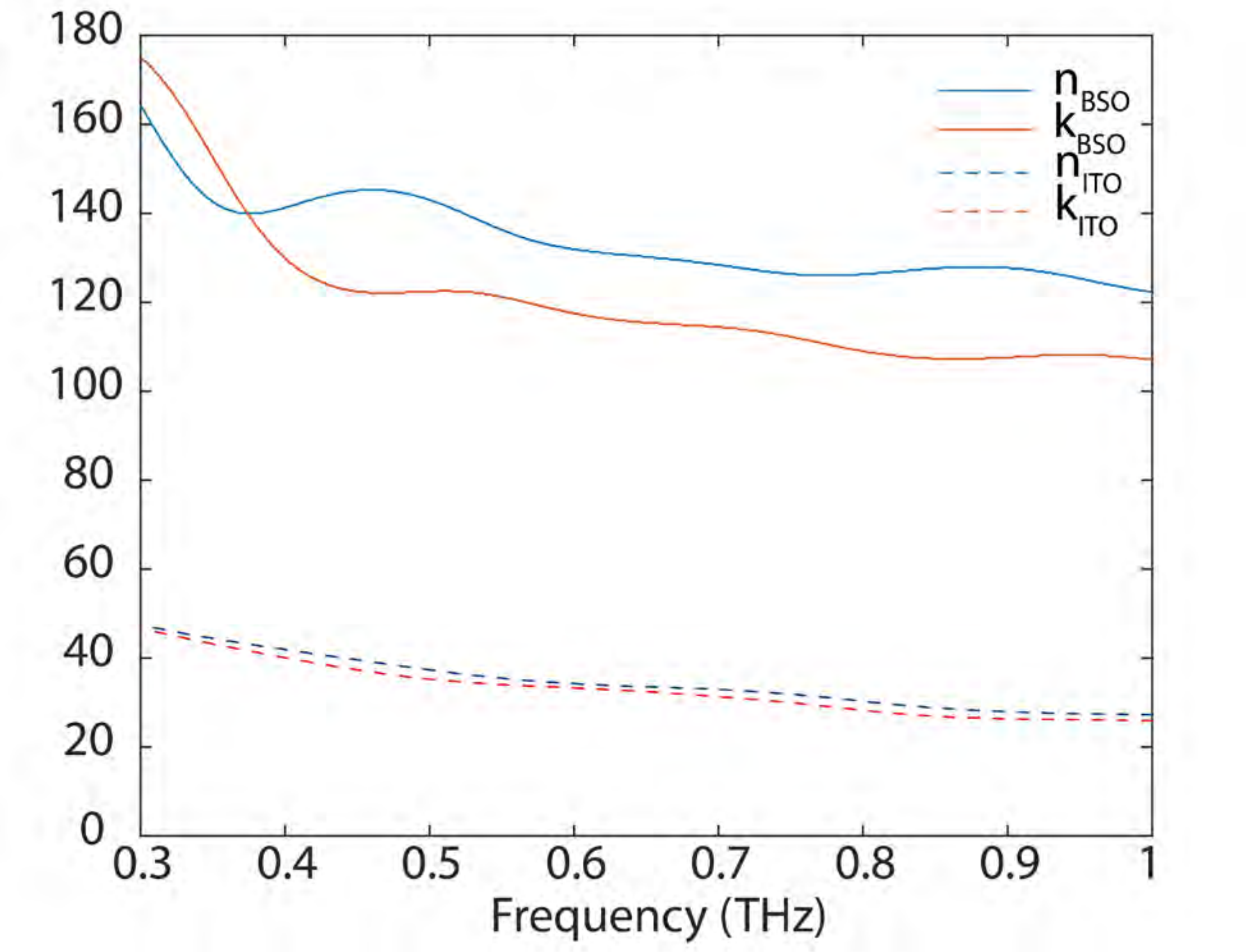

***Figure 3.*** Real($n$) and imaginary($k$) part of refractive index vs. frequency for Sample #1. These parameters are directly extracted from the complex TDS transmission. In addition to the extracted values for our La-doped BSO sample (49.3nm thick film), data reported in the literature [9] for an ITO film (345 nm thick) is also depicted in the plot. In both cases a metal-like response is observed. The conductivity levels observed in the analyzed BSO samples are on par with the best results reported in the literature for ITO [27], and larger than those for the samples studied by terahertz spectroscopy in Ref. [9].

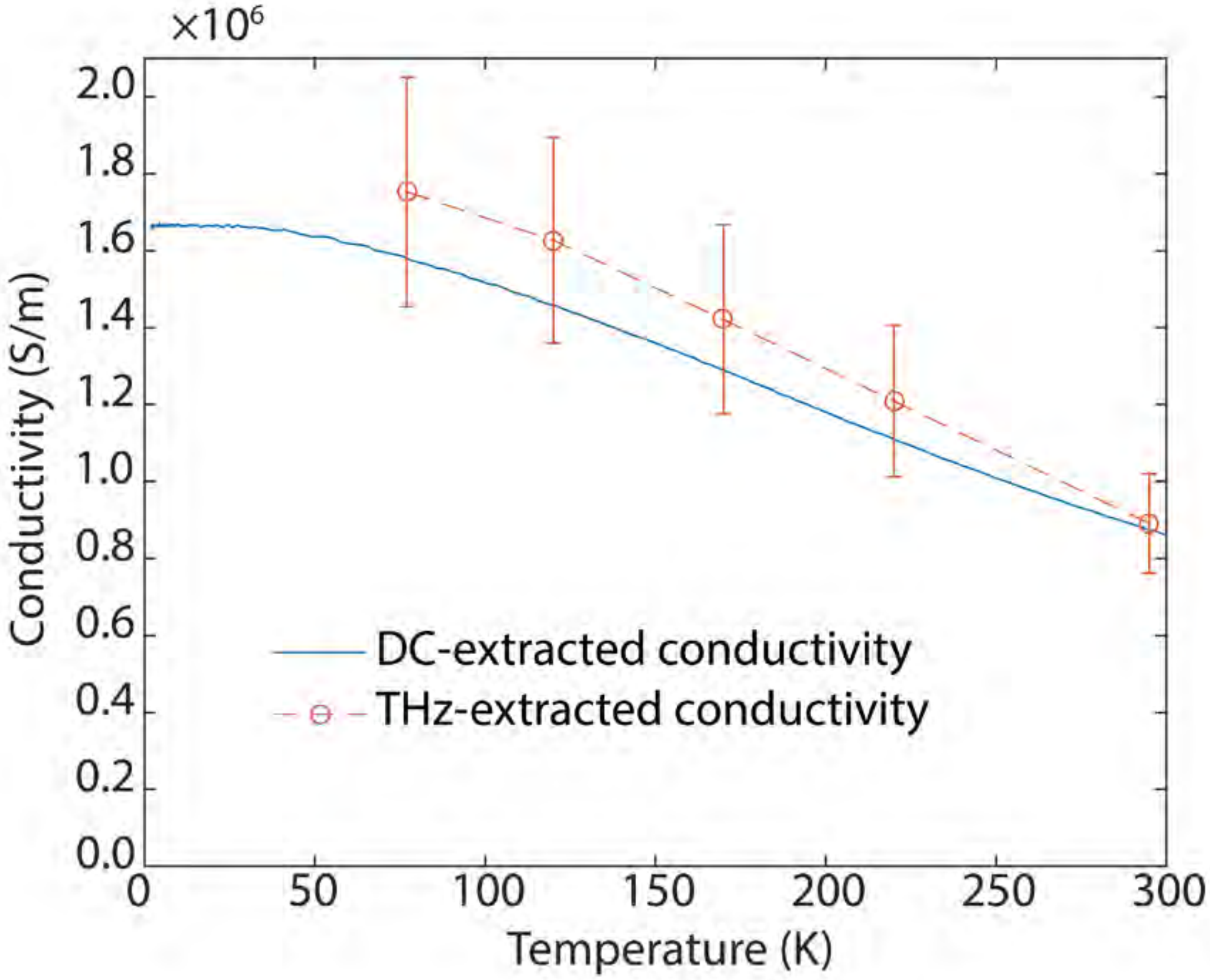

***Figure 4.*** THz-extracted and DC-extracted conductivity vs. temperature for *Sample #1*. THz-extracted conductivity levels statistically agree with those observed in DC measurements; this observation is different to our previous observations in complex oxide 2DEGs, where a much larger THz-extracted conductivity was observed.

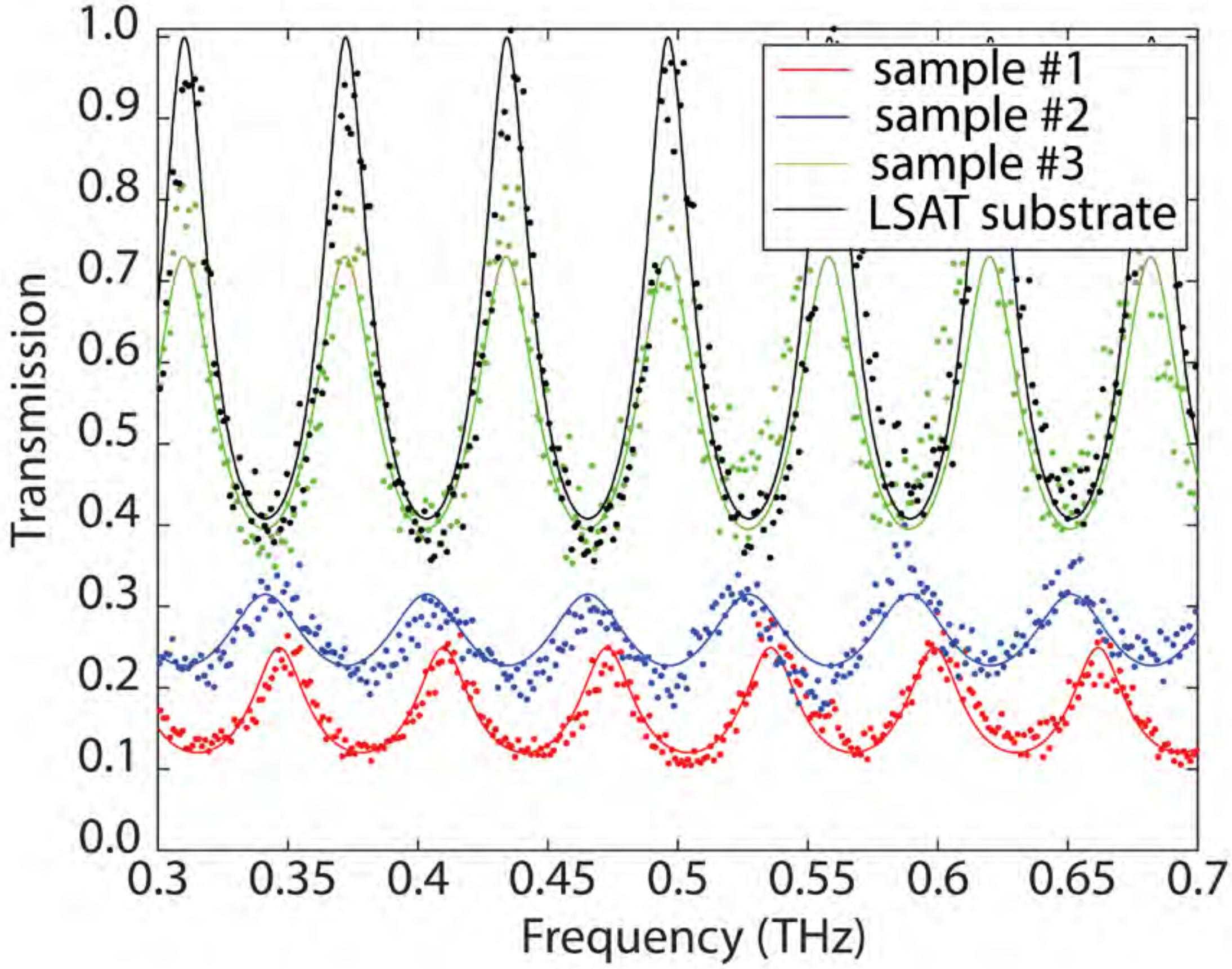

***Figure 5.*** Measured CW terahertz transmission through Samples #1, #2, #3, and a bare LSAT substrate. The solid lines, represent the fitting of the measured data to the model described in *Methods* section.

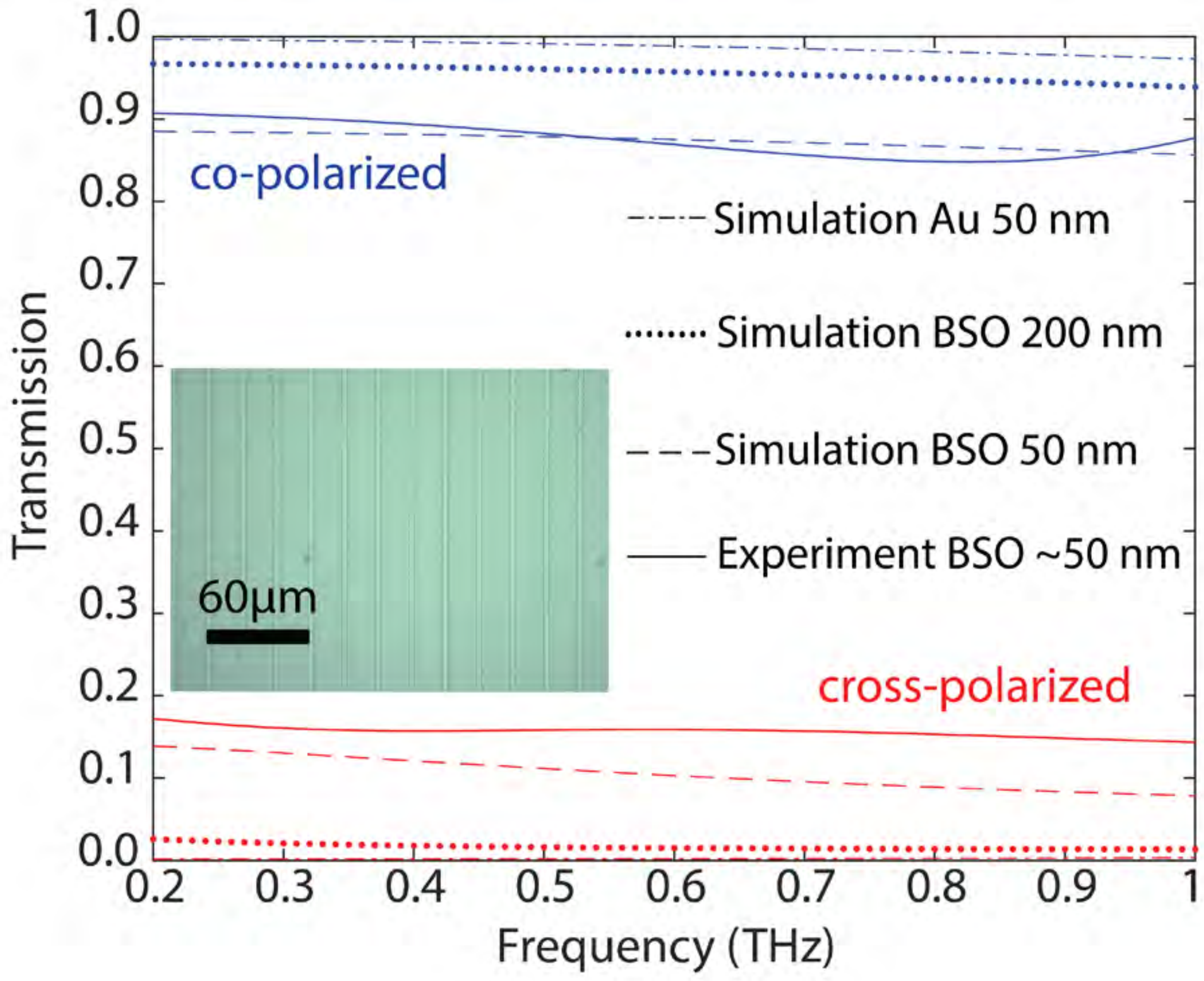

***Figure 6.*** Co-polarized and cross-polarized transmission through polarizer structures. The continuous traces indicate our experimental results. Dotted and dashed curves represent modelled data. The inset shows an optical image of the fabricated sample, which is transparent to visible wavelengths.

# CHAPTER 8

# CONCLUSIONS AND FUTURE WORK

## 8.1 Graphene for THz applications

In the first three chapters of this dissertation, we discussed graphene as an active material to realize devices for THz applications. In this regard, we proposed a deep subwavelength metamaterial based on graphene for THz phase modulation. Our design can improve the efficiency of beam steerers based on phase modulators when compared to other metamaterial phase modulators proposed in the literature. Due to small area covered by the active region of the device, the speed of these devices can also improve with respect to that in previously proposed metamaterials integrated with graphene. In order to provide further insight into the performance levels of such devices, we analyzed their geometrical tradeoffs. For this purpose, we performed a simulation study for two families of deep subwavelength metamaterial, namely multi-spiral resonators and multi-split-ring resonators. As a result of this study we found that there is a specific metal coverage ratio in both families of devices, that leads to the best performance. In these cases, the unit cell to wavelength ratio is around $\lambda/20$.

A long-standing question in the literature is: what should be the role of graphene when developing THz devices? Graphene based devices for THz applications can be divided in two major groups: (i) graphene plasmonic devices, in which graphene is both acting as the plasmonic medium as well as the reconfigurable medium, and (ii) graphene-metal hybrid devices, in which the electromagnetic properties of the device are mainly set by the metallic structure, whereas the reconfigurability is enabled by graphene, which is integrated within the metallic structure. By comparing these two family of devices we found out that due to the poor quality of current large area CVD graphene, which is essential for THz applications, graphene-metal hybrid approaches can achieve much better quality-factors and stronger electromagnetic responses. The carrier momentum relaxation time in CVD grown samples is usually in the range of 100 fs and that leads to Q smaller than 1. However, since the electromagnetic properties of graphene-metal hybrid metamaterials are set by the metal pattern, by using this strategy and use graphene only as a reconfigure medium within the metamaterial, we can achieve much larger quality factors. However, by using this strategy the tunability range on the device is reduced. We demonstrated that there is a trade-off between response strength and tunability.

## 8.2 Other 2D materials beyond graphene

We analyzed 2D materials beyond graphene for THz metamaterial applications. In graphene, due to its Dirac band structure, there is a finite minimum conductivity. Therefore, there is always loss in these devices. By using other 2D materials, having a band gap, we can achieve zero conductivity in the off state. Thus, we can decrease the insertion loss by realizing devices with such a zero off state conductivity. Also, due to the minimum finite absorption in graphene, the active layer cannot be put in close proximity of the metallic pattern. From this perspective, we are losing the possibility of placing the active material on the regions of the device where maximum field enhancement takes place. However, due to the smaller maximum conductivity achievable in $MoS_2$, with respect to that in graphene, the modulation depth observed in our metamaterial devices was smaller than that in graphene-based structures even by using multi-layer $MoS_2$ and by putting the active layer right next to the metamaterial surface.

## 8.3 The power of THz spectroscopy as a technique for studying materials

We also discussed THz spectroscopy as a tool for characterizing 2D materials and TCOs. One of the structures that we studied is the heterostructure between NTO and STO, two complex oxide materials. In this heterostructure, a 2DEG is formed at the interface. We performed a systematic study across samples grown under different conditions, and compared the mobility extracted using DC electrical characterization tools with that extracted from THz measurements. We found that the transport properties of these 2DEGs, as extracted from electrical measurement, are affected by extended effects such as those arising from point defects and dislocations, which lead to a decrease in conductivity of the 2DEG with respect to what is extracted from THz measurements. Therefore, THz spectroscopy is a better estimate for the nanoscale transport properties of these materials. Furthermore, we employed THz spectroscopy to characterize a transparent oxide film (BSO) with record high conductivity (at room temperature and among this family of materials). We extracted the refractive indices, and from the extracted properties we designed and fabricated a THz polarizer that is transparent at visible wavelengths. Our results, show that BSO is an attractive material for the design of THz devices that are transparent in the visible and near-IR range.

## 8.4 Future works

To follow-up our theoretical studies on graphene-based modulators, we demonstrated an alternative geometry to the two discussed in this thesis, which can also improve the phase modulation while keeping a small unit cell to wavelength ratio. Our simulations predict that the proposed device can achieve 2X larger phase modulation than other semiconductor-based THz phase modulators reported in the literature. Moreover, this device can achieve the same transmission amplitude thus same loss level as our pervious proposed MSRR and MSR structures.

In Fig.1.a, a detail of the unit cell is shown; graphene is placed in the center of the unit cell where the metal-arms meet. Furthermore, an optical microscopy image of a fabricated sample is shown in Fig 1.b, although the device is not complete. Finally, in Fig.1.c the simulated amplitude and phase of transmission for different graphene conductivity levels is shown, which indicates a resonance at 0.6THz and ~90° phase modulation [1].

Moreover, we also performed modeling of a beam steerer, by assuming an array of antennae creating a phase gradient in accordance with what is possible in these devices, and same unit cell size. This enabled us to analyze the effect of the unit-cell to wavelength ratio on beam steering performance, thus to demonstrate why it is good to utilize deeply-scaled metamaterials. Fig. 2 depicts the directivity for an infinite antenna array, for different unit-cell to wavelength ratios, the phase gradient in each case is design so to achieve a 45° angle in transmission, and the incident wave is a normally incident plane wave. As demonstrated, with decrease in the unit-cell to wavelength ratio, the array becomes more directive for the designed angle. To improve the model, we can consider the radiation pattern for each element, here the unit cell of the phase modulator. In this preliminary study, we assumed an omnidirectional radiation pattern for each antenna.

The next step is to fabricate and test the devices. In addition, there is also room to come-up with better metamaterial structure geometries that could achieve larger phase modulation as well as higher amplitude of transmission. Either geometric optimization or an exploration of the entire design space by means of random geometry simulations, as performed in Ref. [2] could be employed for this purpose.

In terms of transparent devices, the next step is to design and fabricate a metamaterial based on BSO and to integrate it with active 2D materials such as graphene or TMDCs so to achieve a reconfigurable THz device that is fully transparent to visible and near-IR wavelengths. These devices could find multiple applications in THz systems. For example, as indicated in Fig. 3, if we fabricate the same split ring resonator structure as discussed in Ref. [3] but substitute the gold (metallic part) with 200nm of BSO, then we will obtain a very similar electromagnetic response but in a device being fully transparent. By transferring directly growing $MoS_2$ or $MoTe_2$ (which was found by other of our group members, Prashanth Gopalan, to enable a THz conductivity swing similar to that in graphene by means of optically pumping), we can demonstrate a THz active filter that is fully transparent. This idea can be also used for amplitude modulation, phase modulation, active focusing, and beam steering. We can develop a complete set of devices, with a full array of functionalities by integrating highly-conductive TCOs (as transparent replacements of traditional metals) with 2D materials (as effectively transparent replacements of traditional semiconductors). Furthermore, we are also interested in employing dielectric metamaterials and to combine these with active 2D materials. As first experimental demonstrations of all-dielectric metamaterials at THz frequencies were reported, e.g. [4], the interest in developing these devices for different application has risen. By using dielectric metamaterials, we can overcome the losses arising from using metallic structures, furthermore, these devices could be transparent to other target wavelengths. In this regard, our first goal is to integrate a dielectric metamaterial structure with a reconfigurable medium, e.g. graphene, so to make a reconfigurable device. In order to show the feasibility of this approach, we performed a simulation study so to find the best dimensions of the dielectric geometrical features to achieve maximum absorption modulation. We studied the same cylinder geometry as reported in [4], but employing undoped Si and just looking at the field distribution. In this study, we tried to maximize the electric field at resonance on the top facet of the cylinder. Figure 4 shows a simulation for these silicon cylinder metamaterials integrated with graphene, which is placed on top of the cylinder and whose conductivity is varied. Simulations were performed for two different conductivity levels in graphene. The values employed in this simulation are consistent with those we can typically achieve in CVD graphene. Our results show that by using this approach we can modulate the absorption in the device, from ~20 to ~90% by means of altering the conductivity of graphene. Whereas previous studies showed that dielectric metamaterials can be efficient wavelength selective absorbers, here we are showing that by integrating these with active semiconductors, we can also construct efficient reconfigurable absorbers. In the inset of Fig. 4 we depict an optical image of a fabricated device (without graphene), these devices are still under fabrication thus yet not complete. However, transmission measurements show resonance features in agreement with our simulations. The next steps in this projects are: (a) to transfer graphene in top of the Si cylinder array, (b) to construct a series of samples of different graphene conductivity by means of chemical doping, (c) to measure these and find the THz transmission as a function of graphene conductivity, (d) to check whether the results agree with our simulations, (e) to explore other 2D materials beyond graphene and experimentally demonstrate active tuning of absorption. Finally, it is worth mentioning that many papers have reported on graphene-metal hybrid metamaterials and graphene-only plasmonic metamaterials at THz wavelengths, however, to the best of our knowledge, in the literature there is a complete lack on experimental studies on graphene with all-dielectric metamaterials. Our studies will constitute some of the first demonstrations of these devices and will show that this is also an efficient approach leading to enhance light-matter interaction in graphene.

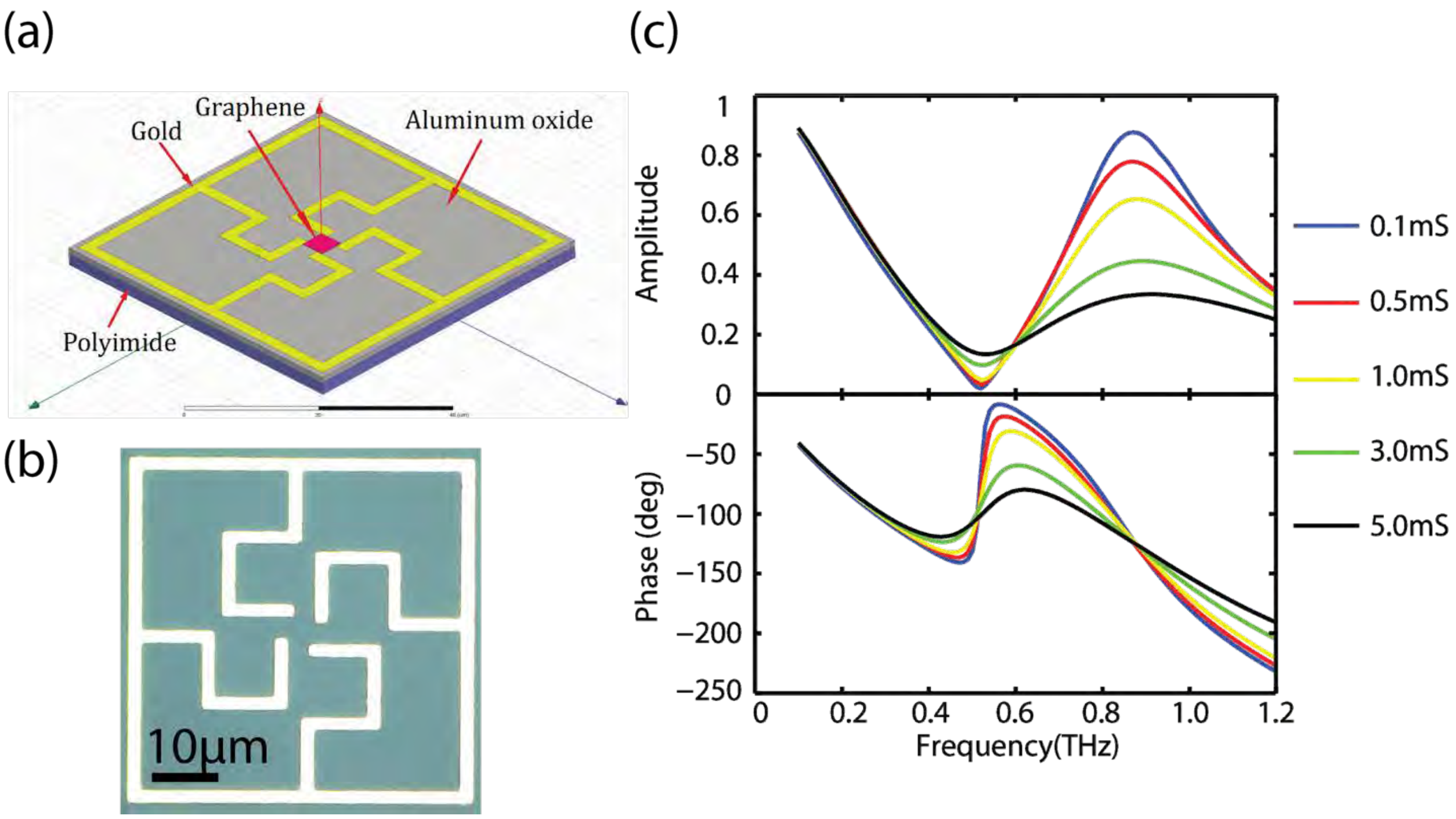

Fig 1. (a) Unit cell structure of the proposed phase modulator, (b) Preliminary work, fabricated unit cell structure, (c) the amplitude and phase of transmission vs. frequency for different conductivity of the graphene sheet.

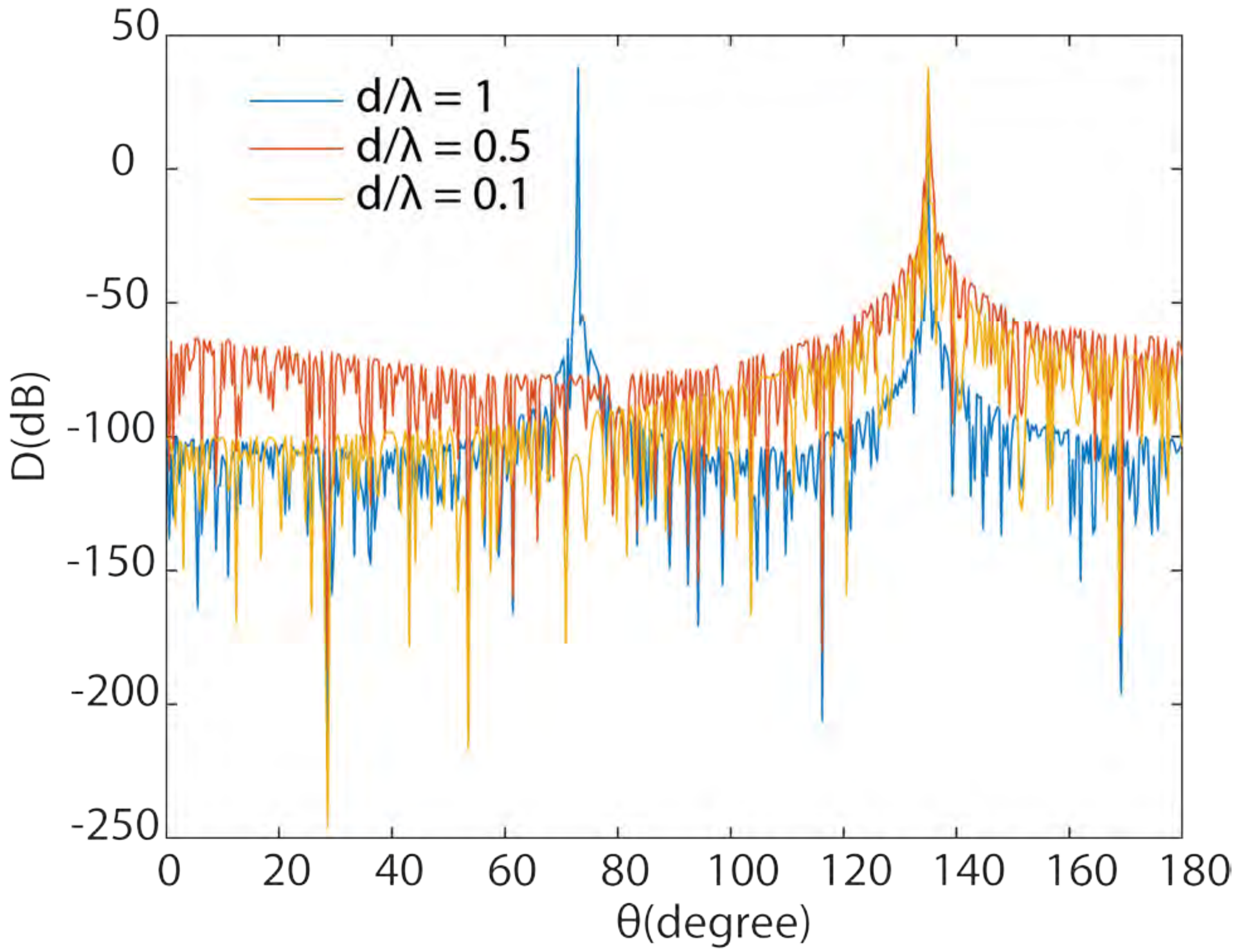

Fig 2. The directivity of an array of omnidirectional antenna with different unit-cell to wavelength ratio. The array is designed so to have the transmission direction at 45°. As indicated, as the ratio becomes smaller the array become more directive.

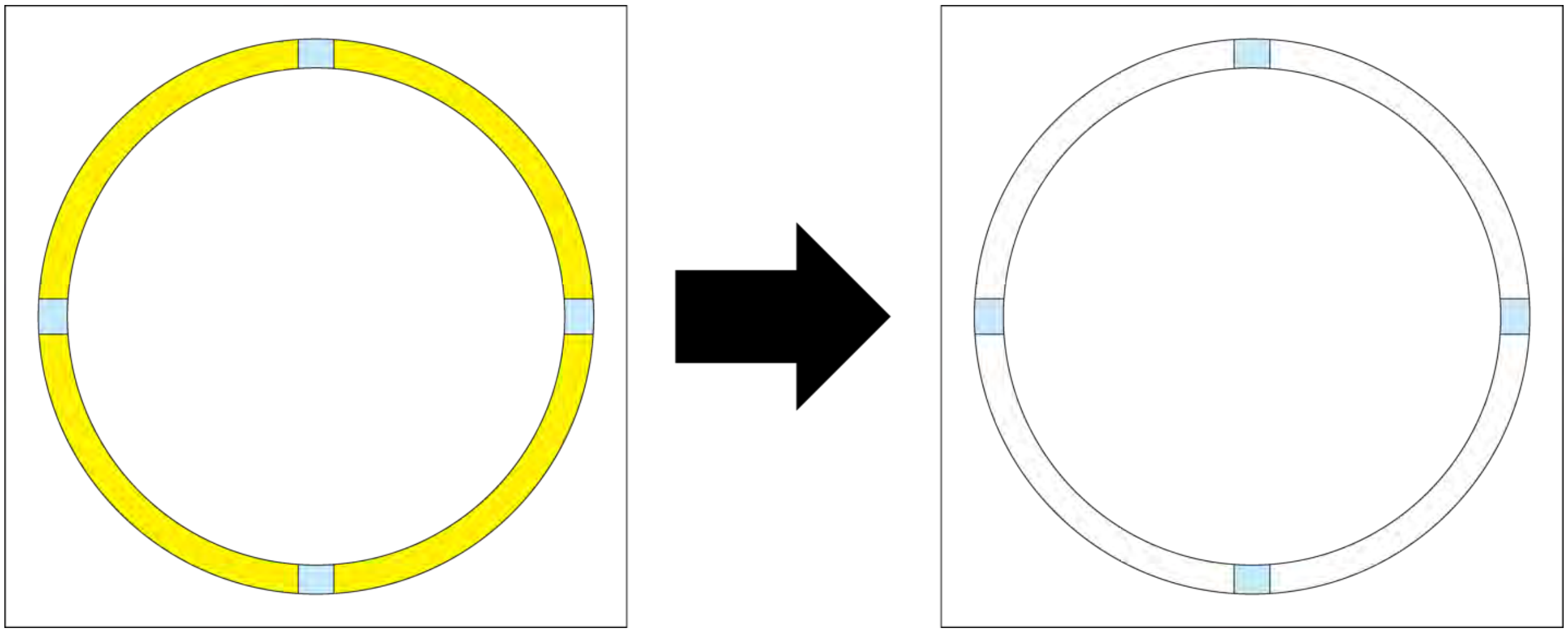

Fig 3. A metal-graphene hybrid metamaterial that operates as a THz filter, from Ref. [3], is shown on the left, gold is used as metal shown in yellow and graphene in the gaps is shown in light blue. Graphene due to its single layer nature is effectively transparent. In the left we show our proposed structure in which the gold is substituted with transparent BSO, thus the final device will be transparent in the visible but operate as reconfigurable filter at THz wavelengths.

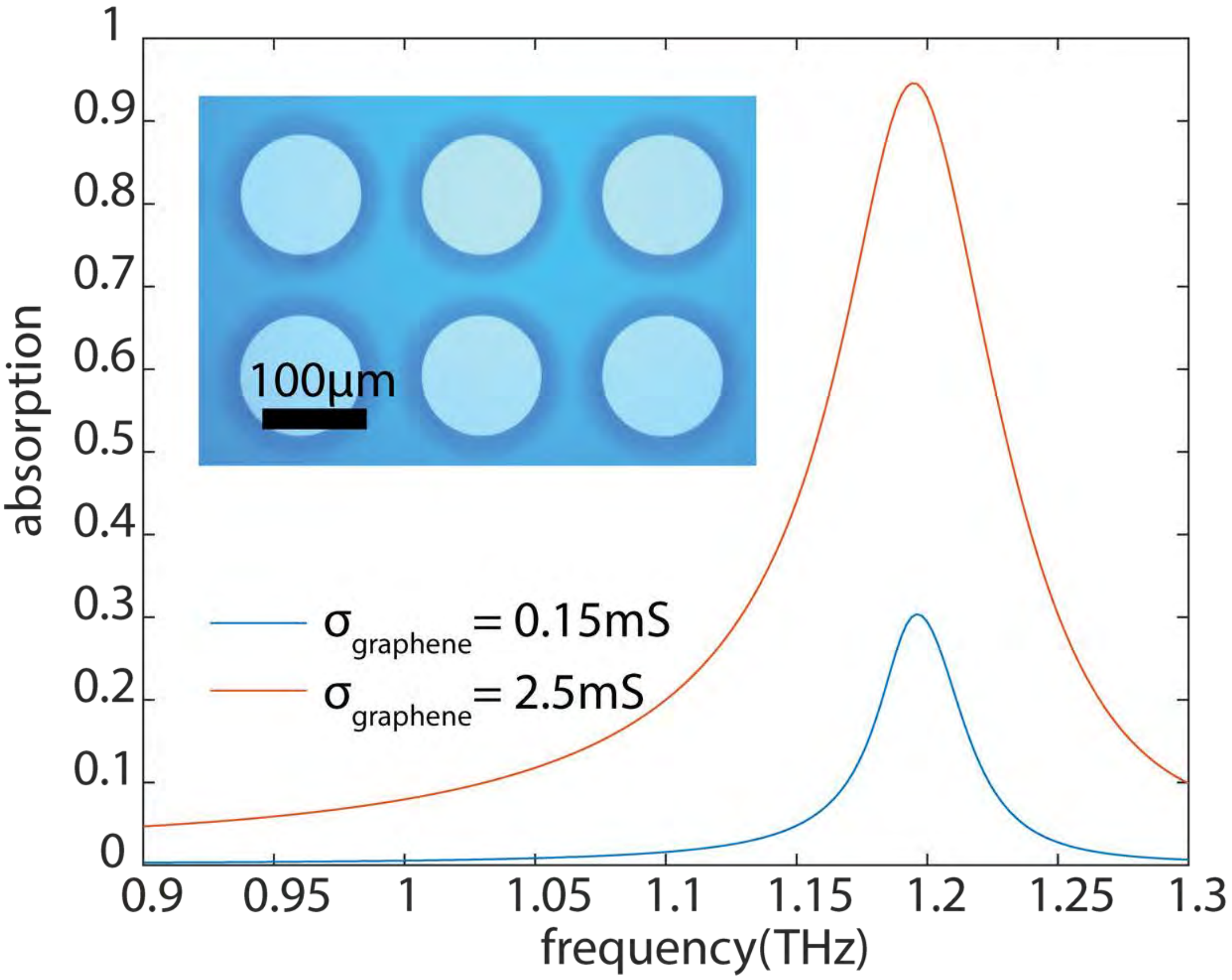

Fig. 4. The simulated absorption for a dielectric metamaterial structure which is integrated with a single layer of graphene placed on top of Si-cylinders. Simulations where carried out for two different graphene conductivities. The inset shows an optical image of the fabricated structure.

APPENDIX I

SUPPLEMENTARY INFORMATION:

GRAPHENE-BASED RECONFIGURABLE TERAHERTZ PLASMONICS AND METAMATERIALS

CONTENTS:

H. Continuous-Wave (CW) terahertz spectroscopy system

FIGURE S8 (*page 15*)

**A. Derivation of the analytical expressions for $\omega_p$, $E$, and $Q$.**

By substituting Eqn. (1) into Eqn. (3) -from the main manuscript-, the following equation is obtained:

$$\frac{T}{T_0} = \frac{1}{\left|1+\frac{Z_0}{1+\sqrt{\varepsilon_S}}F\frac{\sigma_{graphene}(\omega)}{1+\frac{\pi}{2d\varepsilon_0(1+\varepsilon_S)i\omega}\sigma_{graphene}(\omega)}\right|^2}\ . \quad \text{(R1)}$$

Let us introduce the following two parameters ($\alpha$ and $\beta$) so to simplify the notation:

$$\alpha = \frac{Z_0 F}{1+\sqrt{\varepsilon_S}}, \quad \text{(R2)}$$

$$\beta = \frac{\pi}{2d\varepsilon_0(1+\varepsilon_S)}. \quad \text{(R3)}$$

Equation (R1) can thus be rewritten as:

$$\frac{T}{T_0} = \frac{1}{\left|1+\alpha\frac{\sigma_{graphene}(\omega)}{1+\frac{\beta}{i\omega}\sigma_{graphene}(\omega)}\right|^2}. \quad \text{(R4)}$$

Let us define: $f(\omega) = T/T_0(\omega)$, and $g(\omega) = 1/f(\omega)$.

At the plasmonic resonance frequency $f(\omega)$ exhibits a minimum, buy alternatively, when looking at $g(\omega)$, $g(\omega)$ should exhibit a maximum.

From Eqn. (R4), $g(\omega)$ can be expressed as:

$$g(\omega) = \left|1+\alpha\frac{\sigma_{graphene}(\omega)}{1+\frac{\beta}{i\omega}\sigma_{graphene}(\omega)}\right|^2. \quad \text{(R5)}$$

At this stage let us employ Eqn. (2) from the main manuscript and substitute accordingly in Eqn. (R5). By doing this we obtain:

$$g(\omega) = \left|1+\alpha\frac{\frac{\sigma_{DC}}{1+i\omega\tau}}{1+\beta\frac{\sigma_{DC}}{i\omega(1+i\omega\tau)}}\right|^2, \quad \text{(R6)}$$

which can be re-written as:

$$g(\omega) = 1 + \alpha\sigma_{DC}(\alpha\sigma_{DC} + 2)\frac{\omega^2}{(\beta\sigma_{DC} - \omega^2\tau)^2 + \omega^2}. \quad \text{(R7)}$$

Let us define $x = \omega^2$, and take the derivative of $g(x)$ with respect to $x$:

$$\frac{\partial g(x)}{\partial x} = \alpha\sigma_{DC}(\alpha\sigma_{DC} + 2)\left(\frac{1}{[(\beta\sigma_{DC} - x\tau)^2 + x]} - \frac{x[1 - 2\tau(\beta\sigma_{DC} - x\tau)]}{[(\beta\sigma_{DC} - x\tau)^2 + x]^2}\right). \quad \text{(R8)}$$

By looking at the zeros of Eqn. (R8), one can find the value of $x$, $x_0$, at which $g(x)$ exhibits its minimum. It is observed that:

$$(\beta\sigma_{DC} - x_0\tau)^2 + x_0 - x_0(1 - 2\tau(\beta\sigma_{DC} - x_0\tau) = 0, \quad \text{(R9)}$$

so:

$$x_0 = \frac{\beta\sigma_{DC}}{\tau}. \quad \text{(R10)}$$

Since $x = \omega^2$ , and because of the definition of $\beta$ (Eqn. (R3)), the plasmonic resonance frequency is thus given by:

$$\omega_p = \sqrt{\frac{\pi\sigma_{DC}}{2d\varepsilon_0(1 + \varepsilon_s)\tau}}. \quad \text{(4)}$$

This demonstrates Eqn. (4) from the manuscript main text.

At this point let us calculate for $E$:

$$E = 1 - \frac{T}{T_0}\bigg|_{@\omega_p} = 1 - \frac{1}{g(\omega_p)} \quad \text{(R11)}$$

Because of Eqn. (R10), we know that at $\omega = \omega_p$: $\beta\sigma_{DC} - \omega_p^2\tau = 0$

Therefore, by inspecting Eqn. (R7):

$$g(\omega_p) = 1 + \alpha\sigma_{DC}(\alpha\sigma_{DC} + 2)\frac{\omega_p^2}{(\beta\sigma_{DC} - \omega_p^2\tau)^2 + \omega_p^2} = 1 + \alpha\sigma_{DC}(\alpha\sigma_{DC} + 2) = (\alpha\sigma_{DC} + 1)^2 \quad \text{(R12)}$$

$E$ can be now calculated as:

$$E = 1 - \frac{1}{g(\omega_p)} = 1 - \frac{1}{(\alpha\sigma_{DC} + 1)^2}. \quad \text{(R13)}$$

By using the definition of $\alpha$ (Eqn. (R2)), it results:

$$E = 1 - \frac{1}{\left|1+\frac{Z_0 F \sigma_{DC}}{1+\sqrt{\varepsilon_S}}\right|^2}. \qquad (5)$$

This demonstrates Eqn. (5) from the manuscript main text.

So to determine $Q$ it is necessary to find the frequencies that satisfy the condition below:

$$1 - \frac{T}{T_0}\bigg|_{@\omega_1,\omega_2} = \frac{E}{2}. \qquad \text{(R14)}$$

Since:

$$1 - \frac{T}{T_0}\bigg|_{@\omega_1,\omega_2} = 1 - \frac{1}{g(\omega)|_{@\omega_1,\omega_2}} = \frac{E}{2} \quad \Rightarrow \quad g(\omega)|_{@\omega_1,\omega_2} = \frac{2}{2-E}. \qquad \text{(R15)}$$

Since $\beta\sigma_{DC} = \omega_p^2\tau$, we can rewrite $g(\omega)$ as:

$$g(\omega) = \frac{\left(\omega_p^2-\omega^2\right)^2\tau^2+(\alpha\sigma_{DC}+1)^2\omega^2}{\left(\omega_p^2-\omega^2\right)^2\tau^2+\omega^2}. \qquad \text{(R16)}$$

Using Eqn. (5) from the manuscript main text, which we demonstrated a few steps back, we can write: $(\alpha\sigma_{DC}+1)^2 = 1/(1-E)$ and thus by substituting in Eqn. (R16) we obtain:

$$g(\omega) = \frac{\left(\omega_p^2-\omega^2\right)^2\tau^2+\frac{\omega^2}{1-E}}{\left(\omega_p^2-\omega^2\right)^2\tau^2+\omega^2}. \qquad \text{(R17)}$$

Therefore when evaluating at $\omega_1$ and $\omega_2$, the following condition should hold (Eqns. (R15) and R(17)):

$$\frac{\left(\omega_p^2-\omega^2\right)^2\tau^2+\frac{\omega^2}{1-E}}{\left(\omega_p^2-\omega^2\right)^2\tau^2+\omega^2} = \frac{2}{2-E}. \qquad \text{(R18)}$$

By rearranging terms, Eqn. (R18) can be rewritten as:

$$(\omega^2)^2 - \left(2\omega_p^2 + \frac{1}{\tau^2}\frac{1}{1-E}\right)\omega^2 + \omega_p^4 = 0. \qquad \text{(R19)}$$

This equation is of the form: $Ax^2 + Bx + C = 0$, where: $x = \omega^2$.

The sum of the roots is given by:

$$x_1 + x_2 = \omega_1^2 + \omega_2^2 = -\frac{B}{A} = 2\omega p^2 + \frac{1}{\tau^2}\frac{1}{1-E}. \quad \text{(R20)}$$

And the product of roots is given by:

$$x_1 x_2 = \omega_1^2 \omega_2^2 = \frac{C}{A} = \omega_p^4. \quad \text{(R21)}$$

So from Eqns. (R20) and (R21), one can observe that:

$$(\omega_1 - \omega_2)^2 = \omega_1^2 + \omega_2^2 - 2\omega_1\omega_2 = \frac{1}{\tau^2}\frac{1}{1-E}. \quad \text{(R22)}$$

And, therefore:

$$\Delta\omega = |\omega_1 - \omega_2| = \frac{1}{\tau}\frac{1}{\sqrt{1-E}}. \quad \text{(R23)}$$

Hence, the quality factor ($Q$) is given by:

$$Q = \frac{\omega_p}{\Delta\omega} = \omega_p \tau \sqrt{1-E}. \quad (6)$$

This demonstrates Eqn. (6) from the manuscript main text.

It is worth mentioning that Eqns. (5) and (6) are general and hold for any graphene pattern of convex geometry, i.e. squares, rectangles, triangles, etc.

## B. General contour plots for $\omega_p$, $E$, and $Q$.

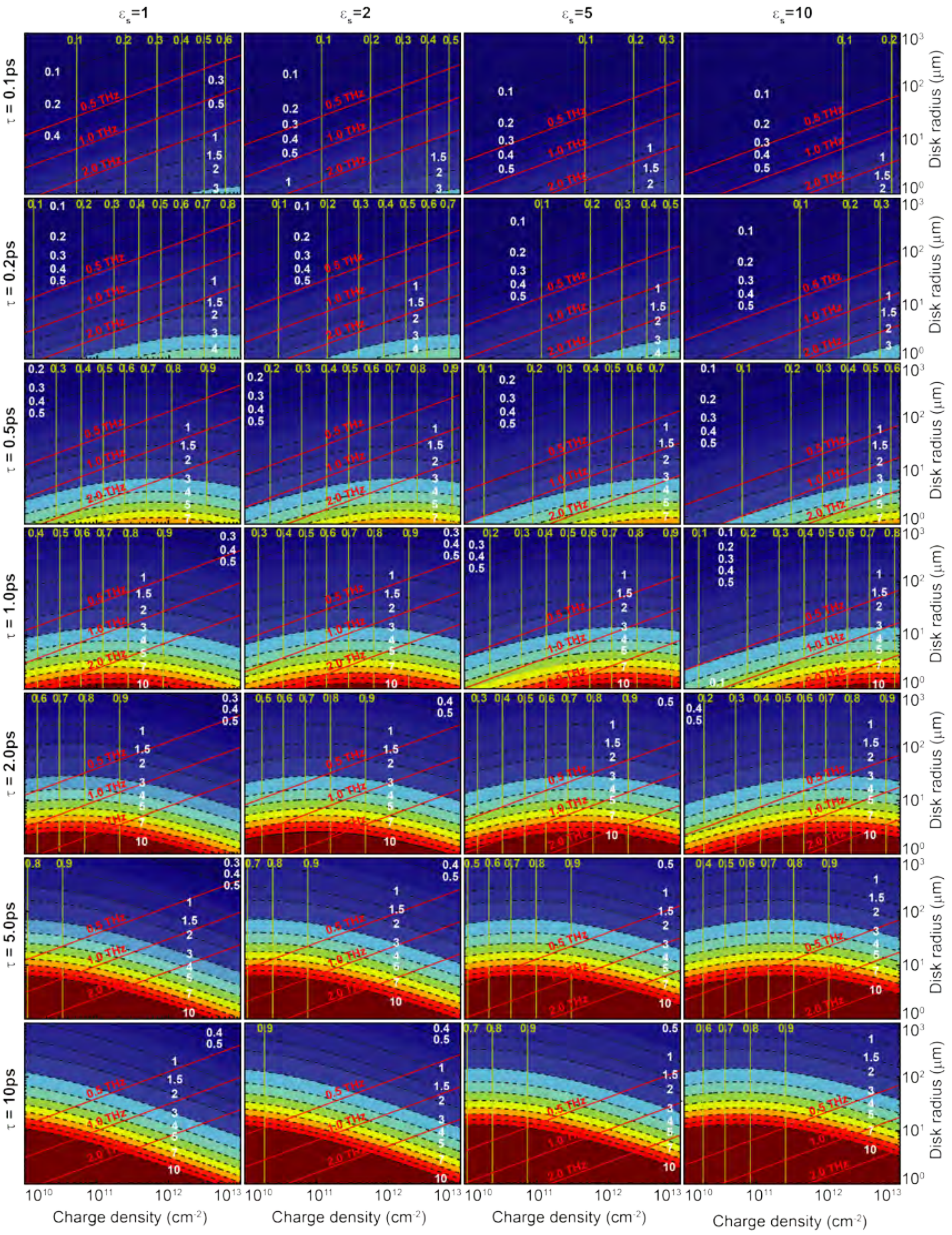

**FIGURE S1.** Contour plots of $Q$ (filled), $E$ (yellow-traces), and $\omega_p$ (red-traces) as a function of charge density and disk radius for different values of $\tau$ and $\varepsilon_s$.

## C. Equivalent Transmission Line Model for the SRR-based graphene/metal hybrid structure

The analyzed SRR-based geometry (*Sample Set #2*), can be modelled using the equivalent circuit model described by **Fig. 2(b)** in the main text. Following the discussion in Ref. [18], graphene is modeled as an impedance of value $Z_g = R_g + i\omega L_g$, where $R_g$ and $L_g$ represent its associated resistance and inductance, respectively, as shown in **Fig. S2**.

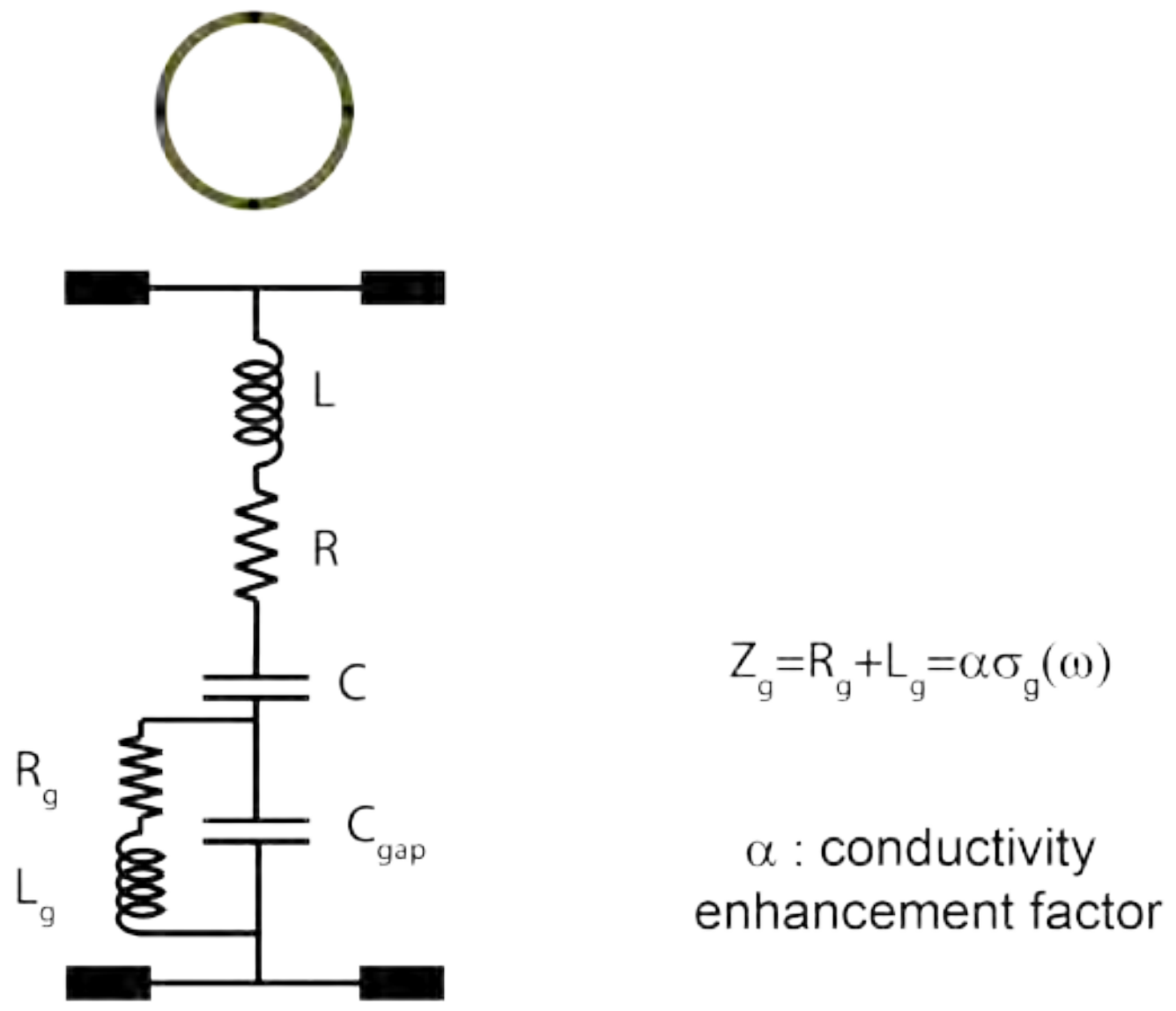

**FIGURE S2.** Equivalent circuit model for the structure analyzed in *Sample Set #2*.

The transmission can be calculated using the following formula:

$$T = \left|\frac{2Z_{in}}{Z_{in}+Z_0}\right|^2.$$

Where $Z_{in}$ represents the input impedance, i.e. $Z_{in} = \{R+i\omega L+1/i\omega C\}//Z_0$. Moreover, as discussed in the main manuscript, $L$ represents the self-inductance of the metal loop, $R$ the metal losses, and the capacitor $C$ results from the separation between adjacent unit cells. Therefore, for the equivalent circuit depicted in the middle panel of **Fig. 2(b)** of the main text, which consists of a closed metallic loop, the transmission is given by:

$$T = \left| \frac{2}{2+\frac{Z_0}{R+i\omega L+\frac{1}{i\omega C}}} \right|^2 .$$

So by fitting the simulated data to this model we find $R = 0\ \Omega$, $C = 1.34$ fF, and $L = 122$ pH. Since this simulation was performed assuming lossless materials (PEC), we expect the resistance to be zero. Moreover, the capacitance and inductance set the resonance-frequency to be at $f = 1/2\pi\sqrt{LC}$ ~ 0.4 THz.

When gaps are added, the capacitance of the structure is increased; the gaps insert a series capacitance $C_{gap}$. We can find the transmission using the following equation, which is based on the model depicted in the right panel of **Fig. 2(b)** on the main text:

$$T = \left| \frac{2}{2+\frac{Z_0}{R+i\omega L+\frac{1}{i\omega C}+\frac{1}{i\omega C_{gap}}}} \right|^2 .$$

By fitting our simulation results to this model, and by using the R, L, C parameters previously found, we find that $C_{gap} = 0.83$ fF. In this case, the resonance corresponds to at $f = 1/2\pi\sqrt{L(C.C_{gap})/(C+C_{gap})}$ ~ 0.65 THz.

When we add graphene into the gaps, the structure can be modeled using the equivalent circuit depicted in the left panel of **Fig. 2(b)** on the main text. The transmittance through the structure is given by:

$$T = \left| \frac{2}{2+\frac{Z_0}{R+i\omega L+\frac{1}{i\omega C}+\frac{R_g+i\omega L_g}{i\omega C_{gap}(R_g+i\omega L_g)+1}}} \right|^2 ,$$

thus:

$$T = \left| \frac{2}{2 + \frac{Z_0}{R + i\omega L + \frac{1}{i\omega C} + \frac{1/\alpha\sigma_{graphene}}{i\omega C_{gap}/\alpha\sigma_{graphene} + 1}}} \right|^2 .$$

In the latter formula, α is a parameter which is used to consider the field enhancement effect in graphene as discussed in Ref. [18]. We employ this model to fit the simulated data for different graphene conductivity values to the equivalent circuit model. To simplify the analysis, an electron momentum relation time (τ) equal to zero is assumed in our simulations as well as in the model. The results of the fitting are depicted in **Fig. S3(a)**. It is observed that the equivalent circuit model is capable of accurately representing the dynamic behavior observed in *Sample Set #2*. Moreover, depicted in **Fig. S3(b)** is a plot of the extracted α versus graphene conductivity. As observed in previous studies, for different metamaterial structures, e.g. Ref. [18], α is found to be independent on the graphene conductivity. In our case we found α ~ 3.4.

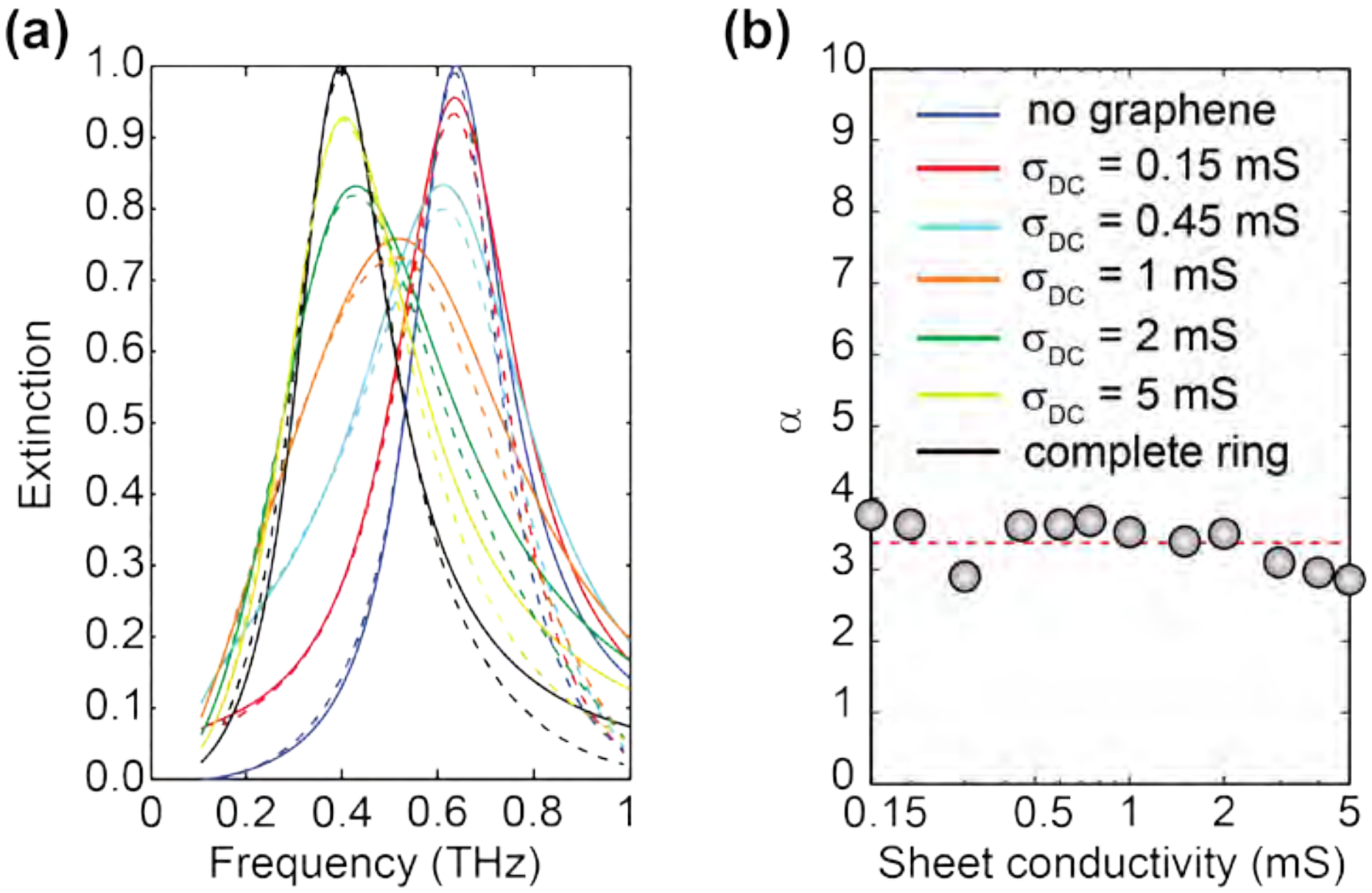

**FIGURE S3. (a)** Simulated (dashed lines) extinction and calculated extinction from the fit to the equivalent circuit model (continuous lines). **(b)** Extracted α for different DC (sheet) conductivity levels in graphene; α ~ 3.4 is found to be independent on the graphene conductivity.

In conclusion, the structure analyzed in Sample Set #2, can be well described by the equivalent circuit model depicted in **Fig. S2** by employing the following parameters: $R$ = 0 Ω, $C$ = 1.34 fF, $L$ = 122 pH, $C_{gap}$ = 0.83 fF, and α ~ 3.4.

### D. Effect of the electron momentum relaxation time on the response of the SRR-based graphene/metal hybrid structures

We performed simulations for the SRR-based graphene/metal hybrid structures (*Sample Set #2*) employing different values for the electron momentum relaxation time (τ). The results of those simulations are depicted in **Fig. S4**. It is observed that for either low graphene conductivity, or high graphene conductivity, the response is (almost) independent of τ. However at moderate conductivities, e.g. 1 mS, deviations start to take place when τ > 100 fs.

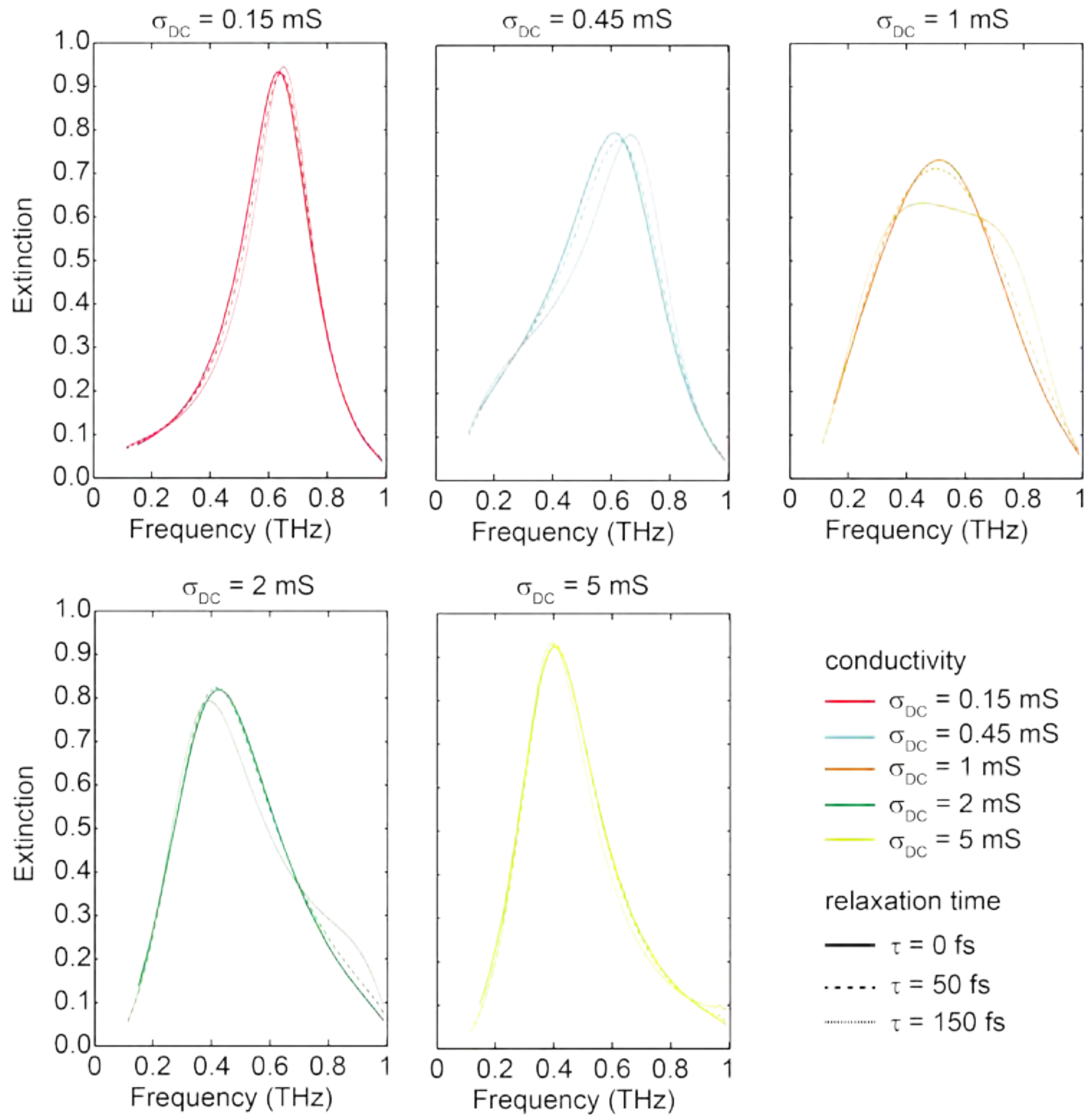

**FIGURE S4.** Simulated extinction for different conductivity of graphene and different electron momentum relaxation times. The solid lines represent τ = 0 fs, the dashed lines τ = 50 fs, and the dotted lines τ = 150 fs, respectively.

## E. Raman spectroscopy of graphene

Raman measurements were conducted employing a WITec (Alpha300S) scanning near field optical microscope (SNOM). These measurements were carried out using a 488 nm linearly polarized excitation source operated in the back scattering configuration. A 20X objective was employed for this measurement, with 1sec integration time and 5mW power on the sample.

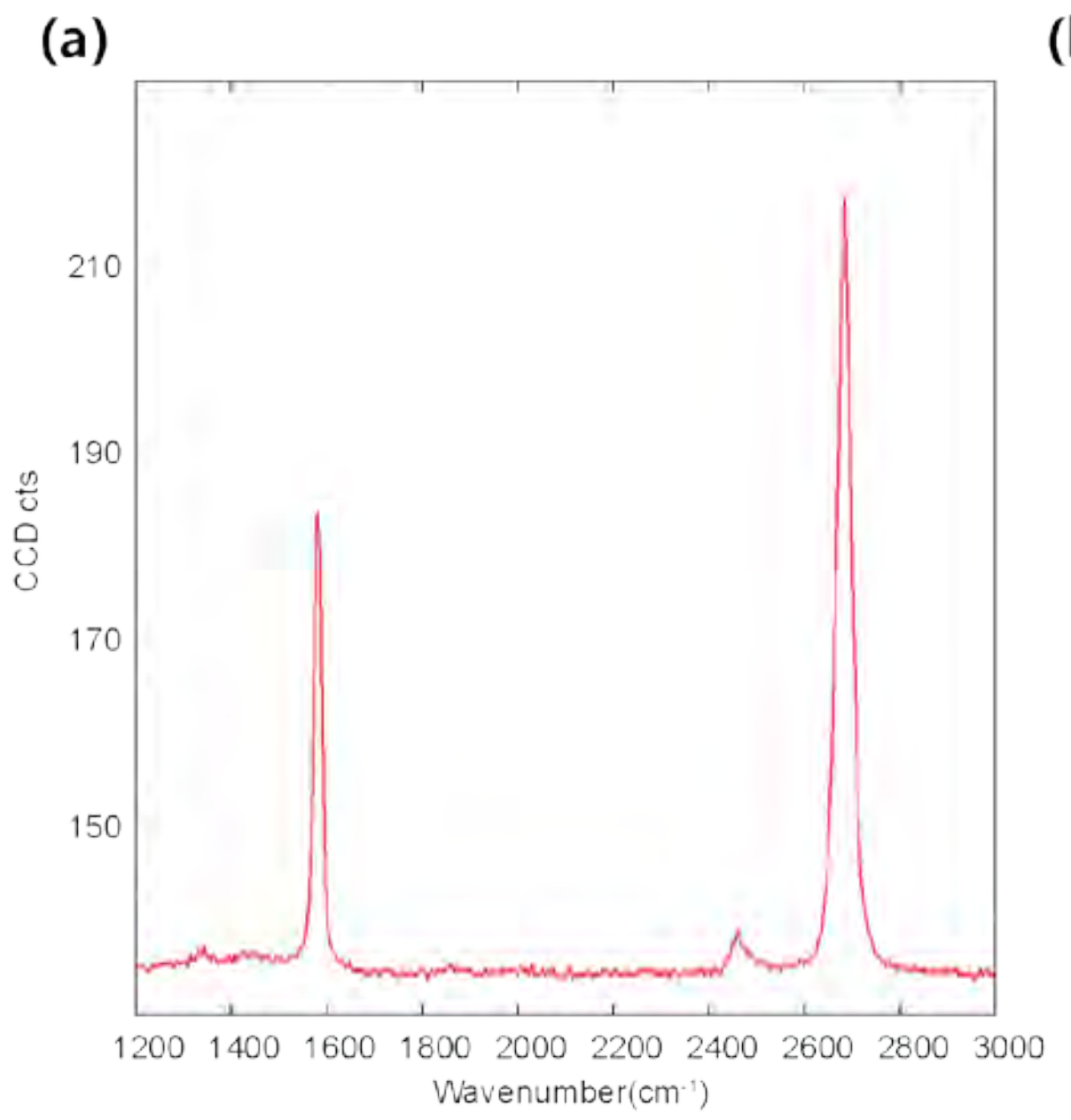

**FIGURE S5. a)** Raman spectroscopy of the commercial graphene transferred on 285 nm $SiO_2$/Si substrate reported by the vendor (Bluestone). **b)** Raman spectroscopy of the one layer transferred film (on PI).

### F. Graphene Drude model parameter extraction

As mentioned in the manuscript body, each sample consists of two adjacent (1cm × 1cm) square regions, containing the structure under test, and an un-patterned graphene control region, respectively. In order to characterize the graphene properties through the control region, two measurements are performed: (*i*) terahertz spectroscopy on the 0.1 to 2 THz spectral range; and (*ii*) FTIR on the 3 to 12 THz spectral range. By using Eqns. (2-3) from the manuscript main text, Eqn. (R24) is obtained, therefore the DC conductivity and momentum relaxation time can be extracted from the transmission measurements by fitting to:

$$\frac{T}{T_0} = \frac{1}{\left|1+\frac{Z_0}{1+\sqrt{\varepsilon_S}}\frac{\sigma_{DC}}{1+i\omega\tau}\right|^2}\,. \qquad \text{(R24)}$$

In Eqn. (R24): $Z_0 = 377\ \Omega$ is the vacuum impedance, and $\varepsilon_s = 3.24$ is the relative permittivity of polyimide [39-40]. Depicted in **Fig. S6** is an example of the fitting, corresponding to *Sample Set #1*, case (*i*).

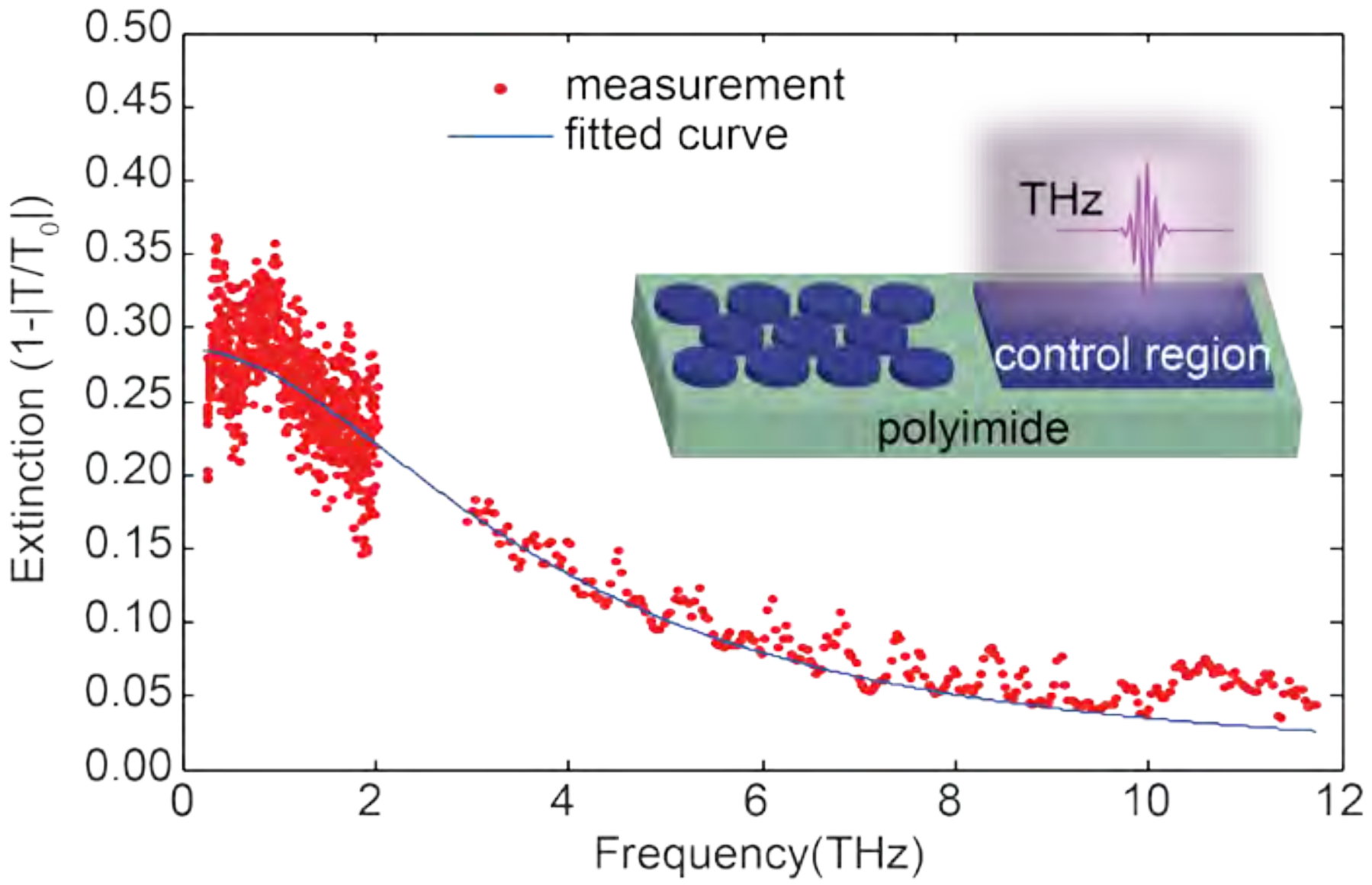

**FIGURE S6.** Measured extinction for a graphene control region as obtained from THz and FTIR measurements, and fit to the analytical expression from where the Drude model parameters are extracted (Eqn. (R24)).

### G. Details of the simulated geometries as set in HFSS

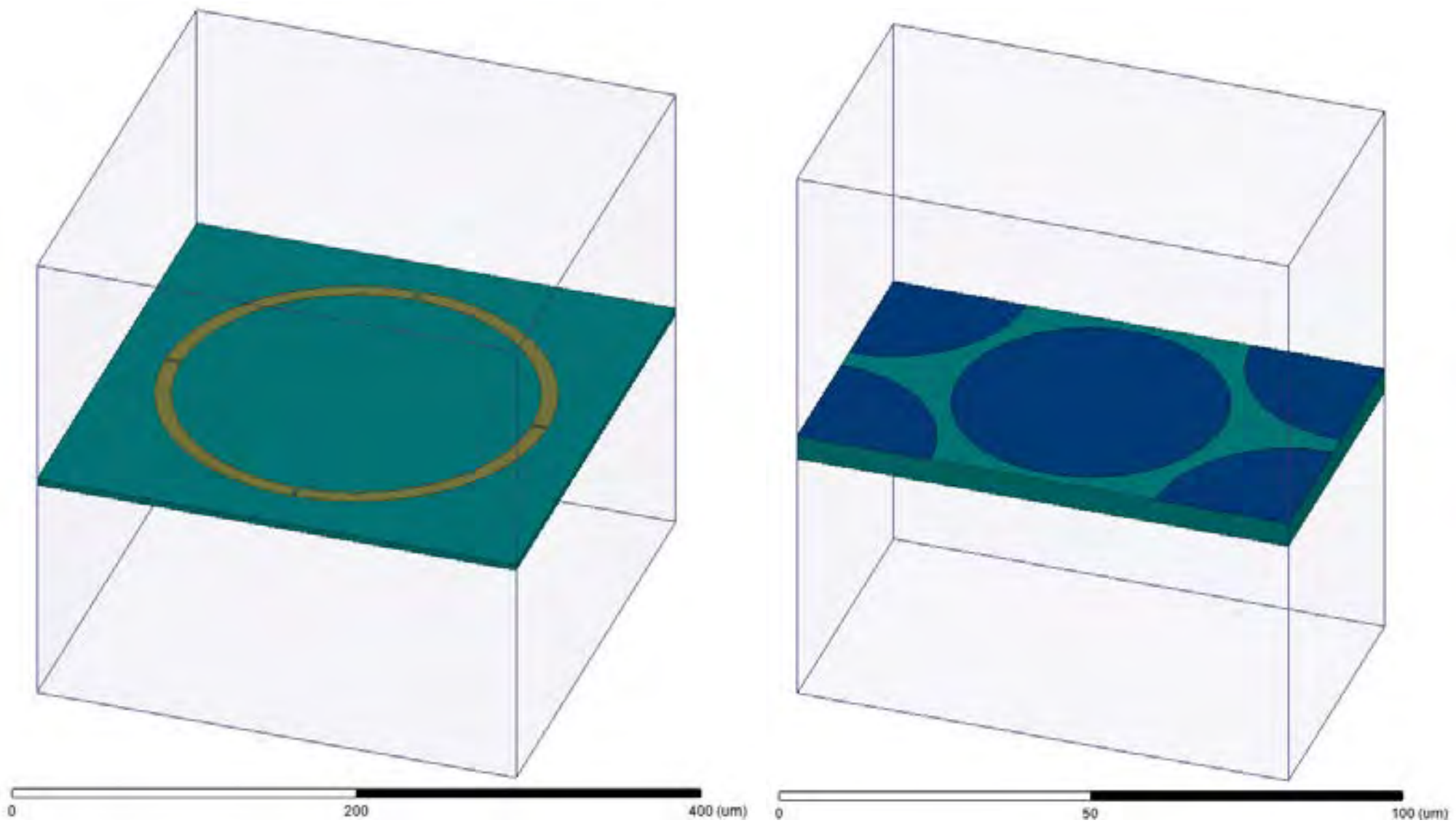

**FIGURE S7.** Detail of the simulation geometries as set in HFSS for (left) the SRR graphene/metal hybrid structure and (right) the graphene-disk plasmonic structure. Periodic boundary conditions were set around the unit cells.

**H. Continuous-Wave (CW) terahertz spectroscopy system**

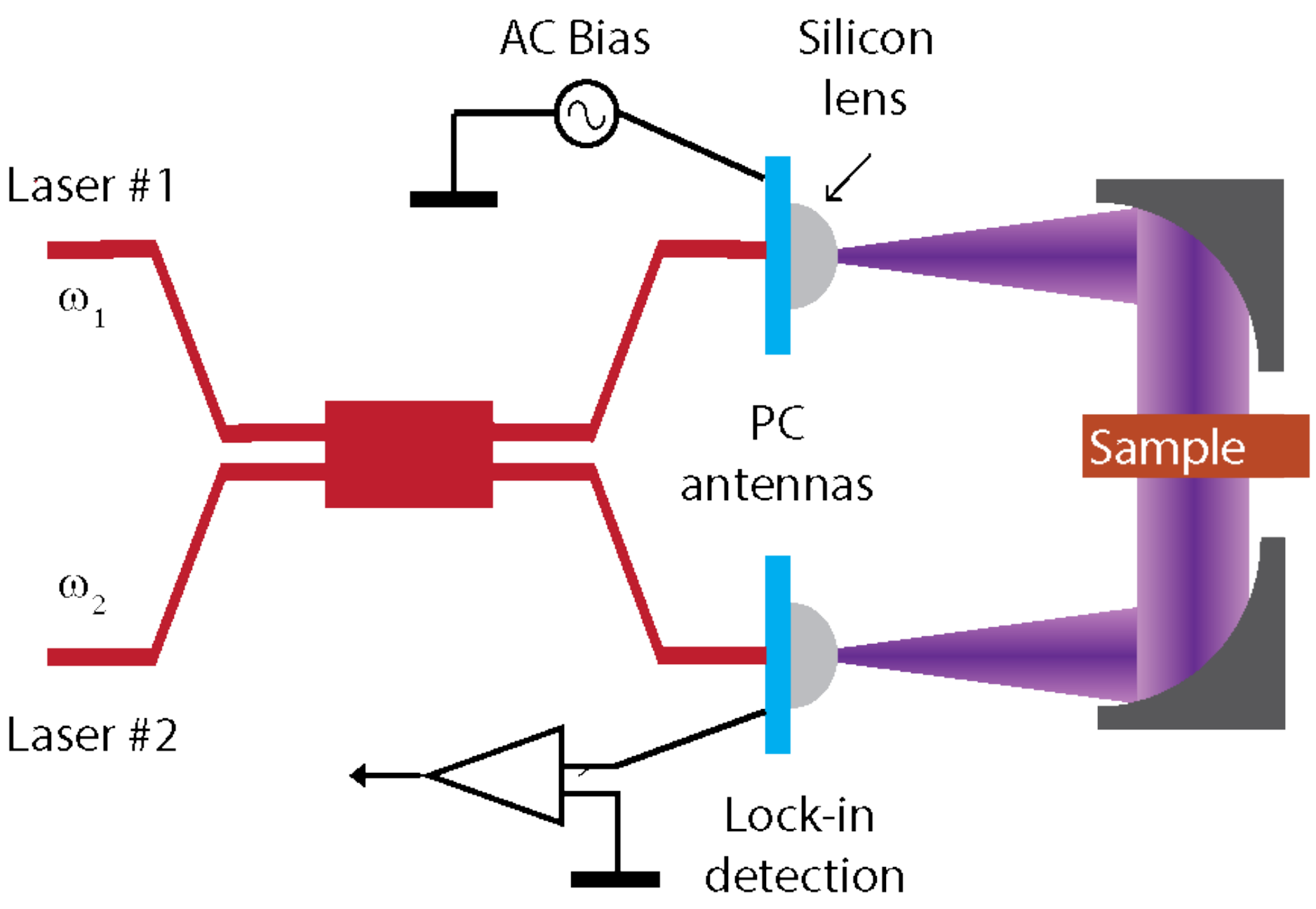

**FIGURE S8.** Schematic diagram of the employed CW THz spectroscopy system.

# APPENDIX II

# FABRICATION TECHNIQUES

## B.1 Wafer cleaning

For wafer cleaning two different process was employed. For cleaning sample that goes into the furnace for oxide or nitrite growth or deposition RCA cleaning procedure is performed. For lithography or polymer deposition nanostrip cleaning process is used.

### B.1.1. RCA cleaning recipe

The RCA cleaning procedure start by RCA2 and then followed by RCA1 cleaning procedure. For RCA 2 procedure the sample is placed in the RCA 2 solution, at 70±5˚C for 10 min. Then followed by 5min DI water rinse. For preparing the RCA 2 solution, hydrogen chloride(HCl) is diluted with water, 6part water, 1part HCl and heat up on a chemical hot plate. As the temperature get to 70˚C 1part hydrogen peroxide(H2O2) is added, and after one or two minutes the solution start making bubbles and it is ready to use. At this temperature the solution is only good to use for half an hour. After the procedure done, solation should be diluted with water and cools down to near room temperature, then pour down it into the drain and all lab wares should be rinse three times with DI water.

For RCA 1 procedure, sample should be placed in the solution for 10 min and rinsed with DI water for 5min afterward and in the end blow dry. For preparing RCA 1 solution, ammonium hydroxide (NH4OH) 1 part with 5 part water is diluted and heat up to 70±5˚C. When the solution reach the temperature1 part hydrogen peroxide(H2O2) is added and the RCA1 solution is ready to use. The similar procedure is done for cleaning RCA 1.

### B.1.2 Nanostrip cleaning

There is a dedicated tank for nanotstrip cleaning. For cleaning, the sample is placed in the nanostrip solution at 80˚C for 10 min, then rinsed with DI water for 5min and blow dry with N2.

B.2 Polyimide substrate

The polyimide substrate was spin coated and cured and after the process finished peeled off. Here we used PI2611 which is customized for polyimide layer with thickness of 2-11µm. The recipe below is used for usual polyimide layer

- After cleaning substrate (usually glass or silicon) dehydrate at 150˚C for 10 min
- Spin coat with PI 2611(note: it is very viscous)
    - Dispense at 500rpm with ramp 1000r/s for 5sec
    - Spin at 5000rpm with ramp 1000r/s for 35sec
- Soft bake first at 90˚C for 90sec the at 150˚C for 90sec
- Curing at 300˚C for 30min, the temperature is increased gradually from room temperature with ramp of 4˚C/min and cooled down to room temperature with 6˚C/min ramp

The film thickness will be between 3.5-4.5µm. The solution become more viscous and the thickness increases during the time, as the bottle opens. The film thickness is increases in the edges of the sample.

B.3 Photolithography

The lithography is done using two different type of photoresist and for each type the recipes are mentioned. The positive photoresist was the most popular one since it needs less preparation steps but for certain application negative photoresist was needed.

B.3.1 positive photoresist

Two types of positive photoresist were used in the cleanroom, S1813 and 9260. S1813 is used for thinner resist application and 9260 for thick applications.

B.3.1.1. Shiply 1813

I used this resist with two different thickness goal, very thin films and thin films. Very thin photoresist film is needed when the rabbit ear problem is bold in the target application.

The recipe below was used for very thin photoresist film (less than 1.2µm)

- after clean up the wafer, dehydration at 150 ˚C for 10 min
- Spin coating with S1813
  - Dispense at 500rpm with ramp 250r/s for 2 sec
  - Spin at 6000rpm with ramp 1000r/s, for 60sec
- Soft-bake at 110˚C for 90sec
- Exposure at SUSS for 4.5sec (lamp intensity 6.2mW/cm$^2$)
- Develop in AZ 300MIF for 30sec

The recipe below was used for thin photoresist (around 2µm)

- after clean up the wafer, dehydration at 150 ˚C for 10 min
- Spin coating with S1813
  - Dispense at 500rpm with ramp 250r/s for 2 sec
  - Spin at 3000rpm with ramp 1000r/s, for 60sec
- Soft-bake at 110˚C for 90sec
- Exposure at SUSS for 7.5sec (lamp intensity 5.3mW/cm$^2$)
- Develop in AZ 300MIF for 30sec

B.3.1.2. AZ 9260

This photoresist is used for achieving thick photoresist film from 5-20µm. The recipe below is used for patterning.

- after clean up the wafer, dehydration at 150 ˚C for 10 min

- Spin HMDS
  - Dispense at 500rpm with ramp 250r/s for 2 sec
  - Spin at 2400rpm with ramp 1000r/s for 60 sec
- Spin coating with AZ 9260
  - Dispense at 500rpm with ramp 250r/s for 2 sec
  - Spin at 2400rpm with ramp 1000r/s for 60sec
- Soft-bake at 110˚C for 165sec
- Exposure at SUSS for 12sec, 6 times (6.2 mW/cm$^2$)
- Develop in AZ 400 MIF (dilute with DI water 1:4) for 120sec

#### B.3.2 negative photoresist

The negative photoresist that is used in the clean room is AZ 5214 E. This product can be used both as positive and negative resist. By performing the post exposure bake and the flood exposure this photoresist become image reversal. The recipe below is used for patterning

- after clean up the wafer, dehydration at 150˚C for 10 min
- Spin coating with 5214E
  - Dispense at 500rpm with ramp 250r/s for 2 sec
  - Spin at 5000rpm with ramp 1000r/s for 35sec
- Softbake at 110˚C for 90sec
- Exposure at SUSS for 4.5sec (lamp intensity was 6.2 mW/cm$^2$)
- Post-exposure bake 120˚C for 120sec
- Flood exposure 4.5sec (lamp intensity was 6.2 mW/cm$^2$)
- Develop in AZ 400 MIF (dilute with DI water 1:4) for 120sec

B.4 lift-off

For lift-off after evaporation of 100nm gold, immerse the sample in vessel with acetone, leave for 1-2 minutes, the 30sec of sonication is enough for the lift-off process. For Different types films and thickness maybe more sonication is needed. Also samples sputtered may need more sonication time for lift-off process.

B.5 Graphene transfer

For graphene transfer there is one-day preparation before the actual transfer process. For this purpose, first graphene sample grown on copper foil should be cut into the size that is needed for the transfer, with scissors (cleaned with wipe). Copper foils are very fragile so the whole process should be done with extreme caution. After cutting graphene pieces should be cleaned, with soaking the pieces in to the acetone for 15 min, and after that just dried by putting them in the ambient and let the acetone evaporate. The graphene sample should become as flat as possible, it can be achieved by putting the sample between two clean glass slides and push them together. Afterward sample is pasted to a clean glass slide using kapton tapes (Note: keep the side of the copper foil with the graphene that needs to be used on top). The area that is covered with kapton tape need to be very small since the graphene on those area is wasted, also all the sides should be cover to prevent the spin coat material to reach the other side of the copper foil.

Then graphene sample is spin coated with PMMA (950 PMMA C2) with speed of 500rpm, ramp 500rpm for 60 sec. Next the sample is left under fume hood for one night at least to get dry, we do not want to solidify the PMMA with heating since the flexibility of the PMMA film is essential for the transfer process. Next step is to remove the kapton tape and flip over the sample on to another clean glass slide and use kapton tape again to paste it to the glass slide. In CVD graphene process on copper, graphene is formed on both side of the copper foil. So the graphene film on the unwanted side should be etched by using O2 plasma. The graphene etch recipe in the oxford

80(RIE) tool use the following steps:

- Pump down to $0.9\times10^{-4}$Torr
- Purge N2 with 80sccm for 5 min
- Purge O2 with 90sccm and pressure 25 mTorr and RF power 150W for 15sec
- Purge and vent

After the graphene etching, the sample is ready for transfer. First the copper film should be etched by removing the kapton tape and flip the sample and place the uncovered side of the copper in the copper etchant solution at 40˚C. After all the copper etched, it can be recognized when copper color cannot be seen with eye and the sample is fully transparent. The etching process can take from 15minutes to 1hour, depend on the thickness of the foil and the etchant etch rate. The etch rate of the solution decrease during the time after the bottle is opened. Then using the clean silicon wafer graphene sample is moved from the copper etchant vessel to another vessel filled with DI water, and after 5 minutes most of the copper etchant in the sample get dissolve in DI water. For removing the copper etchant fully, the sample can be moved to another DI water vessel and, but usually the first vessel is enough. Then by merging the desired substrate and putting the graphene sample on the desired spot the transfer process is done. (Note: due to the hydrophobic nature of some sample this step can be tricky). After taking out the sample from DI water and it should be kept at room temperature for at least 1hour, so the water leaves the area between the graphene and sample (to avoid cracks in the graphene sample due to bubbles). Then the sample is heat up to 120˚C for 10 minutes. After that the PMMA layer can washed with merging the sample in the acetone or using a pipette and pouring acetone on the sample.

## B.6 Mask Fabrication

For mask fabrication the Heidelberg MicroPG 101, is used with 12mW power and 50% intensity.

After the exposure, the mask is developed in developer 352, developer 351(diluted in water 1:5), or AZ 1:1 for 30sec. Then the chrome film is etched in the chrome etchant for 90sec to 2 minutes, the etching should be stopped when the mask become completely transparent. After that the photoresist is washed by merging the mask into the acetone and followed by an oxygen plasma for 5 minute.